\pdfoutput=1
\RequirePackage{ifpdf}
\ifpdf % We are running pdfTeX in pdf mode
\documentclass[pdftex]{sigma}
\else
\documentclass{sigma}
\fi

\numberwithin{equation}{section}
\numberwithin{figure}{section}

\newtheorem{Theorem}{Theorem}[section]
\newtheorem{Corollary}[Theorem]{Corollary}
\newtheorem{Lemma}[Theorem]{Lemma}
\newtheorem{Proposition}[Theorem]{Proposition}
{\theoremstyle{definition}

\newtheorem{Remark}[Theorem]{Remark}
}

\newcommand{\e}{\epsilon}

\newcommand{\mc}{\mathcal}

\newcommand{\wt}{\widetilde}
\newcommand{\ov}{\overline}

\newcommand{\lt}{\left}
\newcommand{\rt}{\right}

\newcommand{\LeadsTo}{\Longrightarrow}
\newcommand{\EquivTo}{\Longleftrightarrow}
\newcommand{\MapsTo}{\longmapsto}

\newcommand{\p}{\partial}

\newenvironment{pmat}[1]{\left(\begin{matrix}{#1}}{\end{matrix}\right)}
\newenvironment{vmat}[1]{\left|\begin{matrix}{#1}}{\end{matrix}\right|}

\newcommand{\pM}[4]{\begin{pmat}  #1 & #2 \\ #3 & #4 \end{pmat}}
\newcommand{\vM}[4]{\begin{vmat}  #1 & #2 \\ #3 & #4 \end{vmat}}

\newcommand{\V}[2]{\begin{pmat}  #1 \\ #2\end{pmat}}
\newcommand{\Vin}[2]{\tr{\lt(\begin{array}{@{}c@{\;}c@{}} #1 & #2\end{array}\rt)}}
\newcommand{\M}[4]{\begin{pmat}  #1 & #2 \\ #3 & #4 \end{pmat}}

\newcommand{\bpm}{\begin{pmatrix}}
\newcommand{\epm}{\end{pmatrix}}
\newcommand{\bvm}{\begin{vmatrix}}
\newcommand{\evm}{\end{vmatrix}}

\newcommand{\secref}[1]{Section~\ref{#1}}
\newcommand{\apdref}[1]{Appendix~\ref{#1}}

\newcommand{\thmref}[1]{Theorem~\ref{#1}}
\newcommand{\lemref}[1]{Lemma~\ref{#1}}
\newcommand{\propref}[1]{Proposition~\ref{#1}}

\newcommand{\N}{\mathbb{N}}
\newcommand{\Z}{\mathbb{Z}}
\newcommand{\R}{\mathbb{R}}
\newcommand{\C}{\mathbb{C}}

\newcommand{\Rmax}{\R_{\max}}
\newcommand{\uZ}{\mathrm{u}\Z}
\newcommand{\uR}{\mathrm{u}\R}
\newcommand{\uC}{\mathrm{u}\kern-.2pt\C}
\newcommand{\uRe}{\mathop{\mathrm{uRe}}}
\newcommand{\uIm}{\mathop{\mathrm{uIm}}}

\newcommand{\Mat}{\mathrm{Mat}}
\newcommand{\uMat}{\mathrm{uMat}}

\newcommand{\minf}{-\infty}

\newcommand{\uabs}[1]{\lt|#1\rt|_{\oplus}}
\newcommand{\bal}[1]{#1^\bullet}

\newcommand{\bals}{\mathrel{\nabla}}
\newcommand{\Rel}{\mathrel{\mc R}}

\DeclareMathOperator{\adj}{adj}
\DeclareMathOperator{\cof}{cof}
\DeclareMathOperator{\sgn}{sgn}
\def\tr#1{\mathord{\mathopen{{\vphantom{#1}}^t}\!#1}}

\newcommand{\ud}{\stackrel{\mathrm{ud}\:}{\longrightarrow}}

\DeclareMathOperator{\ch}{ch}
\DeclareMathOperator{\sh}{sh}

\begin{document}

\allowdisplaybreaks

\renewcommand{\PaperNumber}{068}

\FirstPageHeading

\ShortArticleName{Ultradiscrete and Noncommutative Discrete sine-Gordon Equations}

\ArticleName{Ultradiscrete sine-Gordon Equation\\ over Symmetrized Max-Plus Algebra,\\ and Noncommutative Discrete\\ and Ultradiscrete sine-Gordon Equations}

\Author{Kenichi KONDO}

\AuthorNameForHeading{K.~Kondo}

\Address{5-13-12-207 Matsubara, Setagaya-ku, Tokyo 156-0043, Japan}

\Email{\href{mailto:incidence_algebra@poset.jp}{incidence\_algebra@poset.jp}}

\ArticleDates{Received January 08, 2013, in f\/inal form October 31, 2013; Published online November 12, 2013}

\Abstract{Ultradiscretization with negative values is a long-standing problem and several attempts have been made to solve it. Among others, we focus on the symmetrized max-plus algebra, with which we ultradiscretize the discrete sine-Gordon equation. Another ultradiscretization of the discrete sine-Gordon equation has already been proposed by previous studies, but the equation and the solutions obtained here are considered to directly correspond to the discrete counterpart.
We also propose a noncommutative discrete analogue of the sine-Gordon equation, reveal its relations to other integrable systems including the noncommutative discrete KP equation, and construct multisoliton solutions by a repeated application of Darboux transformations. Moreover, we derive a noncommutative ultradiscrete analogue of the sine-Gordon equation and its 1-soliton and 2-soliton solutions, using the symmetrized max-plus algebra. As a result, we have a complete set of commutative and noncommutative versions of continuous, discrete, and ultradiscrete sine-Gordon equations.}


\Classification{37K10; 39A12}

\section{Introduction}\label{sec:intro}

\subsection{Ultradiscretization and its problem}

Ultradiscrete integrable systems are integrable systems where independent variables take values in $\Z$, and dependent variables in the max-plus algebra $\Rmax=\R\cup\{\minf\}$. Among them is the famous box-ball system~\cite{TS1990}, represented by the equation
\begin{gather*}
U_n^{t+1}=\min\lt[1-U_n^t, \sum_{k=-\infty}^{n-1}U_k^t-\sum_{k=-\infty}^{n-1}U_k^{t+1}\rt].
\end{gather*}
Def\/ining $S_n^t=\sum\limits_{k=n}^\infty\sum\limits_{l=-\infty}^t U_k^l$, we obtain
\begin{gather}
S_{n+1}^{t+1}+S_n^{t-1}=\max\lt[S_{n+1}^{t-1}+S_n^{t+1}-1,S_n^t+S_{n+1}^t\rt], \label{intro:udKdV}
\end{gather}
which is \textit{ultradiscretization} of the discrete KdV equation
\begin{gather}
(1+\delta)\sigma_{n+1}^{t+1}\sigma_n^{t-1}=\delta\sigma_{n+1}^{t-1}\sigma_n^{t+1}+\sigma_n^t\sigma_{n+1}^t. \label{intro:dKdV}
\end{gather}

Ultradiscretization~\cite{TTMS1996} is a systematic procedure to obtain ultradiscrete systems from discrete systems. The fundamental formula of the procedure is
%\begin{subequations}
\begin{gather*}
\lim_{\e\to+0}\e\log\big(e^{A/\e}+e^{B/\e}\big) =\max[A,B], \qquad
\lim_{\e\to+0}\e\log\big(e^{A/s}\cdot e^{B/s}\big) =A+B.
\end{gather*}
%\end{subequations}
This may be understood as a transformation of addition into $\max$ operation and of multiplication into addition. Setting $\delta=e^{-1/\e}$ in~\eqref{intro:dKdV} and applying ultradiscretization, we obtain~\eqref{intro:udKdV}.

The problem is, however, that ultradiscretization cannot be applied to subtraction, which is of course necessary in many discrete integrable systems. The reason is as follows. If one wants to def\/ine an ultradiscrete version of subtraction, it is natural to solve the linear equation
\begin{gather*}
\max[x,a]=b
\end{gather*}
for $x\in\Rmax$. This has no solution when $a>b$, and therefore subtraction cannot be def\/ined in general.

Several attempts~\cite{IGRS2006,KL2006,MimuraEA2009,ON2005,Ormerod,YNA2006} have been made to solve this problem. We focus on the symmetrized max-plus algebra~\cite{MaxPlus, BCOQ}, denoted by $\uR$ in this paper. This algebra is an extension of $\Rmax$ and looks natural in the sense it traces the construction of $\Z$ from $\N^2$. Linear algebra over $\uR$ is also possible, and ultradiscretization with $\uR$ is presented in~\cite{DeSchutter}. These theories of $\uR$ are mainly developed in the f\/ield of discrete event systems and seems little known to the f\/ield of integrable systems.

\subsection{Contents of the paper}

The discrete sine-Gordon equation~\cite{DJMIII, Hirota1977}
%\begin{subequations}
\begin{gather*}
(1-\delta)\tau_l^m\tau_{l+1}^{m+1} =\tau_{l+1}^m\tau_l^{m+1}-\delta\sigma_{l+1}^m\sigma_l^{m+1}, \\
(1-\delta)\sigma_l^m\sigma_{l+1}^{m+1} =\sigma_{l+1}^m\sigma_l^{m+1}-\delta\tau_{l+1}^m\tau_l^{m+1}
\end{gather*}
%\end{subequations}
has not been ultradiscretized until recent years because soliton solutions include subtraction or even complex numbers. The f\/irst attempt is made by Isojima et al.~\cite{Isojima2004, IS2009} where a $\tau$-only trilinear equation is exploited to exclude subtraction. Here we propose another method to ultradiscretize the sine-Gordon equation which utilizes $\uR$. The equation and the solutions are ultradiscretized keeping subtraction and complex numbers in a highly direct fashion.

Noncommutative integrable systems have been drawing more interest in the last two decades. It is dif\/f\/icult to point out the f\/irst appearance of such systems, but the noncommutative KdV equation is already mentioned in~\cite{Lax1968}. The f\/irst discrete noncommutative integrable system is probably the noncommutative discrete KP equation~\cite{Kondo2011, Nimmo2006}. Along this line, we propose a~noncommutative discrete analogue of the sine-Gordon equation, explore relations to other integrable systems, and construct multisoliton solutions by the Darboux transformation. Moreover, we also propose a noncommutative ultradiscrete analogue of the sine-Gordon equation and explicitly derive 1-soliton and 2-soliton solutions by ultradiscretization with $\uR$. As a result, we have a complete set of commutative and noncommutative versions of continuous, discrete, and ultradiscrete sine-Gordon equations.

The rest of the paper is organized as follows.
In~\secref{sec:dsG}, the discrete sine-Gordon equation and 1-soliton and 2-soliton solutions are reviewed. Special solutions such as the traveling-wave and kink-antikink solutions are presented probably for the f\/irst time. In~\secref{sec:udsG}, an ultradiscrete analogue of the sine-Gordon equation is proposed and the solutions are obtained. Because of ultradiscretization with~$\uR$, correspondence between the discrete and ultradiscrete systems are direct, which is also supported by f\/igures.

In~\secref{sec:ncdsG}, a noncommutative discrete analogue of the sine-Gordon equation is proposed. A relation to other integrable systems including the noncommutative discrete KP equation is explained, and multisoliton solutions are constructed by a repeated application of Darboux transformations. In~\secref{sec:ncudsG}, a noncommutative ultradiscrete analogue of the sine-Gordon equation is proposed and 1-soliton and 2-soliton solutions are derived. Also f\/igures of solutions for both equations are displayed.

In~\secref{sec:conclusion}, concluding remarks and discussions are presented.

\section{Discrete and ultradiscrete sine-Gordon equations}

In this section, we f\/irst review the discrete sine-Gordon equation~\cite{DJMIII, Hirota1977} and several results around it. Explicit calculation of the traveling-wave, kink-antikink, kink-kink, and breather solutions are probably presented for the f\/irst time.

Then, we propose an ultradiscrete analogue of the sine-Gordon equation. The solutions are obtained in two ways: by calculations completely inside $\uR$, and by ultradiscretization with $\uR$. The correspondence between the discrete and ultradiscrete systems is quite clear. Similarity of prof\/iles of solutions is also visually conf\/irmed by f\/igures. Our formulation is dif\/ferent from Isojima et al.~\cite{Isojima2004}.

%-----------------------------------------------------------------------------
\subsection{Discrete sine-Gordon equation}\label{sec:dsG}
%-----------------------------------------------------------------------------

We review the three representations of the discrete sine-Gordon equation, their connection to the sine-Gordon equation, and some solutions for the sine-Gordon equation.

For any function $f=f(l,m)$ over $\Z^2$, def\/ine shift operations by
%\begin{subequations}
\begin{gather*}
f_l=f_l(l,m)=f(l+1,m), \qquad f_m=f_m(l,m)=f(l,m+1).
\end{gather*}
Inverse operations are denoted by
\begin{gather*}
f_{\ov l}=f(l-1,m), \qquad f_{\ov m}=f(l,m-1).
\end{gather*}
%\end{subequations}
Let $\tau=\tau(l,m)$, $\sigma=\sigma(l,m)$ be functions $\Z^2\to\C$. Date et al.~\cite{DJMIII} gave the discrete sine-Gordon equation (dsG) in the following form
\begin{subequations}
\begin{gather}
(1-\delta)\tau\tau_{lm} =\tau_l\tau_m-\delta\sigma_l\sigma_m, \label{dsG:dsG1}\\
(1-\delta)\sigma\sigma_{lm} =\sigma_l\sigma_m-\delta\tau_l\tau_m, \label{dsG:dsG2}
\end{gather}
\end{subequations}
where $\delta\in\C^\times$ is a parameter with a small absolute value. The vacuum solution
\begin{gather*}
\tau=\sigma=1
\end{gather*}
is the simplest solution, other than the null solution $\tau=\sigma=0$. Calculating the cross product of the both sides of \eqref{dsG:dsG1}, \eqref{dsG:dsG2}, we have
\begin{gather*}
\tau\tau_{lm}(\sigma_l\sigma_m-\delta\tau_l\tau_m)=\sigma\sigma_{lm}(\tau_l\tau_m-\delta\sigma_l\sigma_m)
\ \EquivTo \ \frac{\tau_{lm}\sigma_m}{\sigma_{lm}\tau_m}-\frac{\tau_l\sigma}{\sigma_l\tau}+\delta\lt(\frac{\sigma_m\sigma}{\tau_m\tau}-\frac{\tau_{lm}\tau_l}{\sigma_{lm}\sigma_l}\rt)=0
\end{gather*}
and thus
\begin{gather}
\frac{w_{lm}}{w_m}-\frac{w_l}{w}+\delta\lt(\frac{1}{w_mw}-w_{lm}w_l\rt)=0, \label{dsG:w}
\end{gather}
where $w$ is def\/ined by
\begin{gather*}
w=\frac{\tau}{\sigma}.
\end{gather*}
If we introduce $u$ def\/ined by
\begin{gather*}
u=\frac{2}{i}\log w,
\end{gather*}
we have
\begin{gather}
e^{i(u_{lm}-u_m)/2}-e^{i(u_l-u)/2}+\delta\big(e^{i(-u_m-u)/2}-e^{i(u_{lm}+u_l)/2}\big)=0 \notag \\
\qquad \EquivTo \ \sin\lt(\frac{u_{lm}-u_l-u_m+u}{4}\rt)=\delta\sin\lt(\frac{u_{lm}+u_l+u_m+u}{4}\rt). \label{dsG:u}
\end{gather}
Each of \eqref{dsG:w} and \eqref{dsG:u} is also called the discrete sine-Gordon equation, where \eqref{dsG:u} is the original form discovered by Hirota~\cite{Hirota1977}.

\begin{Remark}
By the non-autonomous transformation $w'=w^{(-1)^m}$, \eqref{dsG:w} becomes
\begin{gather*}
\frac{w'_m}{w'_{lm}}-\frac{w'_l}{w'}+\delta\lt(\frac{w'_m}{w'}-\frac{w'_l}{w'_{lm}}\rt)=0 \ \EquivTo \ \frac{w'_m-\delta w'_l}{w'_{lm}}=\frac{w'_l-\delta w'_m}{w'}.
\end{gather*}
This is known as a discrete analogue of the modif\/ied KdV equation~\cite{NC1995}, and its ultradiscretization is also known~\cite{QCS2001}.
\end{Remark}

Assume $u$ is also a function $u(x,y)$ of continuum variables $x,y\in\R$ and has an expansion
\begin{gather*}
u(x+r,y+s)=u+(ru_x+su_y)+\frac{1}{2}\lt(r^2u_{xx}+2rsu_{xy}+s^2u_{yy}\rt)+\cdots,
\end{gather*}
where $u_x=\p u/\p x$, etc. Connect $l$, $m$ to $x$, $y$ via the Miwa transformation
\begin{gather*}
u(x,y;l,m)=u(x+la,y+mb)
\end{gather*}
where $a,b\in\R^\times$ are parameters. Then we have
\[
u_{lm}-u_l-u_m+u=abu_{xy}+\text{(higher-order terms of $a$, $b$)}.
\]
Setting $\delta=ab$ and taking the limit $a,b\to0$ of \eqref{dsG:u} successively, we obtain
\begin{gather*}
\lim_{a,b\to0}\frac{1}{ab}\sin\lt(\frac{u_{lm}-u_l-u_m+u}{4}\rt) =\lim_{a,b\to0}\frac{1}{ab}\sin\lt(\frac{abu_{xy}+\cdots}{4}\rt)=\frac{1}{4}u_{xy}, \\
\lim_{a,b\to0}\sin\lt(\frac{u_{lm}+u_l+u_m+u}{4}\rt) =\sin u,
\end{gather*}
and thus the (continuous) sine-Gordon equation
\begin{gather*}
u_{xy}=4\sin u.
\end{gather*}
This is known to have following special types of solutions (see, for example,~\cite{GEK}): the traveling-wave solution
\begin{gather}
u=4\arctan\exp\lt(\pm\frac{x-vy}{\sqrt{1-v^2}}\rt), \label{sG:traveling}
\end{gather}
the kink-antikink solution
\begin{gather}
u=4\arctan\lt(\frac{\sinh\frac{vy}{\sqrt{1-v^2}}}{v\cosh\frac{x}{\sqrt{1-v^2}}}\rt), \label{sG:ka}
\end{gather}
the kink-kink solution
\begin{gather}
u=4\arctan\lt(\frac{v\sinh\frac{x}{\sqrt{1-v^2}}}{\cosh\frac{vy}{\sqrt{1-v^2}}}\rt), \label{sG:kk}
\end{gather}
and the breather solution
\begin{gather}
u=4\arctan\lt(\frac{\sqrt{1-\omega^2}}{\omega} \frac{\sin\omega y}{\cosh\sqrt{1-\omega^2}\,x}\rt), \label{sG:breather}
\end{gather}
where $v$, $\omega$ are constants.

\subsubsection{1-soliton and 2-soliton solutions}

Isojima et al.~\cite{Isojima2004} have given the following conditions for $\tau$ and $\sigma$ to be a 1-soliton or 2-soliton solution. As a 1-soliton solution, assume
\begin{gather}
\tau=1+f, \qquad \sigma=1-f, \qquad f=cp^lq^m \label{dsG:1sol}
\end{gather}
where $c,p,q\in\C^\times$ are constants. By substitution, the dispersion relation
\begin{gather}
(1-\delta)(1+pq)=(1+\delta)(p+q) \ \EquivTo \ q=\frac{(1+\delta)p-(1-\delta)}{(1-\delta)p-(1+\delta)} \label{dsG:disprel1}
\end{gather}
is found to be a necessary and suf\/f\/icient condition. As a 2-soliton solution, assume
\begin{gather*}
\tau=1+f_1+f_2+\alpha f_1f_2, \qquad \sigma=1-f_1-f_2+\alpha f_1f_2, \qquad f_j=c_jp_j^lq_j^m , %\label{dsG:2sol}
\end{gather*}
where $\alpha\in\C^\times$ is a constant. This time, the pair of the dispersion relation
\begin{gather*}
(1-\delta)(1+p_jq_j)=(1+\delta)(p_j+q_j) \ \EquivTo \ q_j=\frac{(1+\delta)p_j-(1-\delta)}{(1-\delta)p_j-(1+\delta)} %\label{dsG:disprel}
\end{gather*}
and the relation
\begin{gather}
\alpha=-\frac{p_1-p_2}{1-p_1p_2}\frac{q_1-q_2}{1-q_1q_2}=\lt(\frac{p_1-p_2}{1-p_1p_2}\rt)^2 \label{dsG:coupling_const}
\end{gather}
is a necessary and suf\/f\/icient condition.

\begin{figure}[t]
\centering
\includegraphics[width=6.5cm]{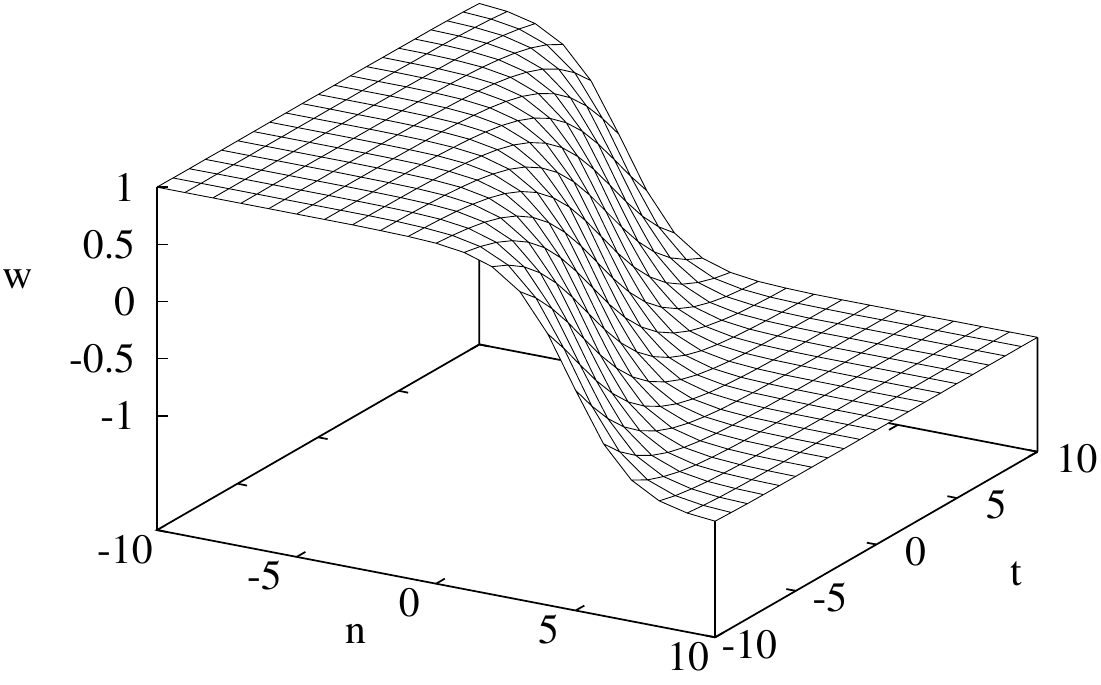} \qquad
\includegraphics[width=6.5cm]{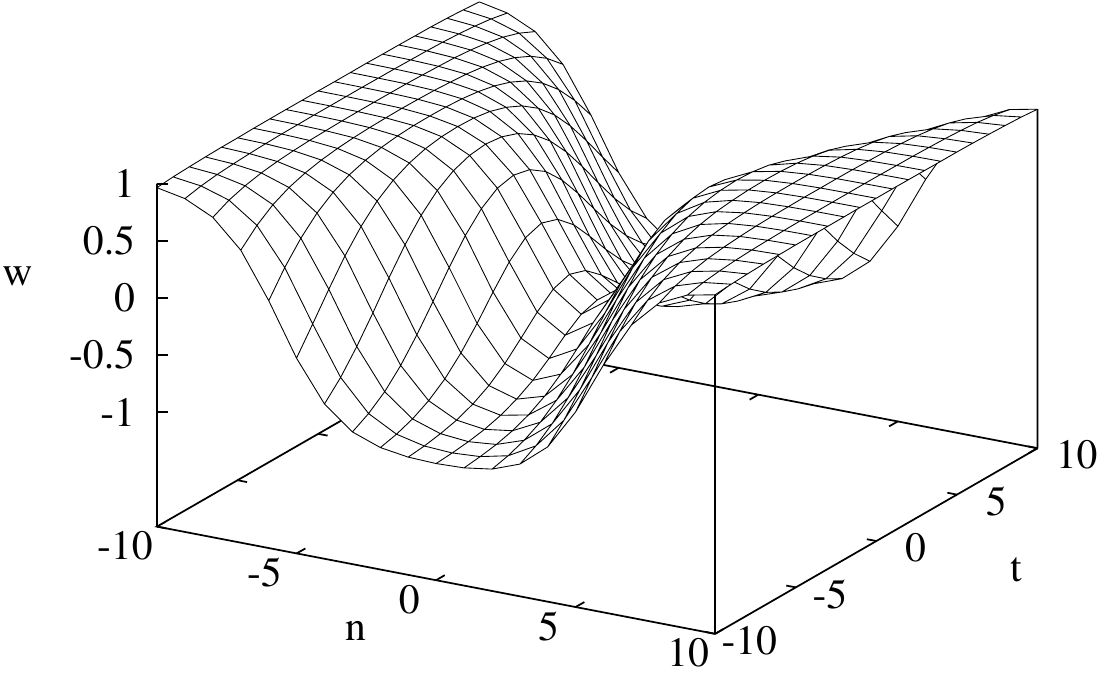}

\caption{1-soliton solution (left) and 2-soliton solution (right) for dsG.}
\label{dsG:fig:r}
\end{figure}

Fig.~\ref{dsG:fig:r} shows the 1-soliton solution with
\[
\delta=0.04, \qquad c=-1, \qquad p=2,
\]
and the 2-soliton solution with
\[
\delta=0.04, \qquad c_1=c_2=-2.125, \qquad p_1=q_2=2,
\]
in the light-cone coordinates
\begin{gather}
(n, t)=\lt(\frac{l+m}{2}, \frac{l-m}{2}\rt) \ \EquivTo \ (l,m)=(n+t, n-t). \label{lightcone}
\end{gather}

\subsubsection{Traveling-wave solution}\label{sec:tr}

In Sections \ref{sec:tr}--\ref{sec:br}, we give solutions for \eqref{dsG:dsG1} and \eqref{dsG:dsG2} which correspond to the continuous counterparts \eqref{sG:traveling}--\eqref{sG:breather}. These do not seem to be previously presented in the literature, including Hirota~\cite{Hirota1977} and Isojima et al.~\cite{Isojima2004}.

\begin{figure}[t]
\centering
\includegraphics[width=6.5cm]{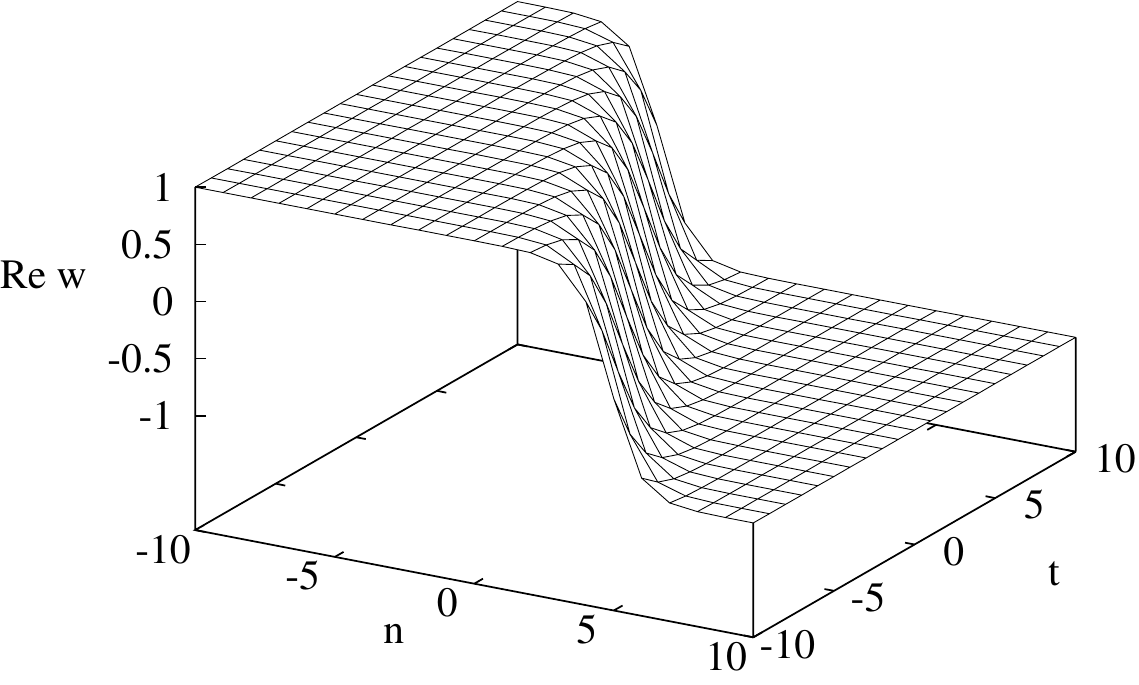} \qquad
\includegraphics[width=6.5cm]{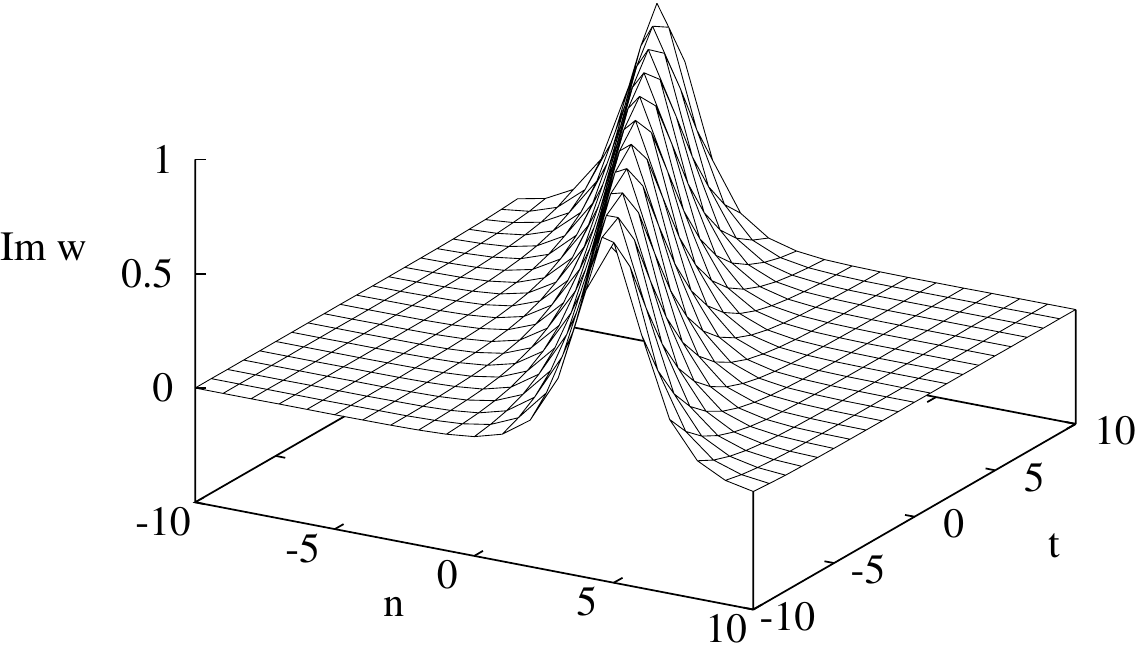}

\includegraphics[width=6.5cm]{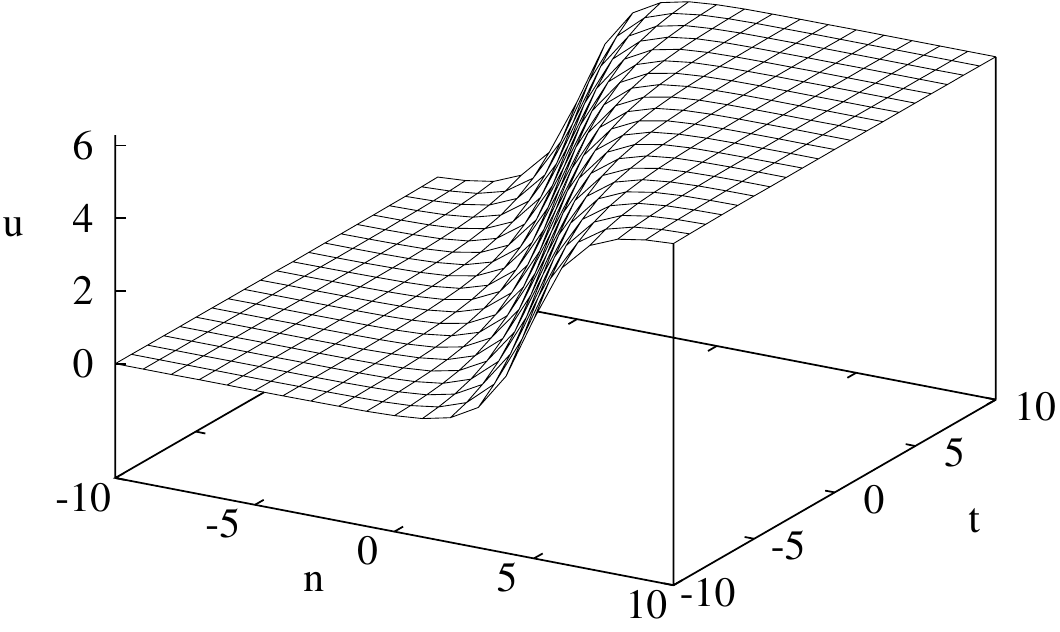}

\caption{Traveling-wave solution for dsG.}
\label{dsG:fig:tr}

\end{figure}

Replacing $c$ by $ic$ in the 1-soliton solution, we obtain
\begin{gather}
w=\frac{1+ic(pq)^n\lt(pq^{-1}\rt)^t}{1-ic(pq)^n\big(pq^{-1}\big)^t}, \qquad  u=4\arctan\lt(c(pq)^n\lt(pq^{-1}\rt)^t\rt) \label{dsG:traveling}
\end{gather}
in the light-cone coordinates. This corresponds to the traveling-wave solution \eqref{sG:traveling} for the sine-Gordon equation (if we restrict $\delta,c,p\in\R^\times$). Fig.~\ref{dsG:fig:tr} shows the solution with
\[
\delta=0.04, \qquad c=1, \qquad p=2.
\]

\subsubsection{Kink-antikink and kink-kink solutions}

\begin{figure}[t]
\centering
\includegraphics[width=6.5cm]{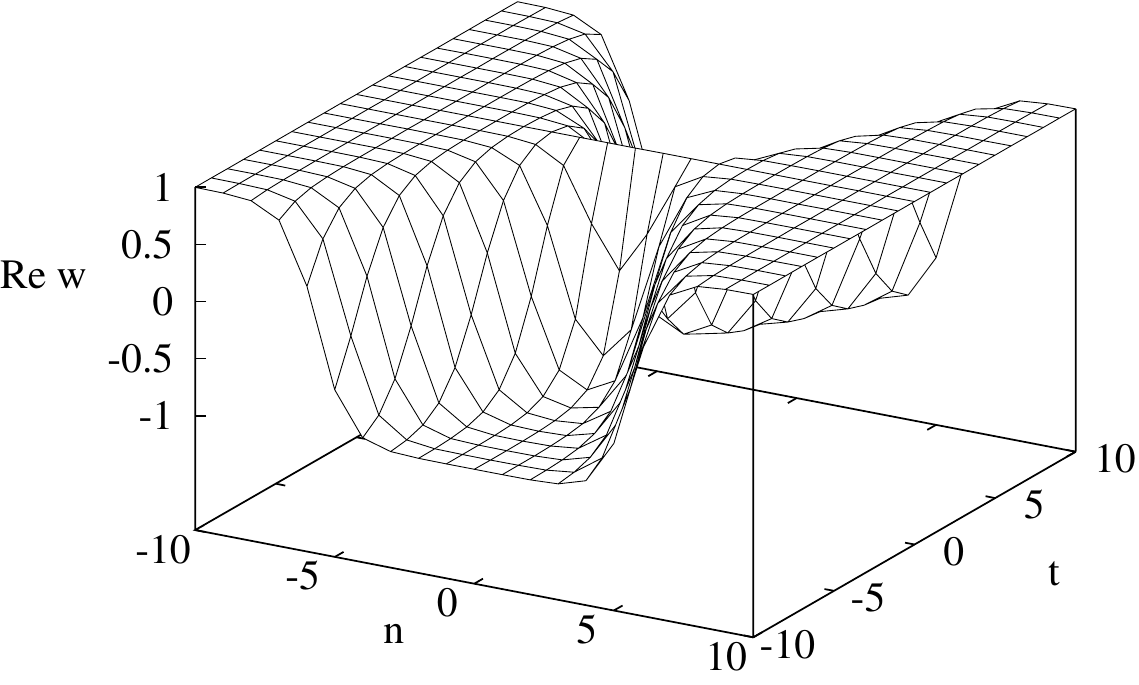} \qquad
\includegraphics[width=6.5cm]{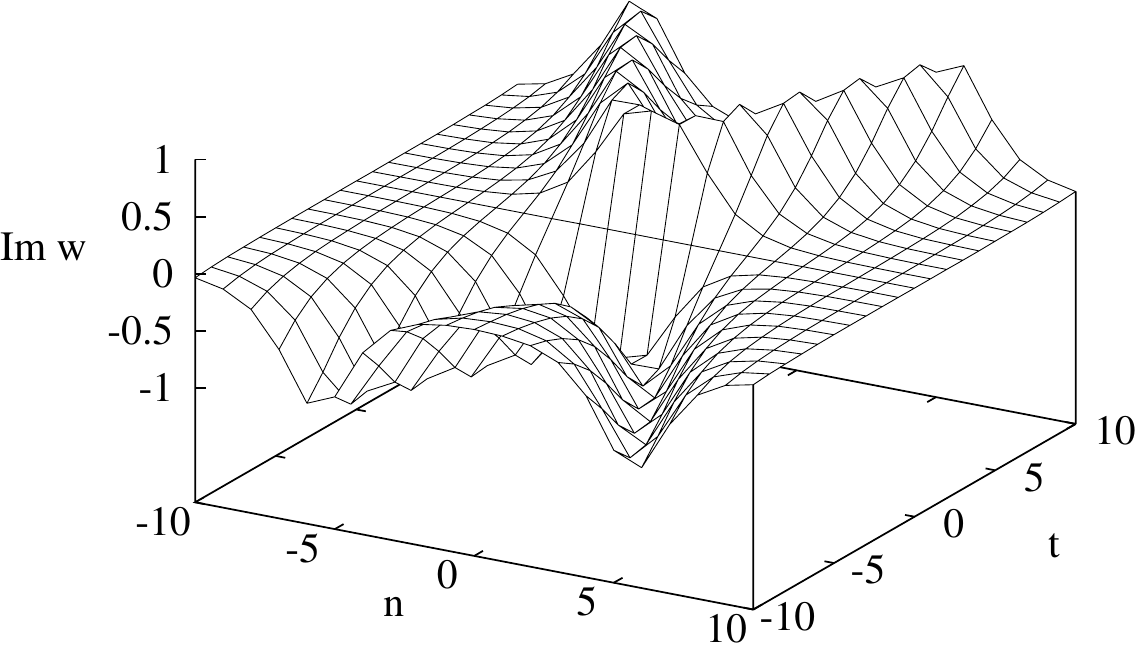} \\

\includegraphics[width=6.5cm]{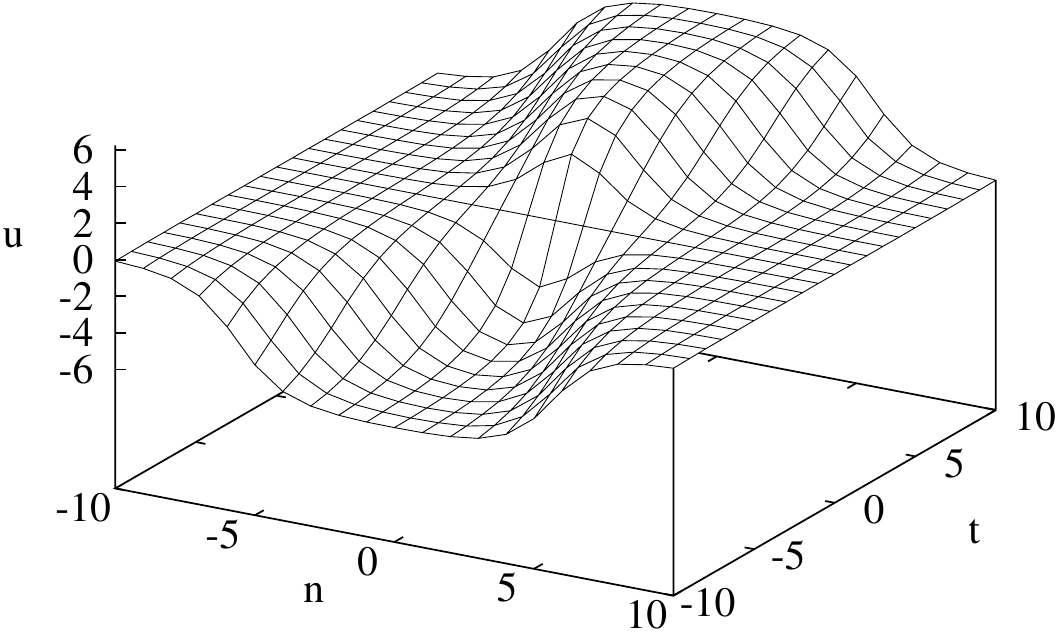}

\caption{Kink-antikink solution for dsG.}
\label{dsG:fig:ka}
\end{figure}

Set $p_1=q_2$
in the 2-soliton solution. Then
\begin{gather}
p_1=\frac{(1+\delta)p_2-(1-\delta)}{(1-\delta)p_2-(1+\delta)} \ \EquivTo \ p_2=\frac{(1+\delta)p_1-(1-\delta)}{(1-\delta)p_1-(1+\delta)} \label{dsG:disprel_p}
\end{gather}
and thus
$p_2=q_1$.
Rewriting in the light-cone coordinates, we have
\begin{gather*}
\tau =1+c_1p_1^{n+t}p_2^{n-t}+c_2p_2^{n+t}p_1^{n-t}+\alpha c_1c_2(p_1p_2)^{2n} \\
\hphantom{\tau}{} =(p_1p_2)^n\lt(\alpha c_1c_2(p_1p_2)^n+(p_1p_2)^{-n}+c_1\lt(p_1p_2^{-1}\rt)^t+c_2\lt(p_1p_2^{-1}\rt)^{-t}\rt), \\
\sigma =(p_1p_2)^n\lt(\alpha c_1c_2(p_1p_2)^n+(p_1p_2)^{-n}-c_1\lt(p_1p_2^{-1}\rt)^t-c_2\lt(p_1p_2^{-1}\rt)^{-t}\rt).
\end{gather*}
We set
\begin{gather*}
\beta=\pm\frac{p_1-p_2}{1-p_1p_2}, \qquad c_1=-c_2=i\beta^{-1}
\end{gather*}
and def\/ine
\begin{gather*}
\ch(p, l)=\frac{p^l+p^{-l}}{2}, \qquad \sh(p, l)=\frac{p^l-p^{-l}}{2}.
\end{gather*}
Then
\begin{gather}
w=\frac{\beta\ch(p_1p_2,n)+i\sh\lt(p_1p_2^{-1},t\rt)}{\beta\ch(p_1p_2,n)-i\sh\lt(p_1p_2^{-1},t\rt)}, \qquad u=4\arctan\lt(\frac{\sh\lt(p_1p_2^{-1}, t\rt)}{\beta\ch(p_1p_2, n)}\rt). \label{dsG:ka}
\end{gather}
This corresponds to the kink-antikink solution \eqref{sG:ka}. Fig.~\ref{dsG:fig:ka} shows the solution with
\[
\delta=0.04, \qquad c_1=-c_2=2.125i, \qquad p_1=2.
\]

\begin{figure}[t]
\centering
\includegraphics[width=6.5cm]{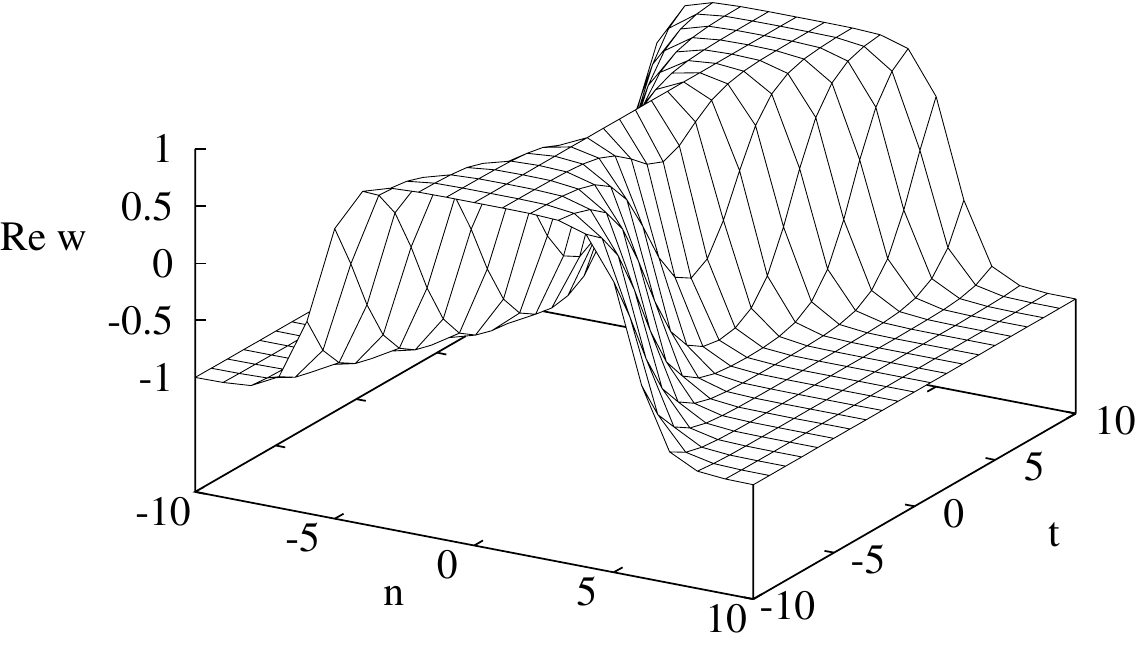} \qquad
\includegraphics[width=6.5cm]{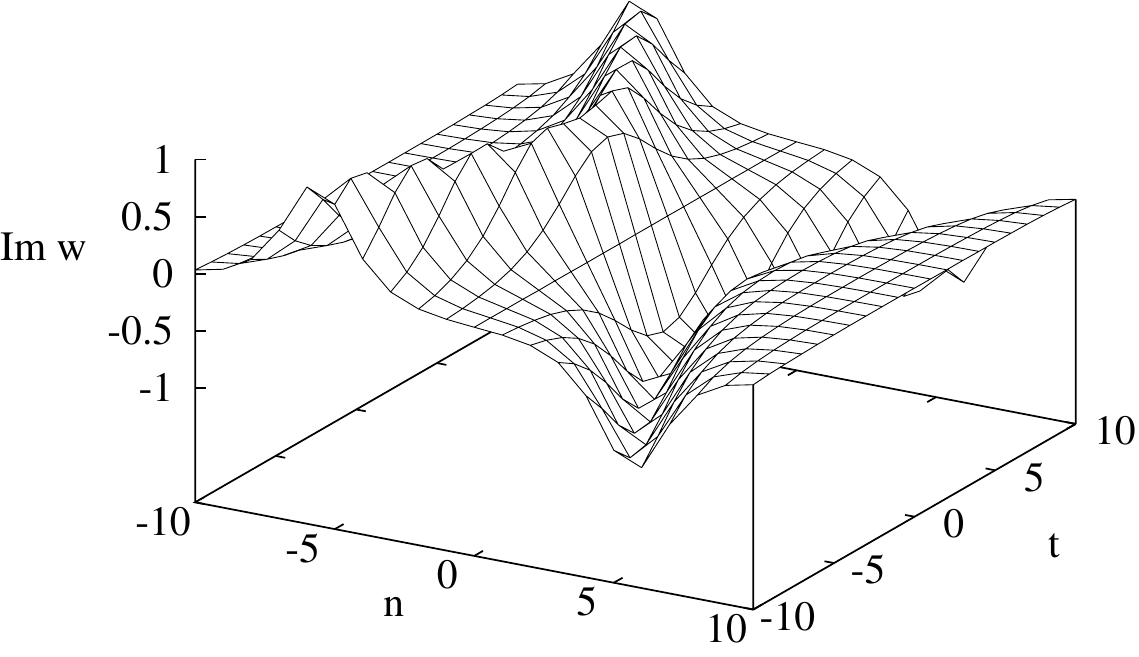}

\includegraphics[width=6.5cm]{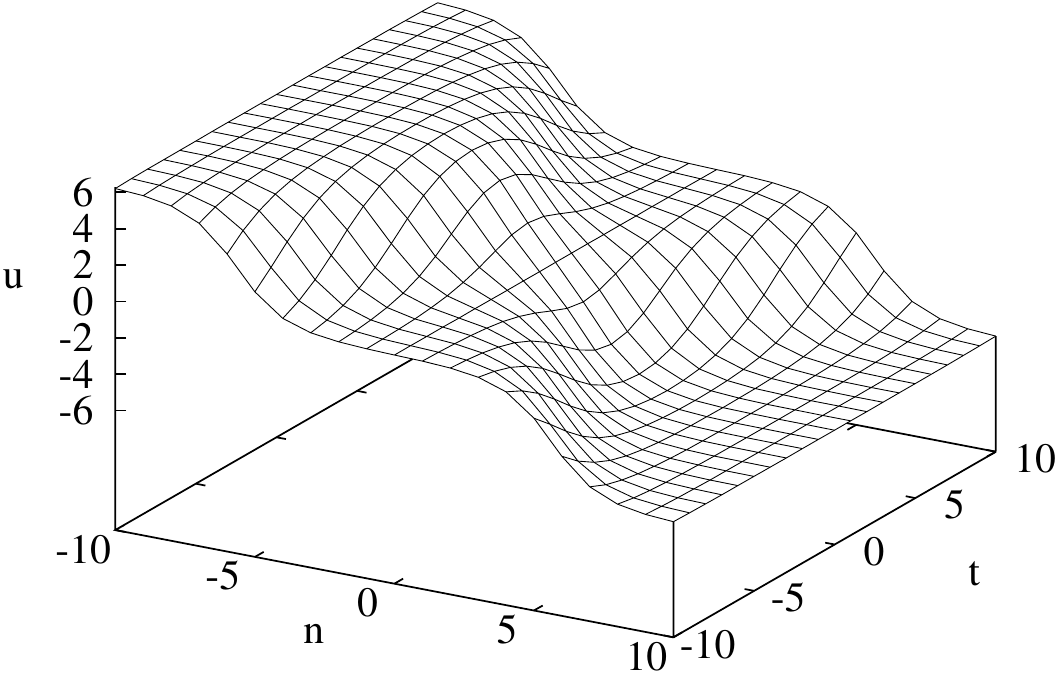}

\caption{Kink-kink solution for dsG.}
\label{dsG:fig:kk}
\end{figure}

Similarly, setting
$
p_1q_2=1
$
gives
\[
p_1^{-1}=\frac{(1+\delta)p_2-(1-\delta)}{(1-\delta)p_2-(1+\delta)} \ \EquivTo \ p_2^{-1}=\frac{(1+\delta)p_1-(1-\delta)}{(1-\delta)p_1-(1+\delta)}
\]
and thus $p_2q_1=1$.
We have
\begin{gather*}
\begin{split}
& \tau =\beta^{-1}(p_1p_2)^t\lt(\beta(p_1p_2)^t+\beta(p_1p_2)^{-t}+i\lt(p_1p_2^{-1}\rt)^n-i\lt(p_1p_2^{-1}\rt)^{-n}\rt), \\
& \sigma =\beta^{-1}(p_1p_2)^t\lt(\beta(p_1p_2)^t+\beta(p_1p_2)^{-t}-i^{-1}\lt(p_1p_2^{-1}\rt)^n+i\lt(p_1p_2^{-1}\rt)^{-n}\rt)
\end{split}
\end{gather*}
for the same $\beta$, $c_1$, $c_2$ def\/ined above and
\begin{gather}
w=\frac{\beta\ch(p_1p_2,t)+i\sh\lt(p_1p_2^{-1},n\rt)}{\beta\ch(p_1p_2,t)-i\sh\lt(p_1p_2^{-1},n\rt)}, \qquad u=4\arctan\lt(\frac{\sh\lt(p_1p_2^{-1}, n\rt)}{\beta\ch(p_1p_2, t)}\rt). \label{dsG:kk}
\end{gather}
This corresponds to the kink-kink solution \eqref{sG:kk}. Fig.~\ref{dsG:fig:kk} shows the solution with
\[
\delta=0.04, \qquad c_1=-c_2=-0.470588i, \qquad p_1=2.
\]

\subsubsection{Breather solution}\label{sec:br}

\begin{figure}[t]
\centering
\includegraphics[width=6.5cm]{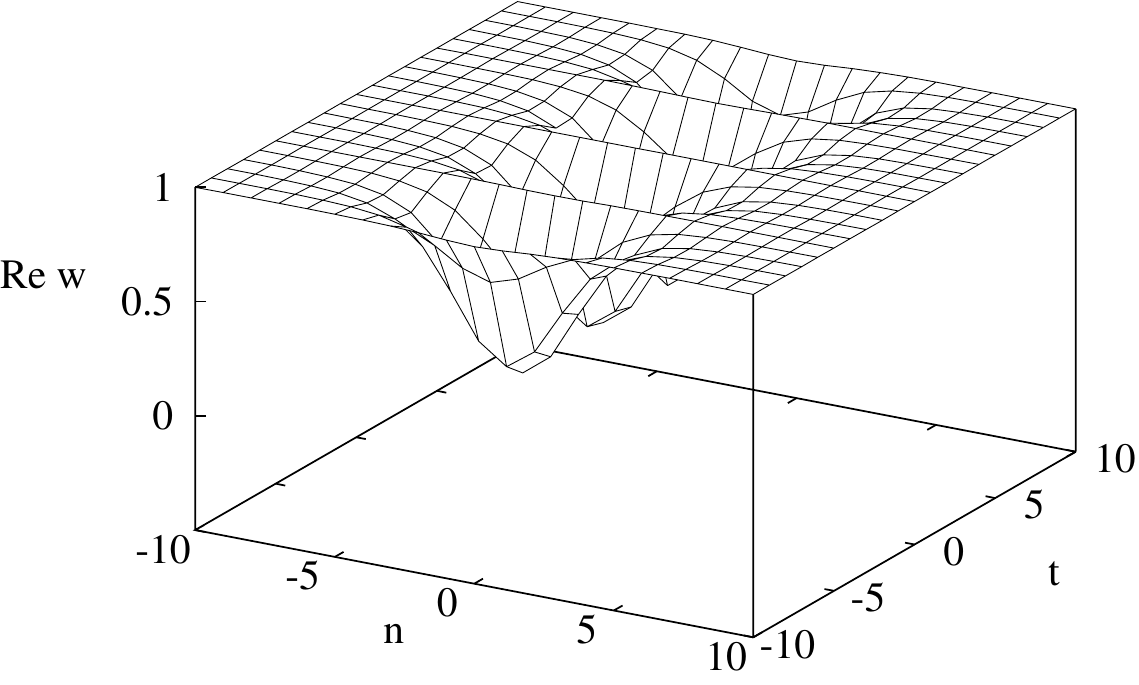} \qquad
\includegraphics[width=6.5cm]{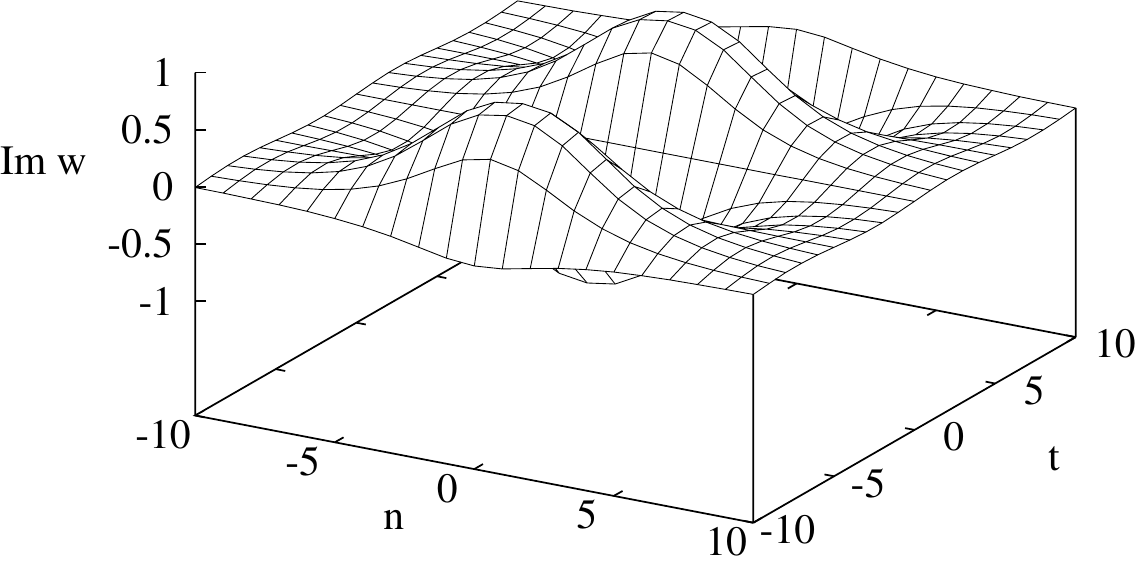}

\includegraphics[width=6.5cm]{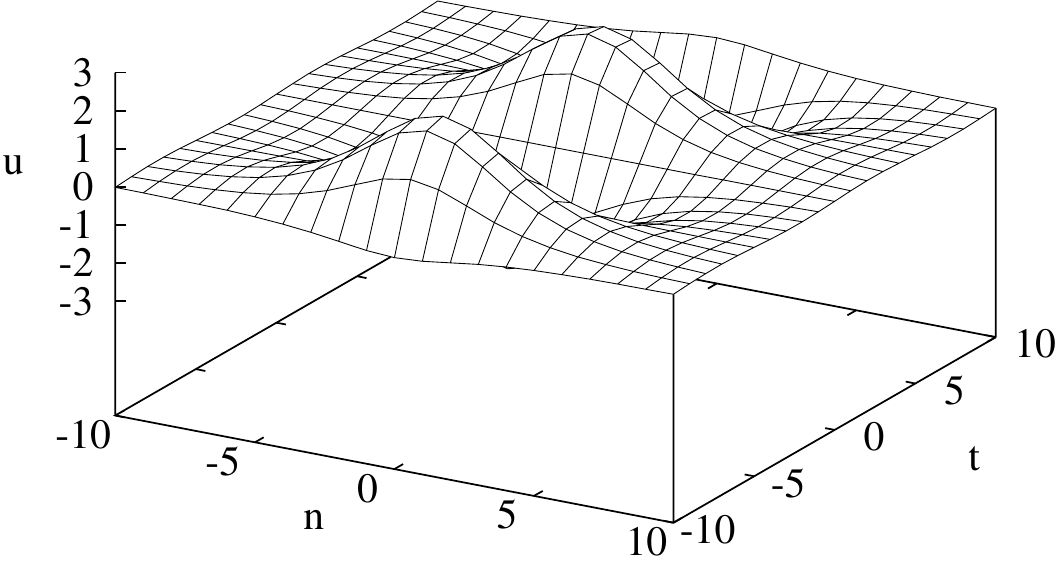}

\caption{Breather solution for dsG.}
\label{dsG:fig:br}
\end{figure}

Consider the kink-antikink solution where $p_1$ and $p_2$ are complex numbers satisfying
\begin{gather*}
p_1p_2\in\R_{>0}, \qquad \lt|p_1p_2^{-1}\rt|=1.
\end{gather*}
Such $p_1$, $p_2$ are complex conjugates of each other. If we write
$p_1=g+hi$, $ p_2=g-hi$
and substitute these into~\eqref{dsG:disprel_p}, we f\/ind $g$, $h$ must satisfy
\begin{gather*}
(1-\delta)\lt(1+g^2+h^2\rt)=2(1+\delta)g \
\EquivTo \ (1-\delta)g^2-2(1+\delta)g+(1-\delta)\lt(1+h^2\rt)=0.
\end{gather*}
As a quadratic equation of $g$, the condition for the existence of real roots is given by
\begin{gather*}
(1+\delta)^2-(1-\delta)^2\lt(1+h^2\rt)\ge0 \ \EquivTo \ h^2\le\lt(\frac{1+\delta}{1-\delta}\rt)^2-1.
\end{gather*}
Such a real number $h$ does exist if $\delta\ge0$, and $g$ is given by
\begin{gather*}
g=\frac{1+\delta}{1-\delta}\pm\sqrt{\lt(\frac{1+\delta}{1-\delta}\rt)^2-\lt(1+h^2\rt)}.
\end{gather*}
Rewriting $p_1=re^{i\theta}$, $p_2=re^{-i\theta}$, we obtain $\beta=i\gamma$ where $\gamma$ is def\/ined by
\[
\gamma=\pm\frac{2r\sin\theta}{1-r^2},
\]
and $\sh\lt(p_1p_2^{-1},t\rt)=i\sin2t\theta$. Thus,
\begin{gather*}
w=\frac{\gamma\ch(r^2,n)+i\sin2t\theta}{\gamma\ch(r^2,n)-i\sin2t\theta}, \qquad u=4\arctan\lt(\frac{\sin2t\theta}{\gamma\ch(r^2,n)}\rt).
\end{gather*}
This corresponds to the breather solution \eqref{sG:breather}. Fig.~\ref{dsG:fig:br} shows the solution with
\[
\delta=0.04, \qquad c_1=-c_2=0.75, \qquad p_1=0.75+0.25i.
\]

\subsection{Ultradiscrete sine-Gordon equation}\label{sec:udsG}

\subsubsection{Ultradiscrete sine-Gordon equation}

In order to ultradiscretize the discrete sine-Gordon equation~\eqref{dsG:dsG1}, \eqref{dsG:dsG2}, we must deal with negative numbers since either or both of $\tau$ and $\sigma$ include subtractions. We adopt ultradiscretization with the symmetrized max-plus algebra~$\uR$. For details, see~\apdref{apd:ud} and references cited there.

We perform ultradiscretization of dsG through the parametrization
\begin{gather}
\delta=\mu_D e^{\wt Ds}, \qquad \wt D<0. \label{udsG:ud_param}
\end{gather}
This can be regarded as an other aspect of continuum limit since $\delta\to0$ as $s\to\infty$. Assuming $\delta\ud D$, $\tau\ud T$, $\sigma\ud S$, we obtain
\begin{subequations}
\begin{gather}
TT_{lm} \bals T_lT_m\ominus DS_lS_m, \label{udsG:T} \\
SS_{lm} \bals S_lS_m\ominus DT_lT_m. \label{udsG:S}
\end{gather}
\end{subequations}
We call the pair \eqref{udsG:T}, \eqref{udsG:S} the ultradiscrete sine-Gordon equation (udsG). The vacuum solution
$
T=S=0$
is the simplest solution, other than the null solution $T=S=\minf$. We can also ultradiscretize~\eqref{dsG:w} to obtain
\begin{gather}
W_{lm}W_m^{-1}\ominus W_lW^{-1}\oplus D\lt(W_m^{-1}W^{-1}\ominus W_{lm}W_l\rt)\bals\minf \label{udsG:W}
\end{gather}
where
$
w\ud W\bals TS^{-1}$.
We also call \eqref{udsG:W} the ultradiscrete sine-Gordon equation. The ultradiscretization of \eqref{dsG:u} is unclear.

\subsubsection{Deterministic time evolution and class of solutions}

It seems sensible to restrict ourselves to the class of signed solutions, that is,
$T,S,W\in\uC^\vee$ for any $(l,m)\in\Z^2$
since it permits basic properties like weak substitution. The null and vacuum solutions are signed solutions.

The problem is that udsG no longer admits time evolution, at least deterministic one, in general, since the balance relation is not equality. For example, if we have
\[
f(t+1)\bals\big(\text{expression including $f(t)$}\big)=\bal{3},
\]
we cannot determine $f(t+1)$ from $f(t)$, since this relation is satisf\/ied whenever $\uabs{f(t+1)}\le3$. Strictly speaking, udsG is not an \textit{equation}.

But in some cases, it actually becomes an equation, or furthermore, a deterministically evolutionary form. Multiplying $T^{-1}$ to \eqref{udsG:T} and $S^{-1}$ to \eqref{udsG:S}, we have
\begin{gather*}
T^{-1}TT_{lm} \bals T^{-1}\lt(T_lT_m\ominus DS_lS_m\rt), \qquad
S^{-1}SS_{lm} \bals S^{-1}\lt(S_lS_m\ominus DT_lT_m\rt).
\end{gather*}
If $T^{-1}T=S^{-1}S=0$ and the right hand sides are signed, we obtain
\begin{subequations}
\begin{gather}
T_{lm} =T^{-1}\lt(T_lT_m\ominus DS_lS_m\rt), \label{udsG:defT} \\
S_{lm} =S^{-1}\lt(S_lS_m\ominus DT_lT_m\rt) \label{udsG:defS}
\end{gather}
\end{subequations}\begin{subequations}
by reduction of balances (see \apdref{apd:ud}). We call \eqref{udsG:defT}, \eqref{udsG:defS} the deterministically evolutionary form of udsG. If we replace~$D$ by $\ominus D$ and restrict ranges of $D$, $T$, $S$ to $\R$ for example, the assumptions are satisf\/ied, and we obtain the completely ordinary-looking ultradiscrete equation:
\begin{gather}
T_{lm} =\max(T_l+T_m, D+S_l+S_m)-T, \label{udsG:posT} \\
S_{lm} =\max(S_l+S_m, D+T_l+T_m)-S. \label{udsG:posS}
\end{gather}
\end{subequations}
Deterministic time evolution is also possible in other settings, which are presented in the fol\-lowing sections.

It might be natural to think we should consider \eqref{udsG:defT}, \eqref{udsG:defS}, or even \eqref{udsG:posT}, \eqref{udsG:posS} only. However, it seems that the former cannot capture the traveling-wave, kink-antikink, and kink-kink solutions. And the latter does not even seem to contain soliton solutions. Therefore, we consider~\eqref{udsG:T}, \eqref{udsG:S} primarily.

\begin{figure}[t]
\centering
\includegraphics[width=6.5cm]{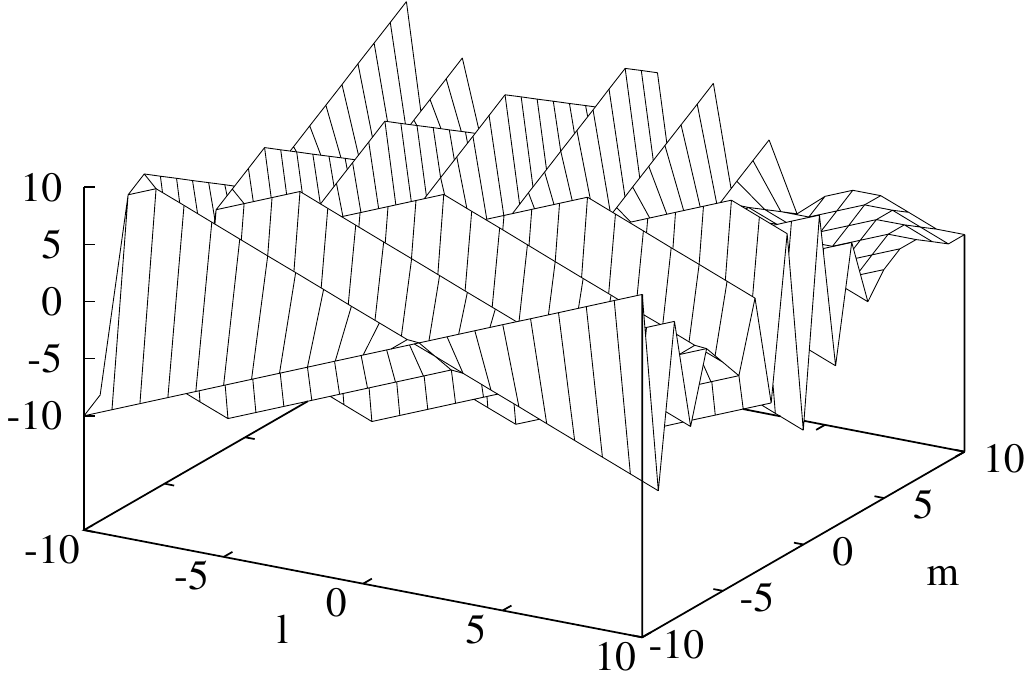}
\caption{Positive $(\R)$ evolution of udsG.}
\label{udsG:fig:pos}
\end{figure}

We show one example of positive $(\R)$ time evolution by \eqref{udsG:posT}, \eqref{udsG:posS} in Fig.~\ref{udsG:fig:pos}. Initial values are set as
\[
T(l,-10)=l, \qquad S(-10,m)=m, \qquad -10\le l,m\le 10
\]
with $D=-1$. The sign (in the sense of $\pm$) of the solution alternates with time.

\subsubsection{1-soliton solution}

Consider a signed solution $T$, $S$ satisfying
\begin{gather}
T\bals0\oplus F, \qquad S\bals 0\ominus F, \qquad F=CP^lQ^m , \label{udsG:1sol}
\end{gather}
where $C\in\uC^\vee$ and $P,Q\in\uR^\otimes$. Weakly substituting these into \eqref{udsG:T}, \eqref{udsG:S}, we have
\begin{subequations}
\begin{gather}
0\oplus PQF^2\oplus(0\oplus PQ)F \bals 0\oplus PQF^2\oplus(P\oplus Q)F, \label{udsG:dr1} \\
0\oplus PQF^2\ominus(0\oplus PQ)F \bals 0\oplus PQF^2\ominus(P\oplus Q)F, \label{udsG:dr2}
\end{gather}
\end{subequations}
where $0\oplus D=0\ominus D=0$ is used. The dispersion relation
\begin{gather}
0\oplus PQ\bals P\oplus Q \label{udsG:disprel1}
\end{gather}
is a suf\/f\/icient condition for \eqref{udsG:dr1}, \eqref{udsG:dr2} to hold, since we can construct them by adding and multiplying same numbers to the both sides of \eqref{udsG:disprel1}. Rewriting \eqref{udsG:disprel1}, we have
\[
(P\ominus 0)Q\bals(P\ominus 0)
\]
and thus $P=0$ or $Q=0$. Obviously, \eqref{udsG:1sol} and \eqref{udsG:disprel1} can be obtained by ultradiscreti\-zing~\eqref{dsG:1sol} and~\eqref{dsG:disprel1}, respectively, through
%\begin{subequations}
\begin{gather*}
c=\mu_{C}e^{\wt Cs}\ud C, \qquad p=\mu_{P}e^{\wt Ps}\ud P, \qquad q\ud Q
\end{gather*}
or
\begin{gather*}
c=\mu_{C}e^{\wt Cs}\ud C, \qquad p\ud P, \qquad q=\mu_{Q}e^{\wt Qs}\ud Q.
\end{gather*}
%\end{subequations}

The solution is, however, not completely determined yet, because the balance relation is not equality as stated before. So we try to utilize reduction of balances. If $C\in\uZ$ is an odd number and $P,Q\in\uZ$ are even numbers, then $F$ is always odd and $0\oplus F$, $0\ominus F$ can never be balanced since $0$ is even. By reduction of balances, we obtain
\begin{gather*}
T=0\oplus F, \qquad S=0\ominus F,
\end{gather*}
and $W$ is also immediately determined since $S^{-1}$ is signed. This solution admits deterministic time evolution since
\[
\uabs{T_lT_m}>\uabs{DS_lS_m}, \qquad \uabs{S_lS_m}>\uabs{DT_lT_m}
\]
and thus
\begin{gather*}
T^{-1}(T_lT_m\ominus DS_lS_m) =T^{-1}T_lT_m\in\uR^\otimes, \\
S^{-1}(S_lS_m\ominus DT_lT_m) =S^{-1}S_lS_m\in\uR^\otimes.
\end{gather*}

\begin{figure}[t]
\centering
\includegraphics[width=6.5cm]{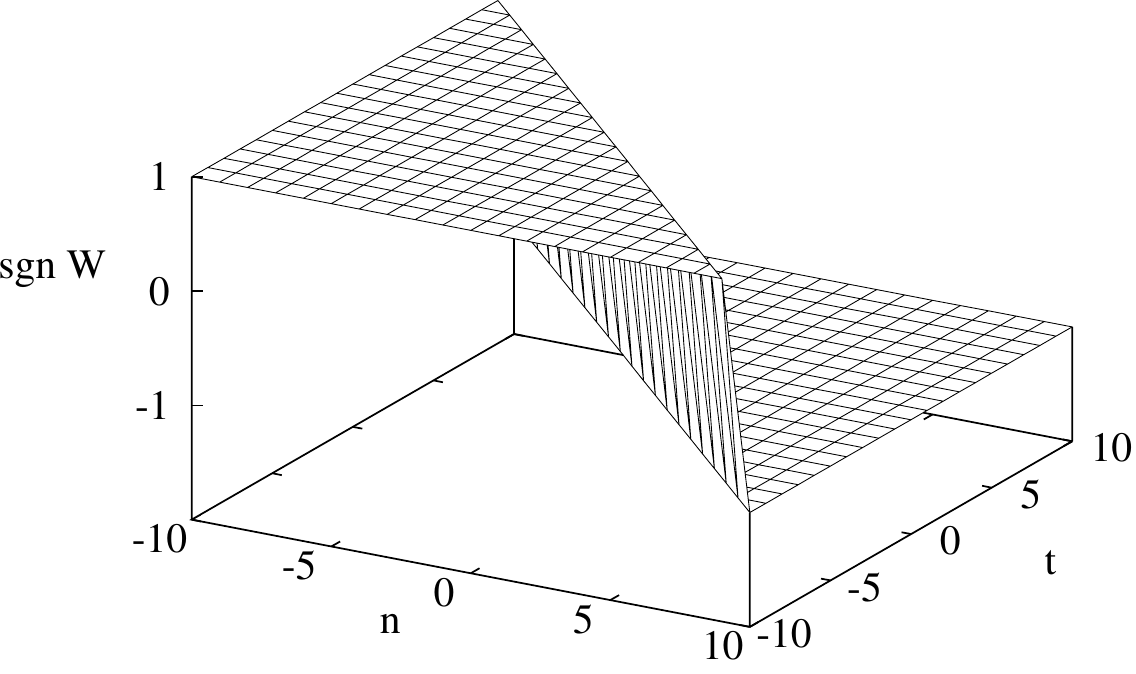} \qquad
\includegraphics[width=6.5cm]{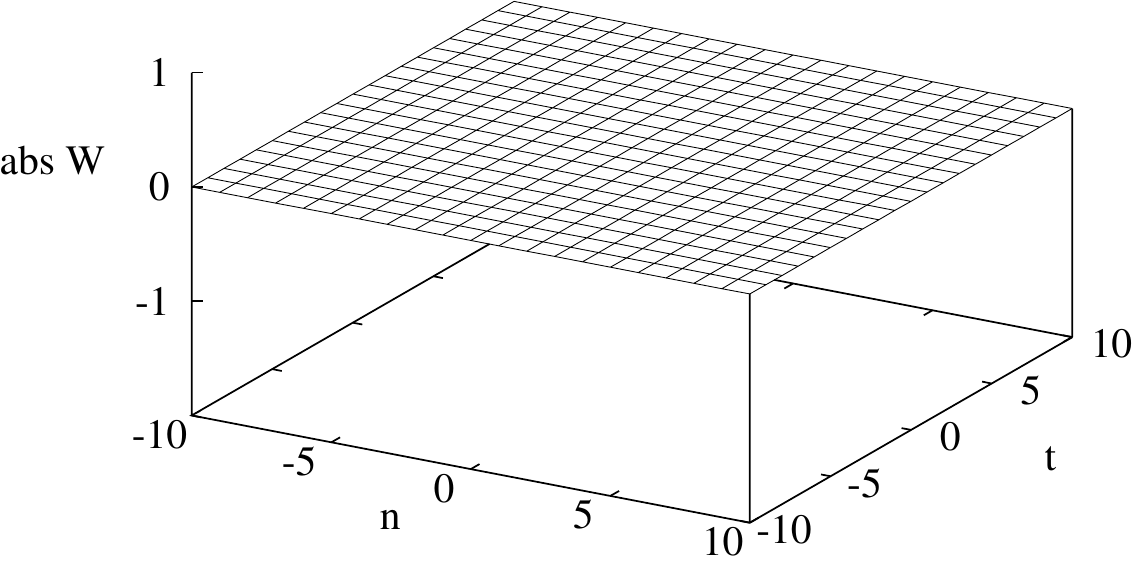}

\caption{Signs (left) and absolute values (right) of 1-soliton solution for udsG.}
\label{udsG:fig:1sol}
\end{figure}

Fig.~\ref{udsG:fig:1sol} shows the solution with
\[
D=-1, \qquad C=\ominus1, \qquad P=2
\]
in the light-cone coordinates~\eqref{lightcone}. It is somehow dif\/f\/icult to depict ultradiscrete numbers in f\/igures; here signs and absolute values are displayed separately, and signs are mapped from $\ominus0$, $\bal{0}$, $\oplus0$ to $-1$, $0$, $1$, respectively (balanced elements do not appear in the f\/igure, though). Observe that the form of the 1-soliton solution $w$ for dsG is preserved in the signs. Absolute values are always~$0$, corresponding to the fact that~$w$ asymptotically behaves as~$\pm1$.

\subsubsection{Traveling-wave solution}

\begin{figure}[t]
\centering
\includegraphics[width=6.5cm]{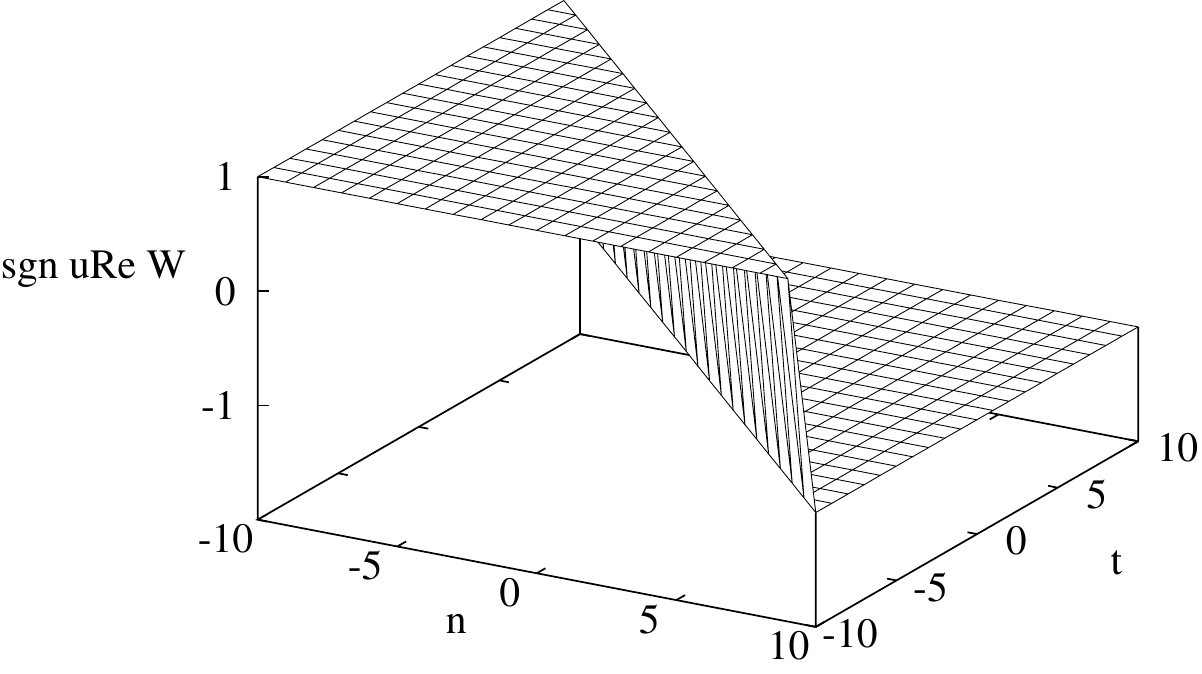} \qquad
\includegraphics[width=6.5cm]{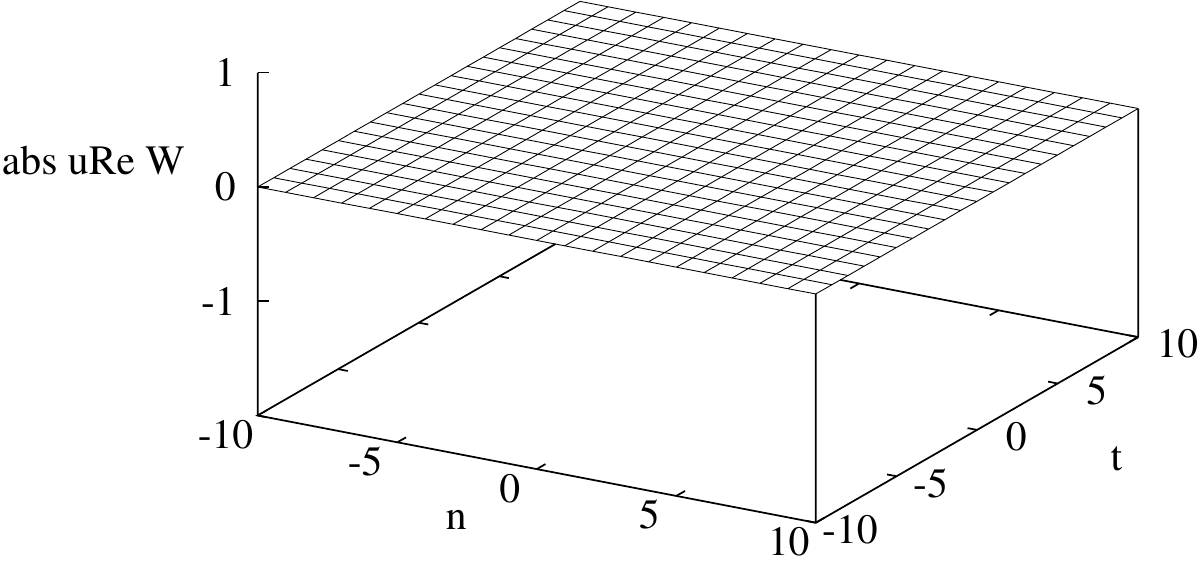}

\includegraphics[width=6.5cm]{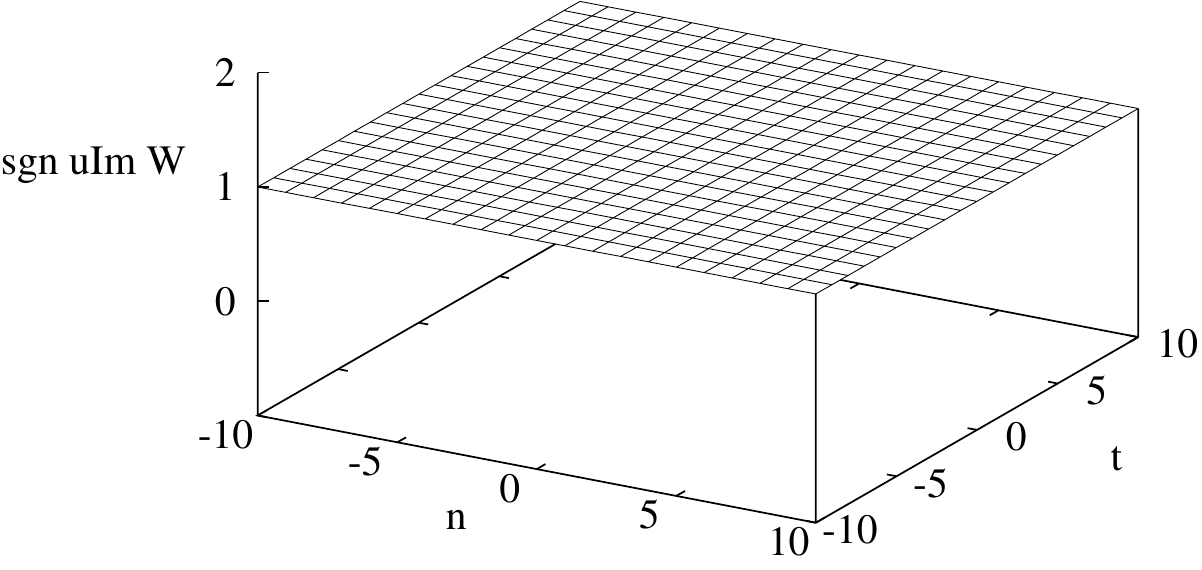} \qquad
\includegraphics[width=6.5cm]{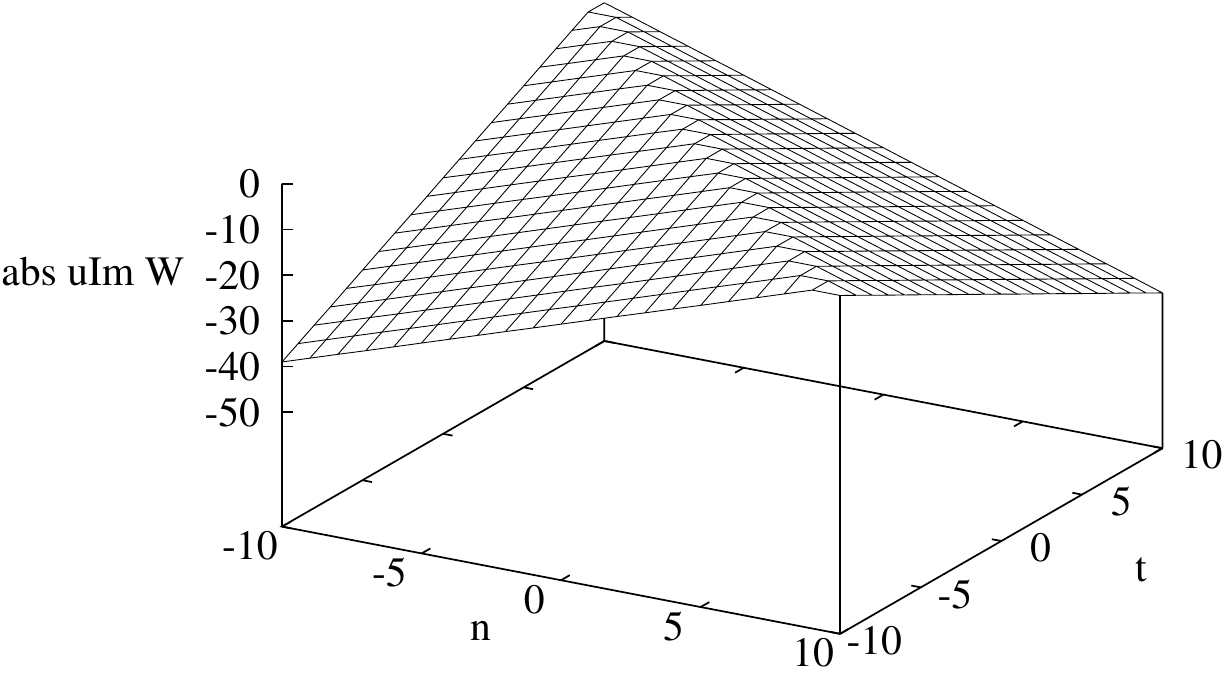}

\caption{Traveling-wave solution for udsG.}
\label{udsG:fig:tr}
\end{figure}

If we replace $C$ by $CI$ and redef\/ine $F=CP^lQ^m$ $(C\in\uR^\otimes)$, we obtain
\begin{gather*}
T=0\oplus FI, \qquad S=0\ominus FI
\end{gather*}
and
\begin{gather*}
W\bals\frac{\lt(0\ominus F^2\rt)\oplus FI}{0\oplus F^2}.
\end{gather*}
We choose odd $C$ and even $P$, $Q$ so that $0\ominus F^2$ is always signed and reduction of balances can be applied. This solution no longer admits deterministic time evolution, but is apparently ultradiscretization of the traveling-wave solution~\eqref{dsG:traveling} for dsG. Fig.~\ref{udsG:fig:tr} shows the solution with
\[
D=-1, \qquad C=1, \qquad P=2.
\]
The $\uRe$ and $\uIm$ parts are displayed separately. The prof\/ile of the traveling-wave solution for dsG is preserved well.

\subsubsection{2-soliton solution}

Assume
\begin{gather}
T\bals 0\oplus F_1\oplus F_2\oplus AF_1F_2, \qquad S\bals 0\ominus F_1\ominus F_2\oplus AF_1F_2, \label{udsG:2sol}
\end{gather}
where $A\in\uR^\otimes$ and $F_j=C_jP_j^lQ_j^m$. We also assume
\[
P_1\ne P_2, \qquad Q_1\ne Q_2.
\]
By substitution, we f\/ind the pair of the dispersion relation
\begin{gather*}
0\oplus P_jQ_j\bals P_j\oplus Q_j %\label{udsG:disprel}
\end{gather*}
and the relation
\begin{gather}
A(0\ominus P_1P_2)(0\ominus Q_1Q_2)\oplus (P_1\ominus P_2)(Q_1\ominus Q_2)\bals\minf \label{udsG:coupling_const}
\end{gather}
is a suf\/f\/icient condition for \eqref{udsG:2sol} to become a solution. Obviously, \eqref{udsG:coupling_const} is ultradiscretization of~\eqref{dsG:coupling_const}. When $P_1=P_2=0$ or $Q_1=Q_2=0$, any $A$ satisf\/ies \eqref{udsG:coupling_const}. When $P_1=Q_2=0$, we have
\begin{gather*}
A(0\ominus P_2)(0\ominus Q_1)\oplus(0\ominus P_2)(Q_1\ominus 0)\bals\minf \ \LeadsTo \ A=0.
\end{gather*}
The case $P_2=Q_1=0$ is similar.

\begin{figure}[t]
\centering
\includegraphics[width=6.5cm]{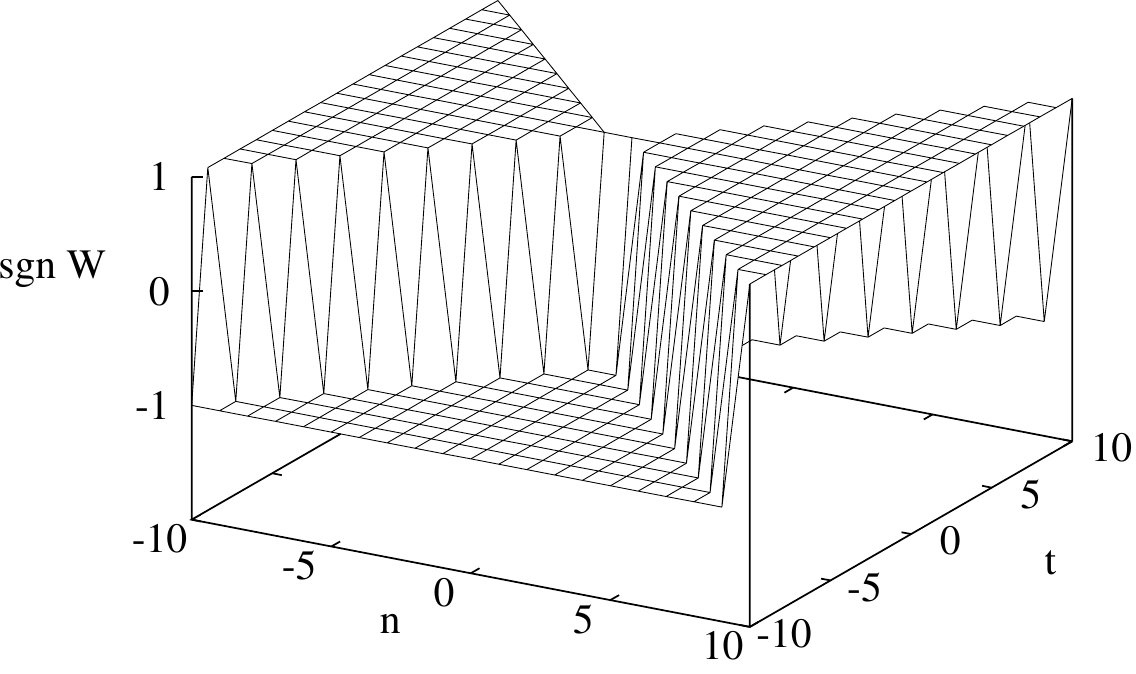} \qquad
\includegraphics[width=6.5cm]{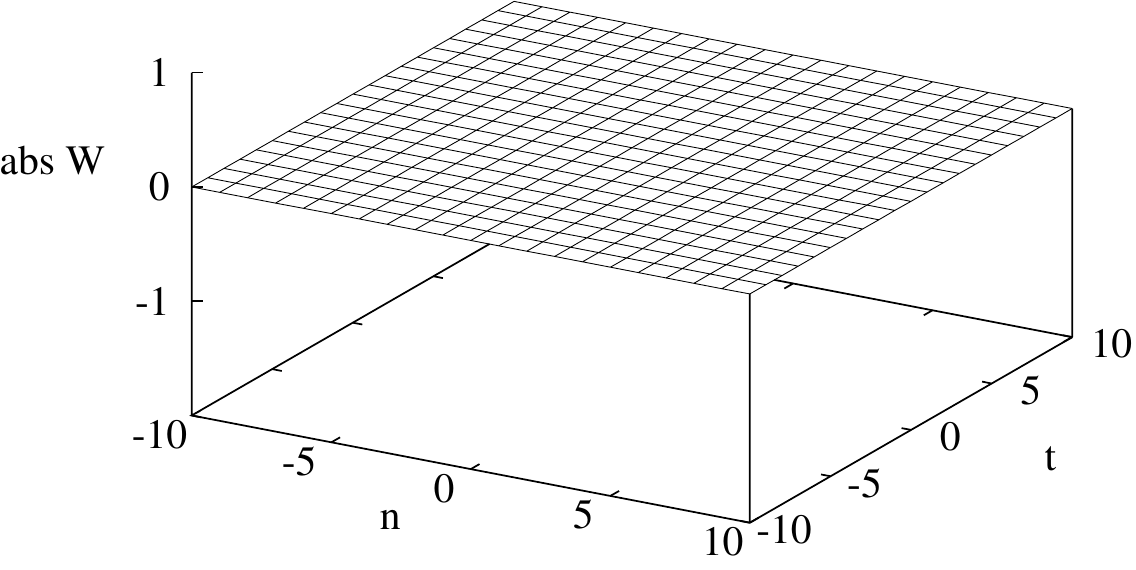}
\caption{2-soliton solution for udsG.}
\label{udsG:fig:2sol}
\end{figure}

We can choose $A,C_j,P_j,Q_j\in\uZ$ such that $0\oplus AF_1F_2$ is always positive, even and $F_1\oplus F_2$ is negative, odd. Then the solution is determined as
\begin{gather*}
T=0\oplus F_1\oplus F_2\oplus AF_1F_2, \qquad S=0\ominus F_1\ominus F_2\oplus AF_1F_2.
\end{gather*}
This admits deterministic time evolution, of course. Fig.~\ref{udsG:fig:2sol} shows the solution with
\[
D=-1, \qquad C_1=C_2=\ominus1, \qquad P_1=Q_2=4.
\]

\subsubsection{Kink-antikink and kink-kink solutions}

If we replace $C_1$ by $C_1I$, $C_2$ by $\ominus C_2I$, and redef\/ine $F_j=C_jP_j^lQ_j^m$ $(C_j\in\uR^\otimes)$ in the 2-soliton solution, we obtain
\begin{gather*}
T\bals 0\oplus AF_1F_2\oplus\lt(F_1\ominus F_2\rt)I, \qquad S\bals 0\oplus AF_1F_2\ominus\lt(F_1\ominus F_2\rt)I
\end{gather*}
and
\begin{gather*}
W\bals\frac{\lt((0\oplus AF_1F_2)^2\ominus(F_1\ominus F_2)^2\rt)\oplus(0\oplus AF_1F_2)(F_1\ominus F_2)I}{(0\oplus AF_1F_2)^2\oplus(F_1\ominus F_2)^2}.
\end{gather*}
We choose $C_j,P_j,Q_j\in\uZ$ such that
%\begin{subequations}
\begin{gather*}
\uabs{F_1} \equiv 1, \qquad \uabs{F_2} \equiv 3 \pmod4
\end{gather*}
or
\begin{gather*}
\uabs{F_1} \equiv 3, \qquad \uabs{F_2} \equiv 1 \pmod4.
\end{gather*}
%\end{subequations}
Then $F_1\ominus F_2$ and $(0\oplus AF_1F_2)^2\ominus(F_1\ominus F_2)^2$ are always signed and the balance relations become equalities.

If we set
$
P_1=Q_2$, $P_2=Q_1$,
we have the kink-antikink solution. Similarly, setting
$P_1=Q_2^{-1}$, $P_2=Q_1^{-1}$
gives the kink-kink solution. These solutions does not admit deterministic time evolution, but are ultradiscretization of \eqref{dsG:ka}, \eqref{dsG:kk}. The ultradiscretization of the breather solution is unclear.

\begin{figure}[t]
\centering
\includegraphics[width=6.5cm]{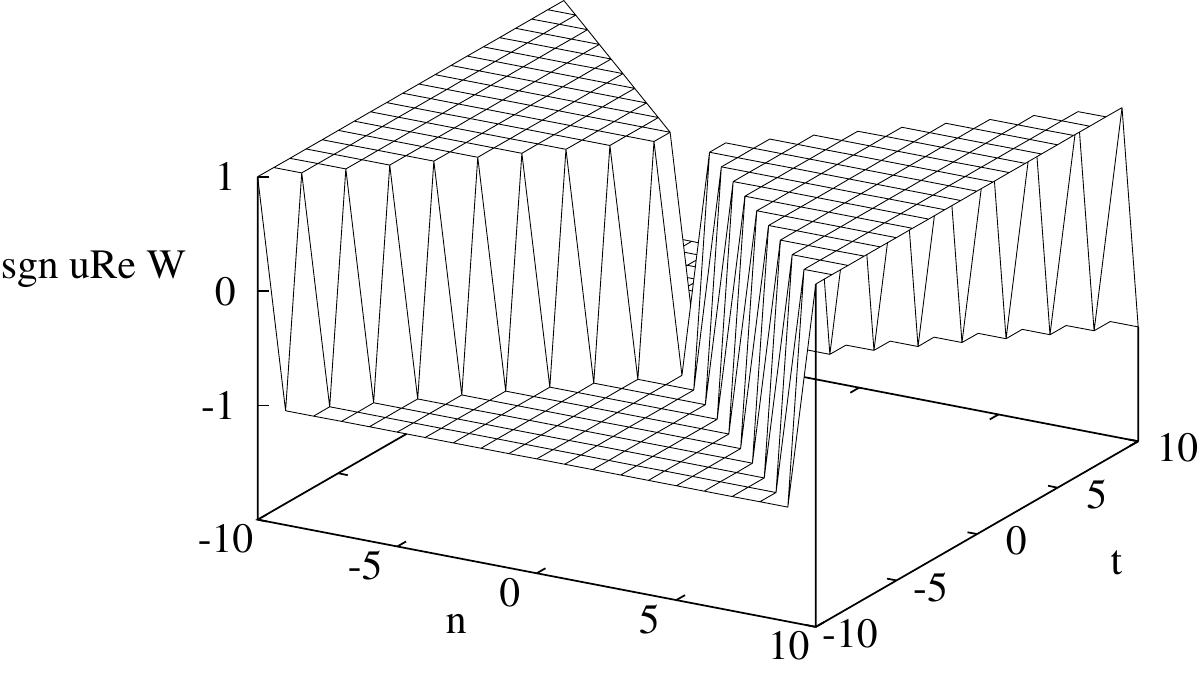} \qquad
\includegraphics[width=6.5cm]{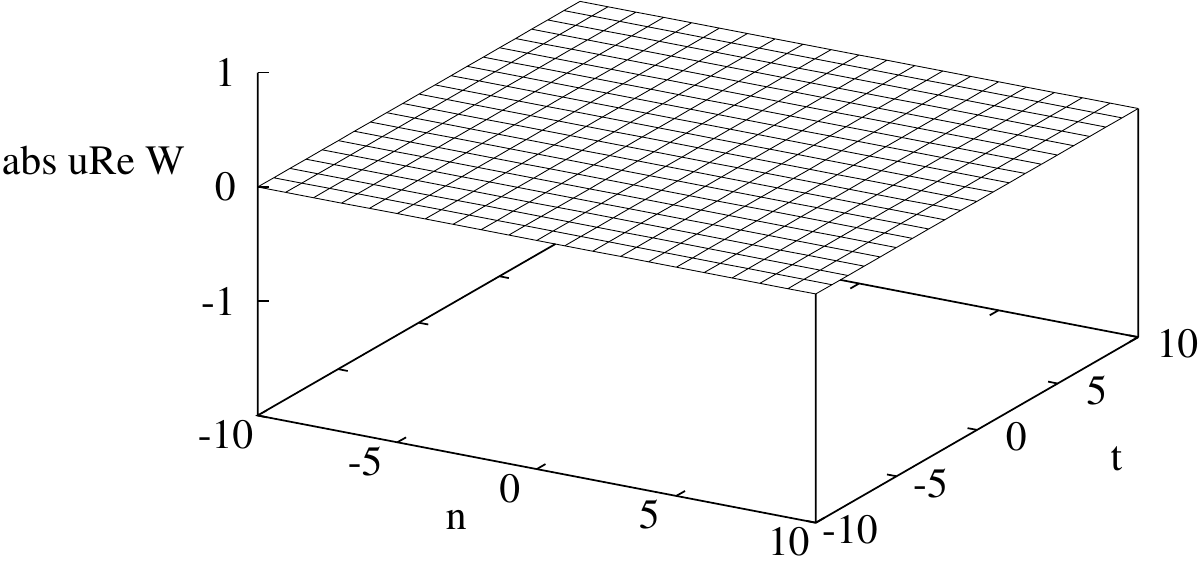}

\includegraphics[width=6.5cm]{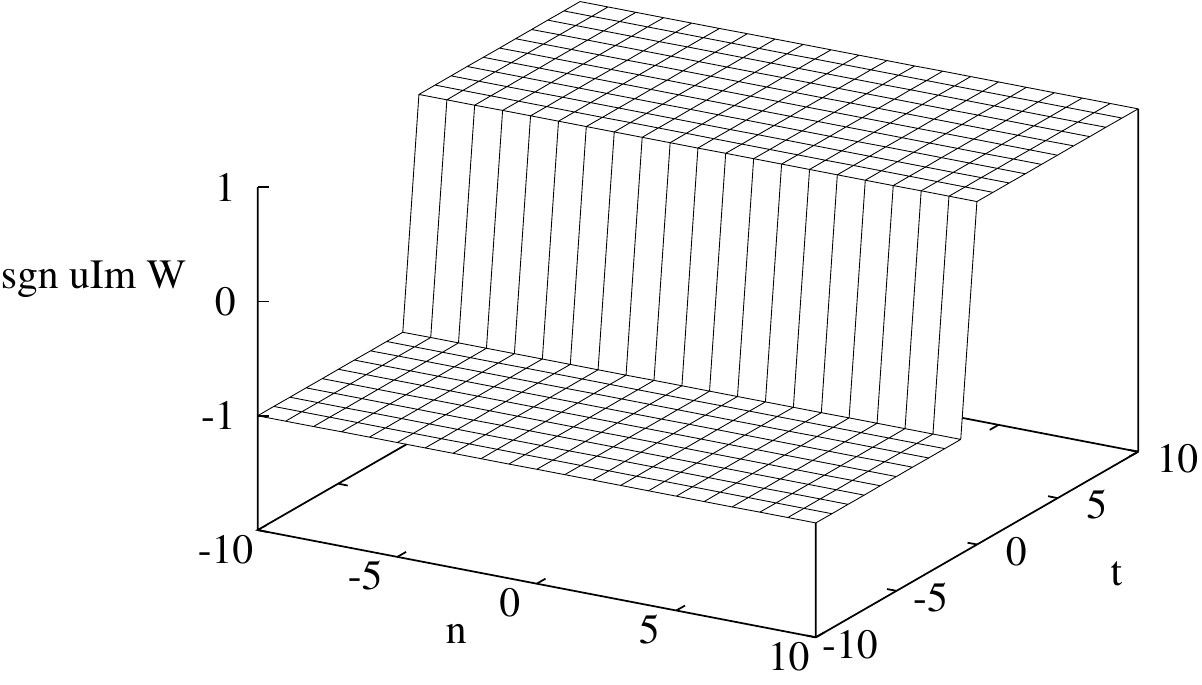} \qquad
\includegraphics[width=6.5cm]{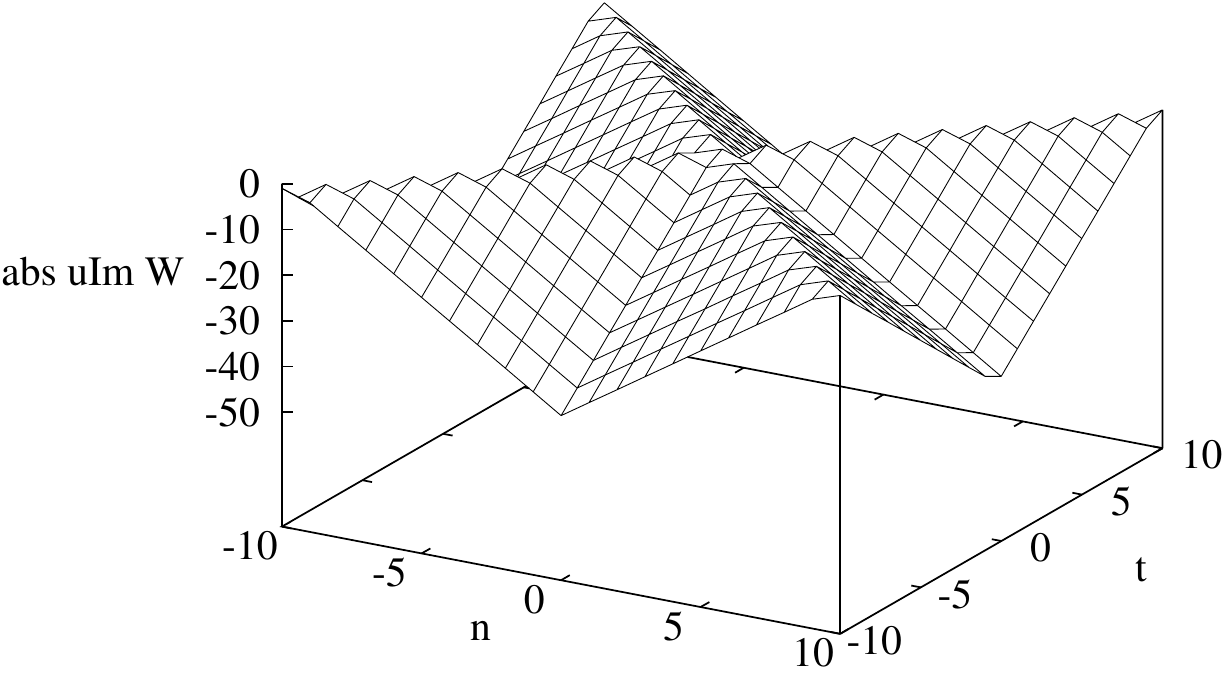}
\caption{Kink-antikink solution for udsG.}
\label{udsG:fig:ka}
\end{figure}

\begin{figure}[t]
\centering
\includegraphics[width=6.5cm]{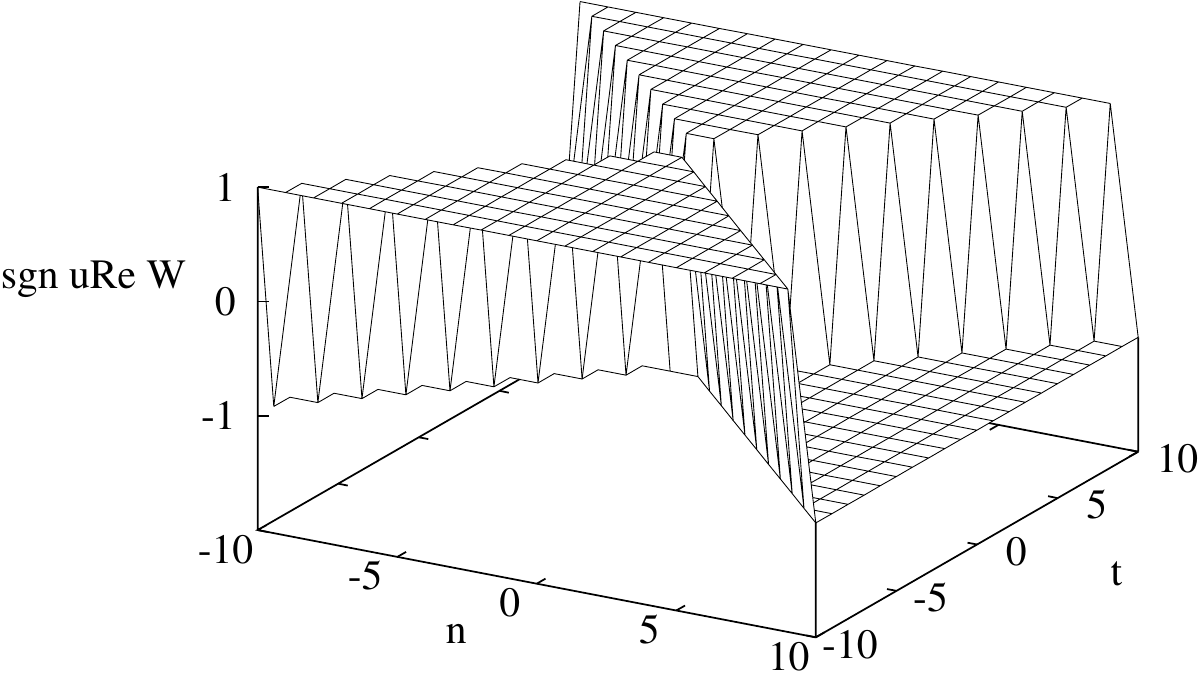} \qquad
\includegraphics[width=6.5cm]{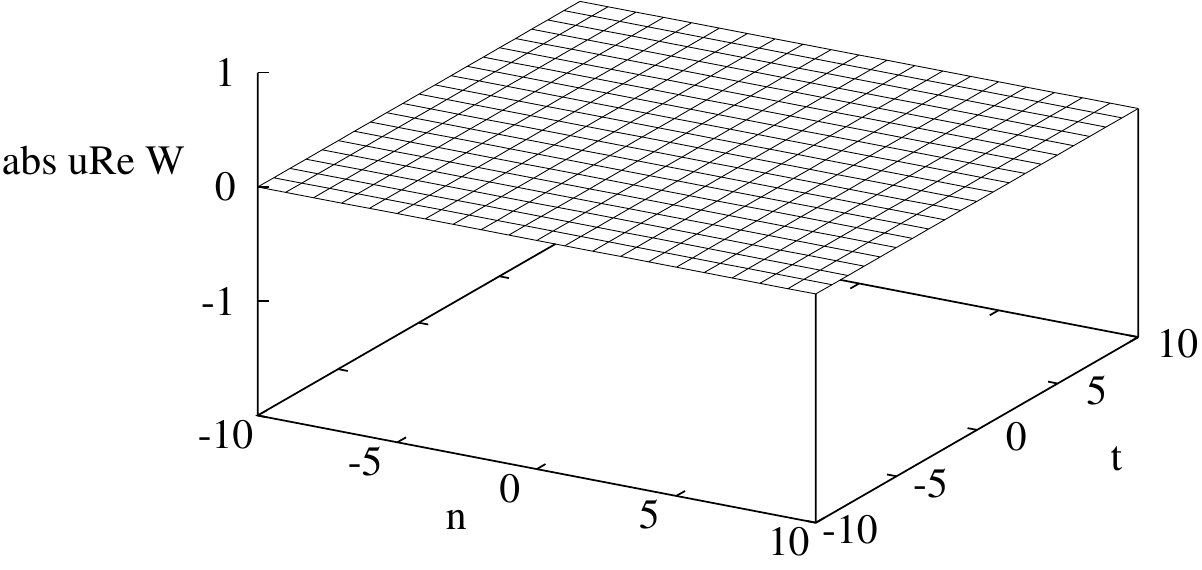}

\includegraphics[width=6.5cm]{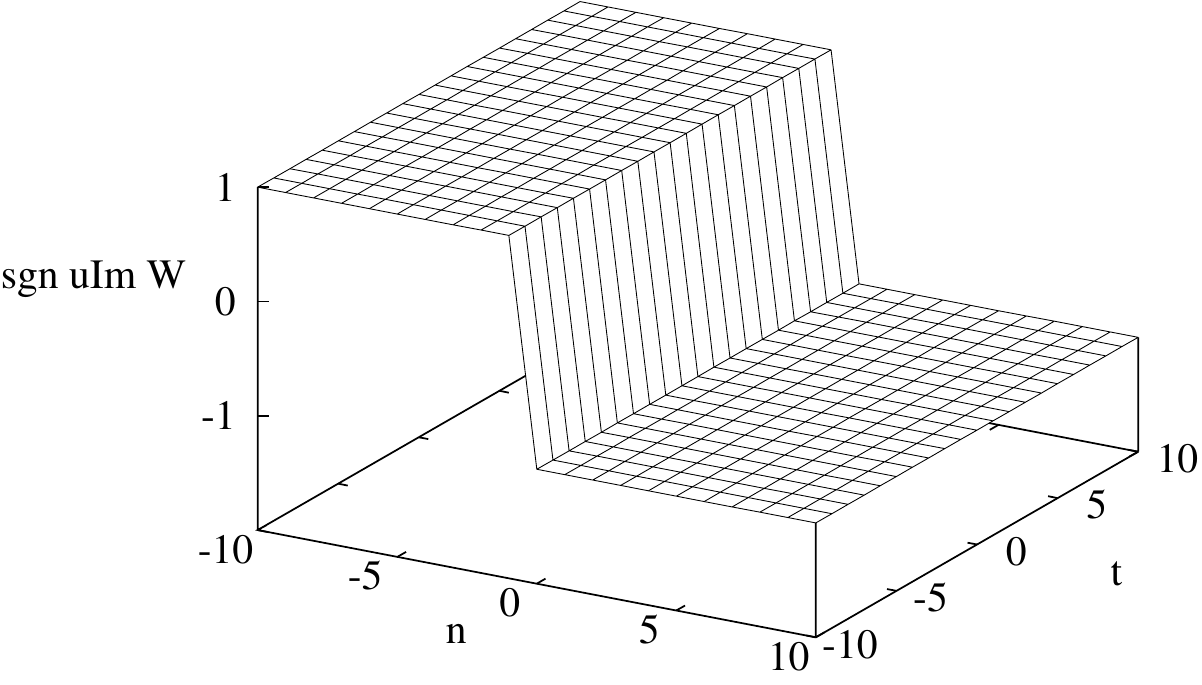} \qquad
\includegraphics[width=6.5cm]{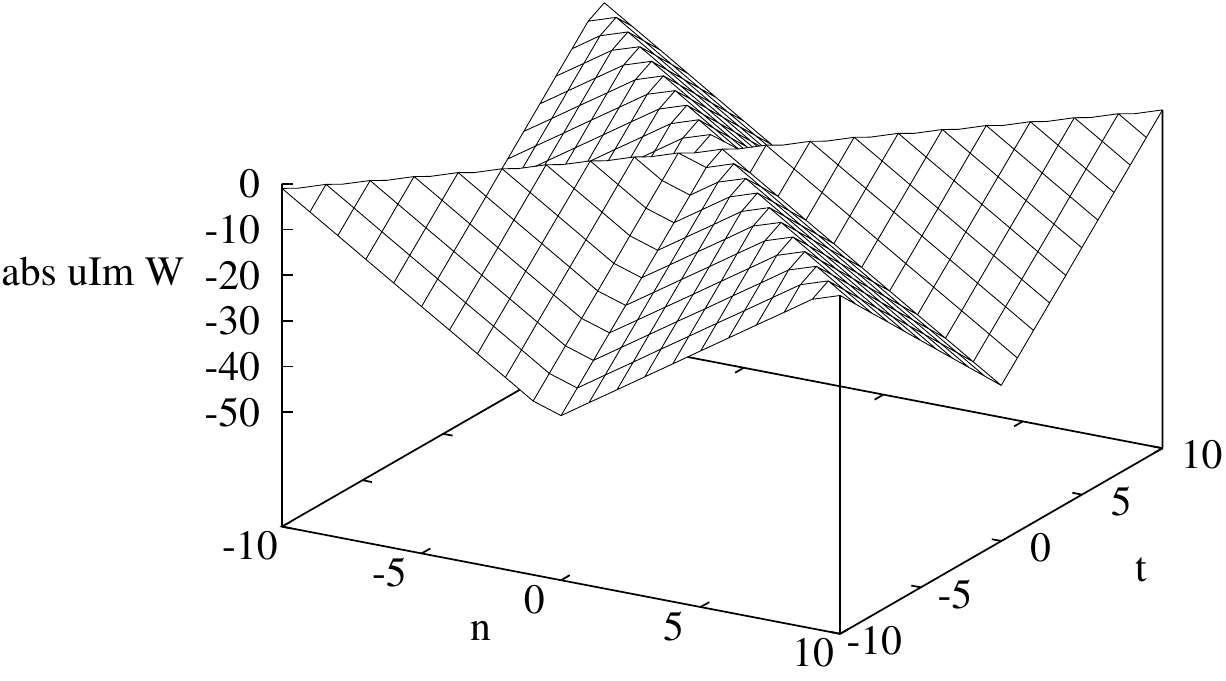}
\caption{Kink-kink solution for udsG.}
\label{udsG:fig:kk}
\end{figure}

Fig.~\ref{udsG:fig:ka} shows the kink-antikink solution with
\[
D=-1, \qquad C_1=1, \qquad C_2=-1, \qquad P_1=Q_2=4,
\]
and Fig.~\ref{udsG:fig:kk} shows the kink-kink solution with
\[
D=-1, \qquad C_1=\ominus1, \qquad C_2=\ominus(-1), \qquad P_1=Q_2^{-1}=4.
\]
Observe that in the $\uIm$ part, two waves approach to each other for $t<0$, collide at $t=0$, and move away from each other for $t>0$. In the kink-antikink solution, the two waves have the same sign and \textit{bump up} by collision. In the kink-kink solution, the two have opposite signs and  \textit{reflect} by collision.

\section{Noncommutative discrete\\ and ultradiscrete sine-Gordon equations}

In this section, we propose a noncommutative discrete analogue of the sine-Gordon equation as a compatibility condition of a certain linear system. This equation reduces to the commutative version once the underlying algebra turns out to be commutative and one simple reduction condition is applied. A reduction from the noncommutative discrete KP equation~\cite{Kondo2011, Nimmo2006} also gives the equation, and continuum limit of the equation gives the noncommutative (continuous) sine-Gordon equation, which is already known in a dif\/ferent context~\cite{Lechtenfeld2005}. We def\/ine the Darboux transformation, which constructs new solutions from old ones, and obtain Casoratian-type solutions by repeating it. Explicitly setting the starting solutions for repetition, we derive so-called multisoliton solutions.

Along the construction of Casoratian-type solutions, quasideterminants~\cite{GGRW} are used, which is a noncommutative extension of determinants. The theory needs some space for explanation, but it is not essential to the main story. Therefore, we only brief\/ly explain the def\/inition and some properties of them in~\apdref{apd:qdet}. For details, see~\cite{GGRW}.

We f\/inally propose a noncommutative ultradiscrete analogue of the sine-Gordon equation. Noncommutative ultradiscrete setting is probably one of the hardest environments for integrable systems to exist, but we manage to obtain 1-soliton and 2-soliton solutions by ultradiscretization.

Notations are slightly changed in this section because of the complexity of expressions we are going to manipulate. Shifts are always indicated after a comma like $f_{,l}$. This is to distinguish indices and shifts. In addition, shift operators $T_l$, $T_m$ are also used:
\begin{gather*}
T_l f=f_{,l}=f(l+1, m), \qquad T_m f=f_{,m}=f(l, m+1).
\end{gather*}
Do not confuse these with the ultradiscretized $\tau$ function of the previous section; in the noncommutative setting, $\tau$ functions do not seem to exist. We also use superscripts for elements of matrices. For example,
\[
w=\lt(w^{\iota\kappa}\rt)=\bpm w^{11} & \cdots & w^{1N} \\ \vdots & \ddots & \vdots \\ w^{N1} & \cdots & w^{NN}\epm.
\]

\subsection{Noncommutative discrete sine-Gordon equation}\label{sec:ncdsG}

\subsubsection{Linear system}

Let $w=w(l,m)$, $v=v(l,m)$ be functions $\Z^2\to\Mat(N,\C)$ and
%\begin{subequations}
\begin{gather*}
B_l =\M{w_{,l}w^{-1}}{-a\lambda}{-a\lambda}{v_{,l}v^{-1}}, \qquad
B_m =\M{1}{-b\lambda^{-1}w_{,m}v^{-1}}{-b\lambda^{-1}v_{,m}w^{-1}}{1},
\end{gather*}
%\end{subequations}
where $a,b,\lambda\in\C^\times$ are parameters. Consider the linear system
\begin{gather}
T_l\V{\phi}{\psi}=B_l\V{\phi}{\psi}, \qquad T_m\V{\phi}{\psi}=B_m\V{\phi}{\psi} \label{ncdsG:linsys}
\end{gather}
for $\phi,\psi:\Z^2\to\Mat(N,\C)$. Denoting entrywise shift operations by $T_mB_l=B_{l,m}$ etc., we have
\begin{gather*}
T_mT_l\V{\phi}{\psi} =B_{l,m}T_m\V{\phi}{\psi}=B_{l,m}B_m\V{\phi}{\psi}, \\
T_lT_m\V{\phi}{\psi} =B_{m,l}T_l\V{\phi}{\psi}=B_{m,l}B_l\V{\phi}{\psi}.
\end{gather*}
These must coincide, so we require the compatibility condition
\begin{gather*}
B_{l,m}B_m=B_{m,l}B_l.
\end{gather*}
This is equivalent to
\begin{subequations}
\begin{gather}
w_{,lm}w_{,m}^{-1}-w_{,l}w^{-1}+ab\lt(v_{,m}w^{-1}-w_{,lm}v_{,l}^{-1}\rt) =0, \label{ncdsG:ncdsG1} \\
v_{,lm}v_{,m}^{-1}-v_{,l}v^{-1}+ab\lt(w_{,m}v^{-1}-v_{,lm}w_{,l}^{-1}\rt) =0. \label{ncdsG:ncdsG2}
\end{gather}
\end{subequations}
We call the pair \eqref{ncdsG:ncdsG1} and \eqref{ncdsG:ncdsG2} the noncommutative discrete sine-Gordon equation (ncdsG).

\begin{Proposition}
When $N=1$, the reduction condition
\begin{gather}
wv=1 \label{ncdsG:red}
\end{gather}
gives the $($commutative$)$ discrete sine-Gordon equation~{\rm \cite{DJMIII, Hirota1977}}
\begin{gather}
\frac{w_{,lm}}{w_{,m}}-\frac{w_{,l}}{w}+ab\lt(\frac{1}{w_{,m}w}-w_{,lm}w_{,l}\rt)=0. \label{ncdsG:dsG}
\end{gather}
\end{Proposition}
\begin{proof}
Under \eqref{ncdsG:red}, \eqref{ncdsG:ncdsG1} is apparently equivalent to \eqref{ncdsG:dsG}. Since
\begin{gather*}
 \mbox{(l.h.s.\ of \eqref{ncdsG:ncdsG1})}\times\lt(w_{,m}w-\frac{1}{w_{,lm}w_{,l}}\rt)
 =w_{,lm}w+\frac{1}{w_{,lm}w}-w_{,l}w_{,m}-\frac{1}{w_{,l}w_{,m}} \\
\hphantom{\mbox{(l.h.s.\ of \eqref{ncdsG:ncdsG1})}\times\lt(w_{,m}w-\frac{1}{w_{,lm}w_{,l}}\rt)}{}
=\mbox{(l.h.s.\ of \eqref{ncdsG:ncdsG2})}\times\lt(\frac{1}{w_{,m}w}-w_{,lm}w_{,l}\rt),
\end{gather*}
\eqref{ncdsG:ncdsG2} is also equivalent to \eqref{ncdsG:dsG}.
\end{proof}

For any $w_0$ satisfying
\begin{gather*}
abw_{0,\ov lm}=\frac{w_{0,l\ov m}}{ab},
\end{gather*}
\eqref{ncdsG:ncdsG1} and \eqref{ncdsG:ncdsG2} are solved by $(w,v)=\big(w_0, abw_{0,\ov lm}\big)$,
which is not an interesting solution. We consider other types of solutions in the rest of this section.

\subsubsection{Reduction from the noncommutative discrete KP equation}

Let $w_i=w_i(n_1,n_2,n_3)$ $(i=1,2,3)$ be functions $\Z^3\to\Mat(N,\C)$. The noncommutative discrete KP equation~\cite{Kondo2011, Nimmo2006} is the set of equations
\begin{gather}
w_{i,j}(c_i-c_j)w_i^{-1}+w_{j,k}(c_j-c_k)w_j^{-1}+w_{k,i}(c_k-c_i)w_k^{-1}=0 \label{ncdsg:ncdKP}
\end{gather}
for any combination of $i,j,k\in\{1,2,3\}$. Here $i$, $j$, $k$ can take same values, and shifts are denoted like
\[
w_{1,2}=w_1(n_1,n_2+1,n_3).
\]
$c_i\in\C^\times$ are parameters taking mutually dif\/ferent values.

We replace $(n_1,n_2,n_3)$ by new coordinates $(n_1',n_2',n_3')=(n_1-n_3,n_2,n_3)$. Shifts are also in new coordinates, and double-shifts are to be used:
\[
w_{1,2}=w_1(n_1',n_2'+1,n_3'), \qquad w_{2,13}=w_2(n_1'+1,n_2',n_3'+1), \qquad \mbox{etc}.
\]
Then, setting
\begin{gather*}
\delta=\frac{c_1-c_3}{c_1-c_2},
\end{gather*}
we can rewrite \eqref{ncdsg:ncdKP} as
\begin{gather*}
 w_{1,2}w_1^{-1}+(\delta-1)w_{2,13}w_2^{-1}-\delta w_{3,1}w_3^{-1}=0, \\
 w_{1,2}w_1^{-1}=w_{2,1}w_2^{-1}, \qquad w_{2,13}w_2^{-1}=w_{3,2}w_3^{-1}, \qquad w_{3,1}w_3^{-1}=w_{1,13}w_1^{-1}.
\end{gather*}

By imposing the reduction condition
\begin{gather*}
w_i(n_1'+2,n_2',n_3')=w_i(n_1',n_2',n_3')
\end{gather*}
and def\/ining $v_i=w_{i,1}$, we obtain
\begin{subequations}
\begin{gather}
w_{1,2}w_1^{-1}+(\delta-1)v_{2,3}w_2^{-1}-\delta v_3w_3^{-1}=0, \label{ncdKP:r1} \\
v_{1,2}v_1^{-1}+(\delta-1)w_{2,3}v_2^{-1}-\delta w_3v_3^{-1}=0, \label{ncdKP:r2} \\
w_{1,2}w_1^{-1}=v_2w_2^{-1}, \label{ncdKP:r3} \\
v_{2,3}w_2^{-1}=w_{3,2}w_3^{-1}, \label{ncdKP:r4} \\
v_3w_3^{-1}=v_{1,3}w_1^{-1}, \label{ncdKP:r5} \\
v_{1,2}v_1^{-1}=w_2v_2^{-1}, \label{ncdKP:r6} \\
w_{2,3}v_2^{-1}=v_{3,2}v_3^{-1}, \label{ncdKP:r7} \\
w_3v_3^{-1}=w_{1,3}v_1^{-1}. \label{ncdKP:r8}
\end{gather}
\end{subequations}

\begin{Proposition}
For any $w_1$, $v_1$ satisfying \eqref{ncdKP:r1}--\eqref{ncdKP:r8},
\[
(w,v)=(w_1,v_1) \qquad (l=n_2', m=n_3')
\]
solves \eqref{ncdsG:ncdsG1} and \eqref{ncdsG:ncdsG2} with $ab=\delta$.
\end{Proposition}
\begin{proof}
Let us rewrite \eqref{ncdKP:r1} using only $w_1$, $v_1$. From \eqref{ncdKP:r5} we immediately have{\samepage
\begin{gather}
w_{1,2}w_1^{-1}+(\delta-1)v_{2,3}w_2^{-1}-\delta v_{1,3}w_1^{-1}=0, \label{ncdKP:tmp0}
\end{gather}
and thus try to rewrite the second term.}

\eqref{ncdKP:r4} implies
\begin{gather}
(\delta-1)v_{2,3}w_2^{-1} =(\delta-1)w_{3,2}w_3^{-1}
 =w_{3,2}v_{3,2}^{-1}\cdot(\delta-1)v_{3,2}v_3^{-1}\cdot v_3w_3^{-1}. \label{ncdKP:tmp1}
\end{gather}
Then, \eqref{ncdKP:r8} implies
\begin{gather}
w_{3,2}v_{3,2}^{-1}=w_{1,23}v_{1,2}^{-1}, \qquad v_3w_3^{-1}=v_1w_{1,3}^{-1},
\end{gather}
and \eqref{ncdKP:r7} implies
\[
(\delta-1)v_{3,2}v_3^{-1}=(\delta-1)w_{2,3}v_2^{-1}.
\]
By \eqref{ncdKP:r2} and \eqref{ncdKP:r8}, this equation becomes
\begin{gather}
(\delta-1)w_{2,3}v_2^{-1} =\delta w_3v_3^{-1}-v_{1,2}v_1^{-1}
 =\delta w_{1,3}v_1^{-1}-v_{1,2}v_1^{-1}. \label{ncdKP:tmp2}
\end{gather}
Combining \eqref{ncdKP:tmp1}--\eqref{ncdKP:tmp2}, we obtain
\begin{gather*}
(\delta-1)v_{2,3}w_2^{-1} =w_{1,23}v_{1,2}^{-1}\cdot\lt(\delta w_{1,3}v_1^{-1}-v_{1,2}v_1^{-1}\rt)\cdot v_1w_{1,3}^{-1}
=\delta w_{1,23}v_{1,2}^{-1}-w_{1,23}w_{1,3}^{-1}.
\end{gather*}

Finally, \eqref{ncdKP:tmp0} becomes
\begin{subequations}
\begin{gather}
w_{1,2}w_1^{-1}-w_{1,23}w_{1,3}^{-1}+\delta\big(w_{1,23}v_{1,2}^{-1}-v_{1,3}w_1^{-1}\big)=0. \label{ncdKP:ncdsG1}
\end{gather}
In the same way, \eqref{ncdKP:r2} becomes
\begin{gather}
v_{1,2}v_1^{-1}-v_{1,23}v_{1,3}^{-1}+\delta\big(v_{1,23}w_{1,2}^{-1}-w_{1,3}v_1^{-1}\big)=0. \label{ncdKP:ncdsG2}
\end{gather}
\end{subequations}
If we set
\[
(w,v)=(w_1,v_1), \qquad (l,m)=(n_2',n_3'), \qquad ab=\delta,
\]
we have
\[
w_{1,2}=w_{,l}, \qquad w_{1,23}=w_{,lm}, \qquad \mbox{etc}.
\]
Therefore, \eqref{ncdKP:ncdsG1} and \eqref{ncdKP:ncdsG2} become \eqref{ncdsG:ncdsG1} and \eqref{ncdsG:ncdsG2}.
\end{proof}

\begin{Remark}
The solution constructed here seems to be only a part of the whole solutions of~\eqref{ncdsG:ncdsG1} and~\eqref{ncdsG:ncdsG2}, since it satisf\/ies extra conditions
\[
w_{1,2}w_1^{-1}\cdot v_{1,2}v_1^{-1}=1, \qquad v_{1,3}w_1^{-1}\cdot w_{1,3}v_1^{-1}=1.
\]
\end{Remark}

\subsubsection{Continuum limit}

Assume $w$ is also a function $w(x,t)$ of continuum variables $x,t\in\R$ and has an expansion
\begin{gather*}
w(x+r,t+s)=w+(rw_x+sw_t)+\frac{1}{2}\lt(r^2w_{xx}+2rsw_{xt}+s^2w_{tt}\rt)+\cdots,
\end{gather*}
where $w_x=\p w/\p x$, etc. Connect $l$, $m$ to $x$, $t$ via the Miwa transformation
\begin{gather*}
w(x,t;l,m)=w(x+la,t+mb).
\end{gather*}
Assume similarly for $v=v(x,t;l,m)$. Then we have
\begin{gather*}
w_{,l} =w+aw_x+\frac{a^2}{2}w_{xx}+\cdots, \\
w_{,lm} =w+(aw_x+bw_t)+\frac{1}{2}\lt(a^2w_{xx}+2abw_{xt}+b^2w_{tt}\rt)+\cdots, \\
w_{,m}^{-1} =w^{-1}-bw^{-1}w_tw^{-1}  -\frac{b^2}{2}\lt(w^{-1}w_{tt}w^{-1}-2w^{-1}w_tw^{-1}w_tw^{-1}\rt)+\cdots, \\
v_{,l} =\cdots,
\end{gather*}
and from \eqref{ncdsG:ncdsG1}, \eqref{ncdsG:ncdsG2}
\begin{gather*}
0 =w_{,lm}w_{,m}^{-1}-w_{,l}w^{-1}+ab\big(v_{,m}w^{-1}-w_{,lm}v_{,l}^{-1}\big) \\
\hphantom{0}{} =ab\big(w_{xt}w^{-1}-w_xw^{-1}w_tw^{-1}+vw^{-1}-wv^{-1}\big)+\mbox{(higher-order terms)}, \\
0 =ab\big(v_{xt}v^{-1}-v_xv^{-1}v_tv^{-1}+wv^{-1}-vw^{-1}\big)+\mbox{(higher-order terms)}.
\end{gather*}
Taking the limit $a,b\to0$ successively, we obtain
\begin{gather*}
w_{xt}w^{-1}-w_xw^{-1}w_tw^{-1}+vw^{-1}-wv^{-1}=0, \\
v_{xt}v^{-1}-v_xv^{-1}v_tv^{-1}+wv^{-1}-vw^{-1}=0.
\end{gather*}
Since $-w^{-1}w_tw^{-1}=(w^{-1})_t$, these are transformed into
\begin{subequations}
\begin{gather}
\big(w_xw^{-1}\big)_t=wv^{-1}-vw^{-1}, \label{ncsG:ncsG1} \\
\big(w_xw^{-1}+v_xv^{-1}\big)_t=0. \label{ncsG:ncsG2}
\end{gather}
\end{subequations}
We call the pair \eqref{ncsG:ncsG1} and \eqref{ncsG:ncsG2} the noncommutative sine-Gordon equation. A quite similar equation with the same name has been derived in a dif\/ferent context~\cite[(3.10)]{Lechtenfeld2005}.

\begin{Proposition}
When $N=1$, the reduction condition
\begin{gather}
wv=1 \label{ncsG:red}
\end{gather}
gives the $($commutative$)$ sine-Gordon equation
\begin{gather}
u_{xt}=4\sin u, \label{ncsG:sG_u}
\end{gather}
where $u$ is defined by
\begin{gather*}
u=\frac{2}{i}\log w.
\end{gather*}
\end{Proposition}

\begin{proof}
Under \eqref{ncsG:red}, \eqref{ncsG:ncsG2} clearly holds. And \eqref{ncsG:sG_u} is immediate from \eqref{ncsG:ncsG1} since
\begin{gather*}
u_{xt}=\frac{2}{i}\frac{w_{xt}w-w_xw_t}{w^2}, \qquad \sin u=\frac{w^2-w^{-2}}{2i}.  \tag*{\qed}
\end{gather*}
\renewcommand{\qed}{}
\end{proof}

\subsubsection{Darboux transformation}

When $(w, v)$ is a solution for \eqref{ncdsG:ncdsG1} and \eqref{ncdsG:ncdsG2}, the column vector $\Vin{\phi}{\psi}$ satisfying the linear system \eqref{ncdsG:linsys} is called the eigenfunction of $(w, v)$ for eigenvalue $\lambda$.

Let $\Vin{\phi_\lambda}{\psi_\lambda}$, $\Vin{\phi_\mu}{\psi_\mu}$ be eigenfunctions of $(w,v)$ for eigenvalues $\lambda$, $\mu$, respectively. Def\/ine the Darboux transformation of $(w, v)$ and $\Vin{\phi_\lambda}{\psi_\lambda}$ by $\Vin{\phi_\mu}{\psi_\mu}$ as
\begin{gather*}
\wt w=\psi_\mu\phi_\mu^{-1}w, \qquad \wt v=\phi_\mu\psi_\mu^{-1}v, \qquad \begin{pmat} {}\wt\phi_\lambda \\ \wt\psi_\lambda\end{pmat}=K\begin{pmat} {}\phi_\lambda  \\ \psi_\lambda\end{pmat}
\end{gather*}
where
\begin{gather*}
K=\begin{pmat}{} {-}\mu\psi_\mu\phi_\mu^{-1} & \lambda \\ \lambda & -\mu\phi_\mu\psi_\mu^{-1}\end{pmat}.
\end{gather*}

\begin{Theorem}
$(\wt w, \wt v)$ is a solution to \eqref{ncdsG:ncdsG1} and \eqref{ncdsG:ncdsG2}, and $\Vin{\wt\phi_\lambda}{\wt\psi_\lambda}$ is an  eigenfunction of $(\wt w, \wt v)$ for eigenvalue $\lambda$.
\end{Theorem}
\begin{proof}
From the linear system \eqref{ncdsG:linsys}, we can write
\begin{gather*}
w_{,l}w^{-1}=(\phi_{\mu,l}+a\mu\psi_\mu)\phi_\mu^{-1}, \qquad v_{,l}v^{-1}=(\psi_{\mu,l}+a\mu\phi_\mu)\psi_\mu^{-1}, \\
w_{,m}v^{-1}=b^{-1}\mu(\phi_\mu-\phi_{\mu,m})\psi_\mu^{-1}, \qquad v_{,m}w^{-1}=b^{-1}\mu(\psi_\mu-\psi_{\mu,m})\phi_\mu^{-1}.
\end{gather*}
Then we have
\begin{gather*}
\wt w_{,l}\wt w^{-1} =\psi_{\mu,l}\big(\psi_\mu^{-1}+a\mu\phi_{\mu,l}^{-1}\big)
 =v_{,l}v^{-1}+a\mu\big(\psi_{\mu,l}\phi_{\mu,l}^{-1}-\phi_\mu\psi_\mu^{-1}\big), \\
\wt v_{,l}\wt v^{-1} =\phi_{\mu,l}\big(\phi_\mu^{-1}+a\mu\psi_{\mu,l}\big)
 =w_{,l}w^{-1}+a\mu\big(\phi_{\mu,l}\psi_{\mu,l}^{-1}-\psi_\mu\phi_\mu^{-1}\big), \\
\wt w_{,m}\wt v^{-1} =b^{-1}\mu\psi_{\mu,m}\big(\phi_{\mu,m}^{-1}-\phi_\mu^{-1}\big)
 =v_mw^{-1}+b^{-1}\mu\big(\psi_{\mu,m}\phi_{\mu,m}^{-1}-\psi_\mu\phi_\mu^{-1}\big), \\
\wt v_{,m}\wt w^{-1} =b^{-1}\mu\phi_{\mu,m}\big(\psi_{\mu,m}^{-1}-\psi_\mu^{-1}\big)
 =w_{,m}v^{-1}+b^{-1}\mu\big(\phi_{\mu,m}\psi_{\mu,m}^{-1}-\phi_\mu\psi_\mu\big),
\end{gather*}
which imply
\begin{gather*}
\big(\wt w_{,l}\wt w^{-1}\big)_{,m}-\wt w_{,l}\wt w^{-1}+ab\big(\wt v_{,m}\wt w^{-1}-\big(\wt w_{,m}\wt v^{-1}\big)_{,l}\big)=0, \\
\big(\wt v_{,l}\wt v^{-1}\big)_{,m}-\wt v_{,l}\wt v^{-1}+ab\big(\wt w_{,m}\wt v^{-1}-\big(\wt v_{,m}\wt w^{-1}\big)_{,l}\big)=0.
\end{gather*}

Def\/ine $\wt B_l$, $\wt B_m$ by
\[
\wt B_l=\begin{pmat} {}\wt w_{,l}{\wt w}^{-1} & -a\lambda \\ -a\lambda & \wt v_{,l}{\wt v}^{-1} \end{pmat}, \qquad \wt B_m=\begin{pmat} {}1 & -b\lambda^{-1}\wt w_{,m}{\wt v}^{-1} \\ -b\lambda^{-1}\wt v_{,m}{\wt w}^{-1}& 1\end{pmat}.
\]
Then we have
\[
K_{,l}B_l=\wt B_lK, \qquad K_{,m}B_m=\wt B_mK
\]
and thus
\begin{gather*}
T_l\lt(K\V{\phi}{\psi}\rt) =K_{,l}B_l\V{\phi}{\psi}=\wt B_l\lt(K\V{\phi}{\psi}\rt), \\
T_m\lt(K\V{\phi}{\psi}\rt) =K_{,m}B_m\V{\phi}{\psi}=\wt B_m\lt(K\V{\phi}{\psi}\rt).  \tag*{\qed}
\end{gather*}
\renewcommand{\qed}{}
\end{proof}

\subsubsection{Multisoliton solutions}

The simplest solution for \eqref{ncdsG:ncdsG1} and \eqref{ncdsG:ncdsG2} is the vacuum solution $(w,v)=(1,1)$. The linear system \eqref{ncdsG:linsys} of the vacuum solution is
%\begin{subequations}
\begin{gather*}
T_l\V{\phi}{\psi} =\M{1}{-a\lambda}{-a\lambda}{1}\V{\phi}{\psi}, \qquad
T_m\V{\phi}{\psi} =\M{1}{-b\lambda^{-1}}{-b\lambda^{-1}}{1}\V{\phi}{\psi},
\end{gather*}
%\end{subequations}
which has two basic solutions
\begin{gather*}
\V{\phi}{\psi}=\V{(1-a\lambda)^l\lt(1-b\lambda^{-1}\rt)^m}{(1-a\lambda)^l\lt(1-b\lambda^{-1}\rt)^m}, \V{(1+a\lambda)^l\lt(1+b\lambda^{-1}\rt)^m}{-(1+a\lambda)^l\lt(1+b\lambda^{-1}\rt)^m}.
\end{gather*}
Let $\lambda_k$ $(k=1,2,\ldots)$ be mutually dif\/ferent eigenvalues and def\/ine
%\begin{subequations}
\begin{gather*}
\begin{split}
& \phi_k =(1-a\lambda_k)^l\lt(1-b\lambda_k^{-1}\rt)^m+(1+a\lambda_k)^l\lt(1+b\lambda_k^{-1}\rt)^mc_k, \\
& \psi_k =(1-a\lambda_k)^l\lt(1-b\lambda_k^{-1}\rt)^m-(1+a\lambda_k)^l\lt(1+b\lambda_k^{-1}\rt)^mc_k,
\end{split}
\end{gather*}
%\end{subequations}
where $c_k\in\Mat(N,\C)$ are parameters introducing noncommutativity. $\Vin{\phi_k}{\psi_k}$ is of course an eigenfunction of the vacuum solution for eigenvalue $\lambda_k$. Repeating the Darboux transformation by $\Vin{\phi_k}{\psi_k}$, we can construct multisoliton solutions.

\begin{figure}[t]
\centering
\includegraphics[width=6.5cm]{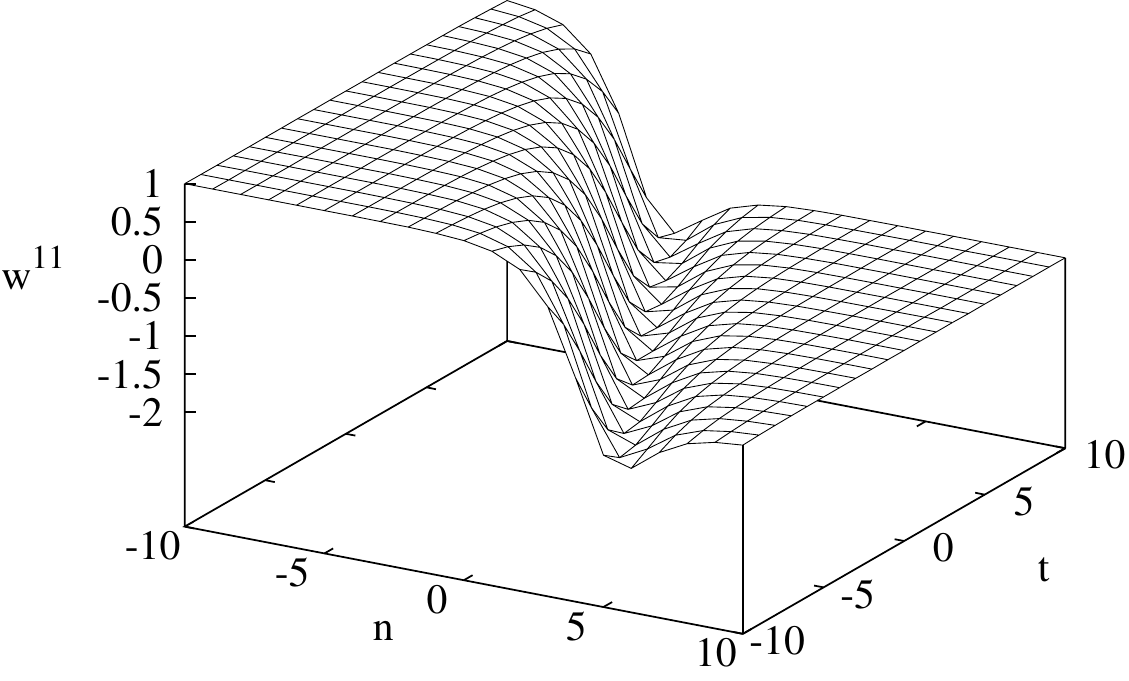} \qquad
\includegraphics[width=6.5cm]{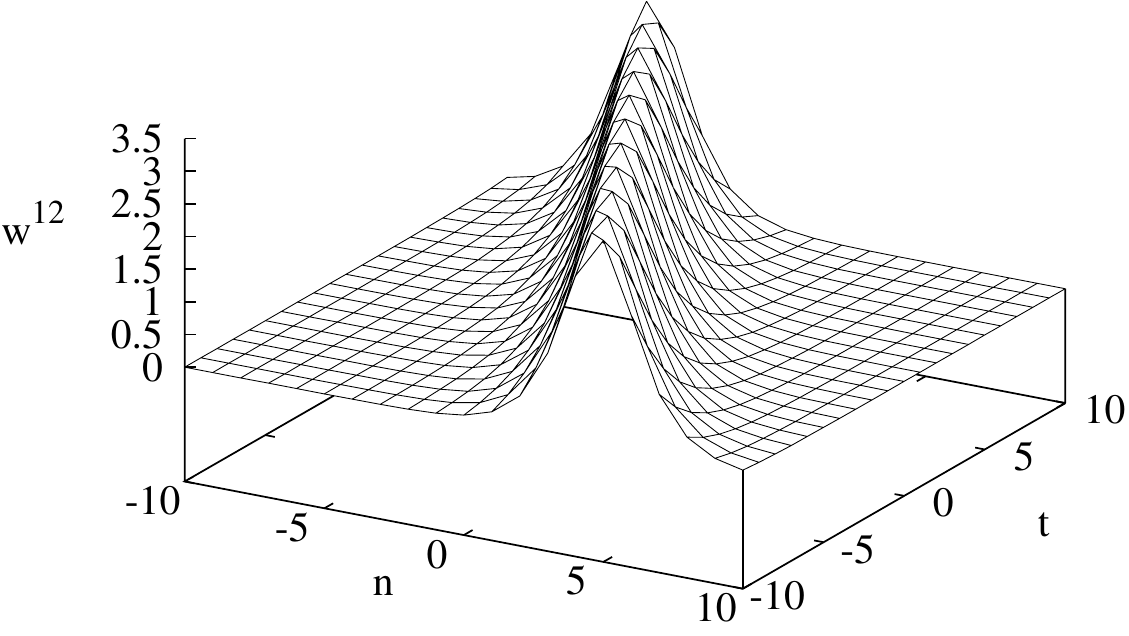}

\includegraphics[width=6.5cm]{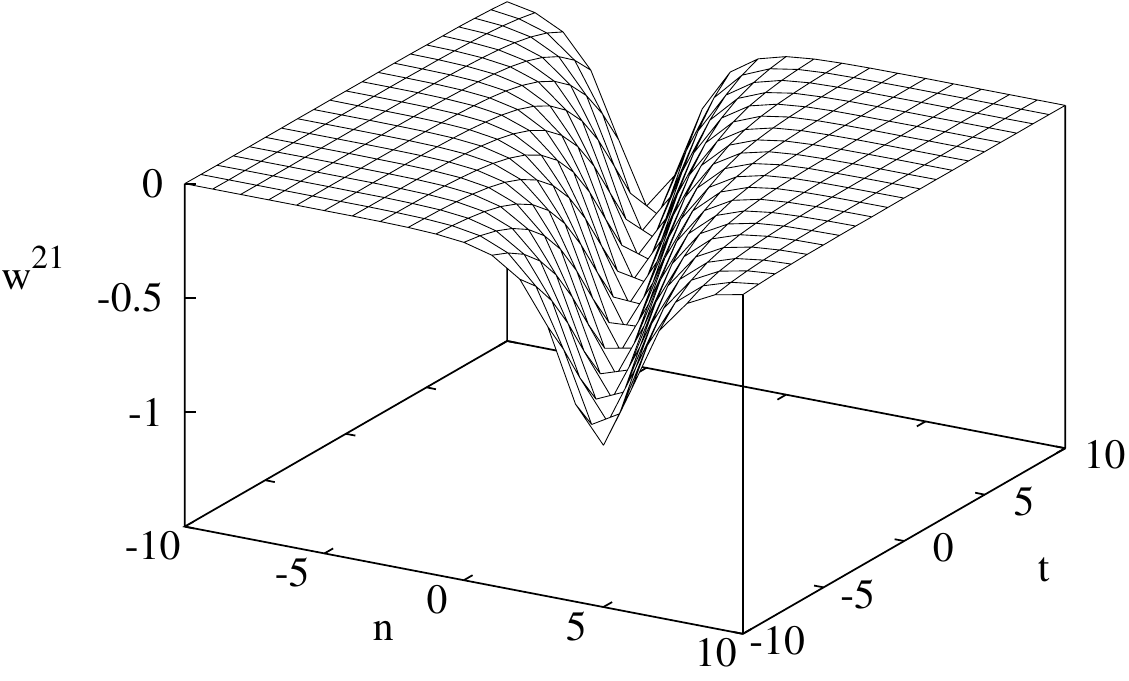} \qquad
\includegraphics[width=6.5cm]{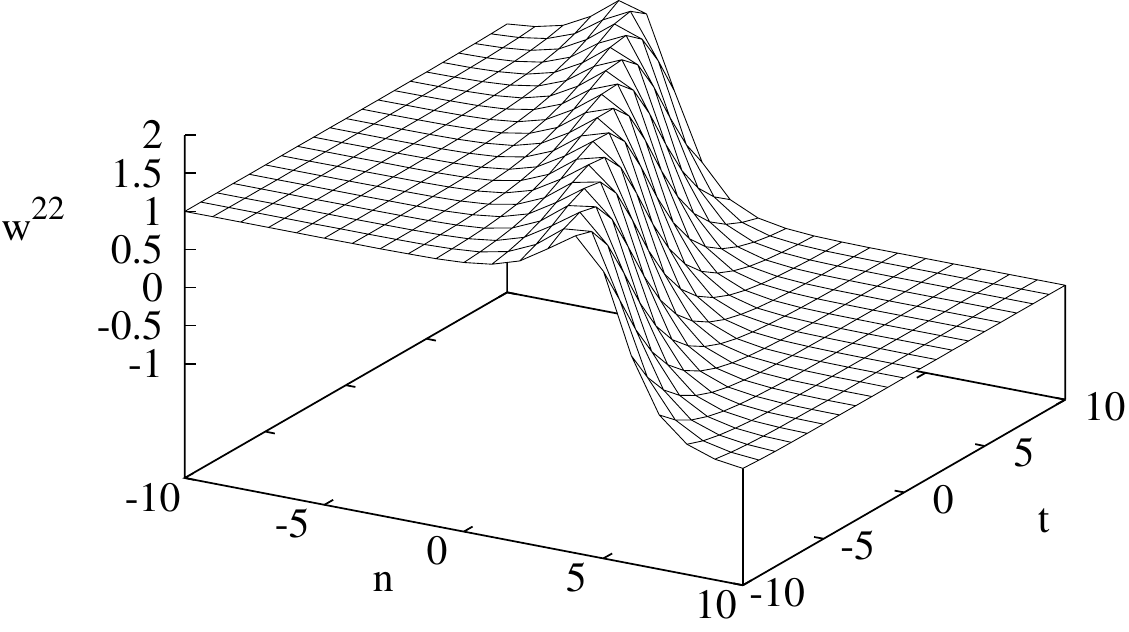}

\caption{1-soliton solution for ncdsG.}
\label{ncdsG:fig:1sol}
\end{figure}

A 1-soliton solution is given by
\begin{subequations}
\begin{gather}
w =\psi_1\phi_1^{-1}=(1-f_1)(1+f_1)^{-1}, \label{ncdsG:1sol1} \\
v =\phi_1\psi_1^{-1}=(1+f_1)(1-f_1)^{-1}, \label{ncdsG:1sol2}
\end{gather}
\end{subequations}
where $f_k$ is def\/ined by
\begin{gather*}
f_k=\lt(\frac{1+a\lambda_k}{1-a\lambda_k}\rt)^l\lt(\frac{1+b\lambda_k^{-1}}{1-b\lambda_k^{-1}}\rt)^mc_k.
\end{gather*}
As a concrete example, Fig.~\ref{ncdsG:fig:1sol} shows the behavior of
\[
w=\pM{w^{11}}{w^{12}}{w^{21}}{w^{22}} \qquad (N=2)
\]
with
\[
a=b=0.2, \qquad c_1=\pM{2}{-4}{1}{-1.5}, \qquad \lambda_1=\frac{5}{3}
\]
in the light-cone coordinates~\eqref{lightcone}.

\begin{figure}[t]
\centering
\includegraphics[width=6.5cm]{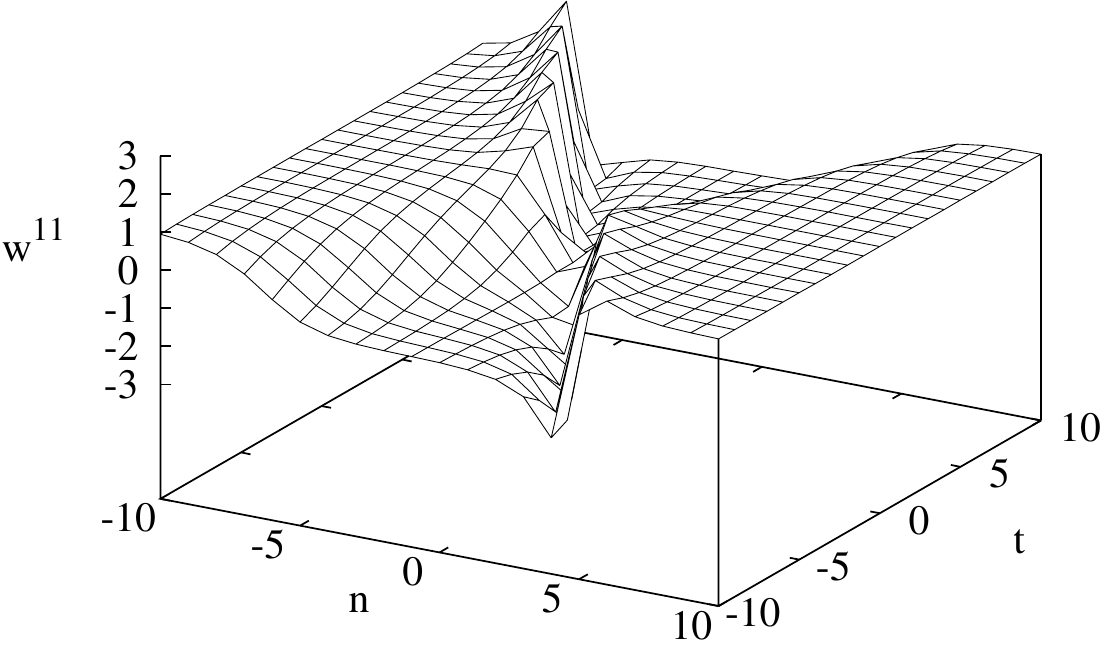} \qquad
\includegraphics[width=6.5cm]{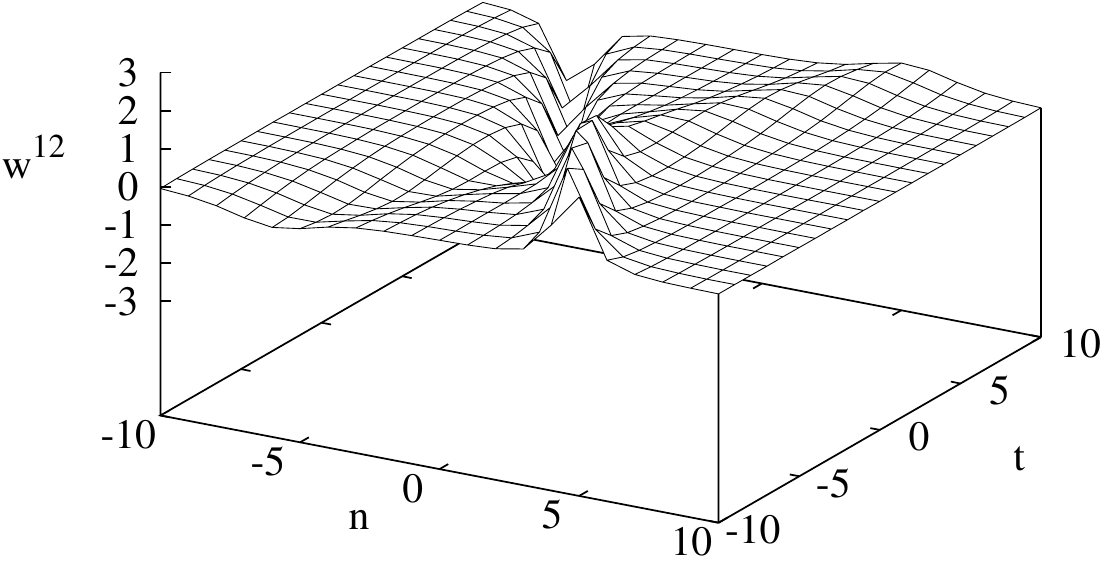}

\includegraphics[width=6.5cm]{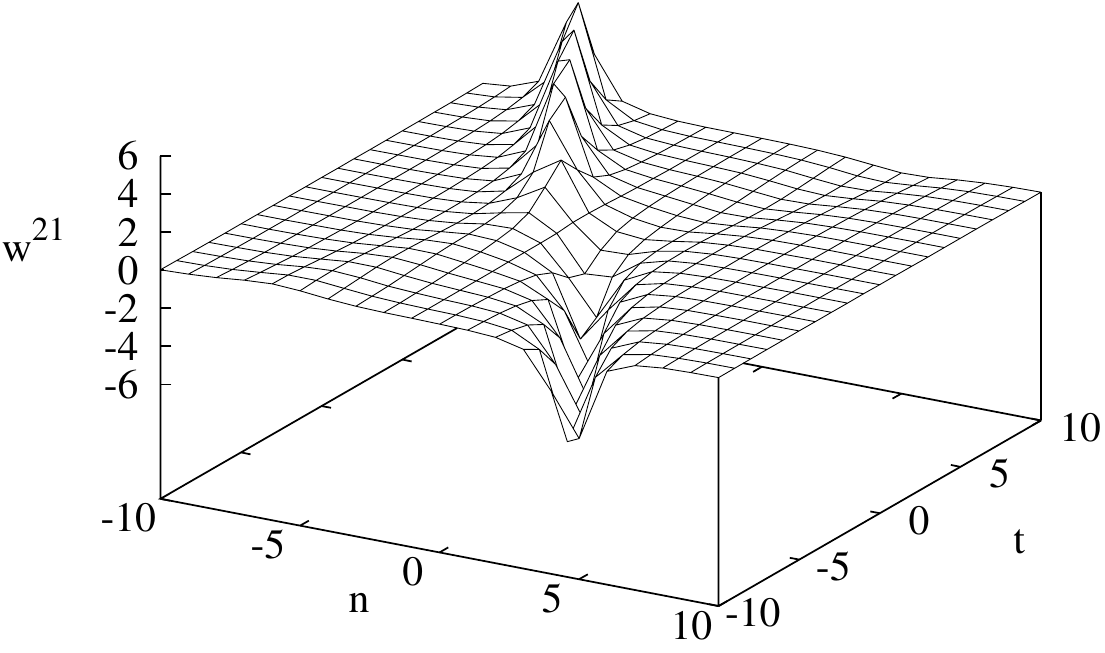} \qquad
\includegraphics[width=6.5cm]{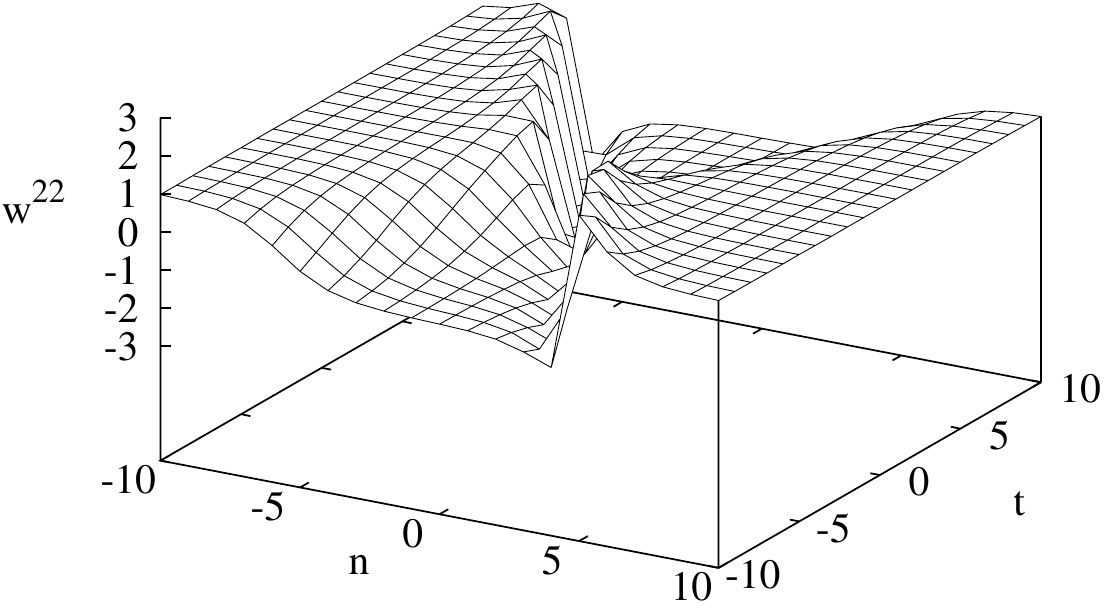}
\caption{2-soliton solution for ncdsG.}
\label{ncdsG:fig:2sol}
\end{figure}

A 2-soliton solution is given by
\begin{subequations}
\begin{gather}
w =\lt(\lambda_2\phi_2-\lambda_1\phi_1\psi_1^{-1}\psi_2\rt)\lt(\lambda_2\psi_2-\lambda_1\psi_1\phi_1^{-1}\phi_2\rt)^{-1}\psi_1\phi_1^{-1} \notag \\
\hphantom{w}{} =\lt(\lambda_2\phi_2\psi_2^{-1}-\lambda_1\phi_1\psi_1^{-1}\rt)\lt(\lambda_2\phi_1\psi_1^{-1}-\lambda_1\phi_2\psi_2^{-1}\rt)^{-1} \notag \\
\hphantom{w}{}=\lt(\lambda_2(1+f_2)(1-f_2)^{-1}-\lambda_1(1+f_1)(1-f_1)^{-1}\rt) \notag \\
\hphantom{w=}{}\times\lt(\lambda_2(1+f_1)(1-f_1)^{-1}-\lambda_1(1+f_2)(1-f_2)^{-1}\rt)^{-1}, \label{ncdsG:2sol1} \\
v =\lt(\lambda_2(1-f_2)(1+f_2)^{-1}-\lambda_1(1-f_1)(1+f_1)^{-1}\rt) \notag \\
\hphantom{v=}{}\times\lt(\lambda_2(1-f_1)(1+f_1)^{-1}-\lambda_1(1-f_2)(1+f_2)^{-1}\rt)^{-1}. \label{ncdsG:2sol2}
\end{gather}
\end{subequations}
Fig.~\ref{ncdsG:fig:2sol} shows the solution with
\[
a=b=0.2, \qquad c_1=\pM{2.5}{-0.8}{2}{1.8}, \qquad c_2=\pM{1.5}{1.2}{-1}{0.5}, \qquad \lambda_1=\lambda_2^{-1}=\frac{5}{3}.
\]

\subsubsection{Casoratian-type solutions}

Let $(w,v)$ be a solution for \eqref{ncdsG:ncdsG1} and \eqref{ncdsG:ncdsG2}, $\Vin{\phi_k}{\psi_k}$ be eigenfunctions of $(w,v)$ for eigenvalues $\lambda_k$ $(k=1,2,\ldots)$, where $\lambda_k$ are mutually dif\/ferent. Def\/ine repetition of the Darboux transformation by
%\begin{subequations}
\begin{gather*}
w^{(n)} =\psi_n^{(n)}\big(\phi_n^{(n)}\big)^{-1}w^{(n-1)}, \\
v^{(n)} =\phi_n^{(n)}\big(\psi_n^{(n)}\big)^{-1}v^{(n-1)}, \\
\phi_k^{(n+1)} =\lambda_k\psi_k^{(n)}-\lambda_n\psi_n^{(n)}\big(\phi_n^{(n)}\big)^{-1}\phi_k^{(n)}, \\
\psi_k^{(n+1)} =\lambda_k\phi_k^{(n)}-\lambda_n\phi_n^{(n)}\big(\psi_n^{(n)}\big)^{-1}\psi_k^{(n)}
\end{gather*}
and
\begin{gather*}
w^{(0)}=w, \qquad v^{(0)}=v, \qquad \phi_k^{(1)}=\phi_k, \qquad \psi_k^{(1)}=\psi_k.
\end{gather*}
%\end{subequations}
For notational convenience, we introduce \textit{reduced} shift operator $T$ def\/ined by
\begin{gather*}
Tf(\phi_1,\psi_1,\phi_2,\psi_2,\ldots)=f(\lambda_1\psi_1,\lambda_1\phi_1,\lambda_2\psi_2,\lambda_2\phi_2,\ldots),
\end{gather*}
where $f(x_1,x_2,\ldots)$ is any rational function of noncommutative variables $x_j$. For example, we have
\[
T\phi_k=\lambda_k\psi_k, \qquad T\psi_k=\lambda_k\phi_k, \qquad T^2\phi_k=\lambda_k^2\phi_k.
\]

\begin{Lemma}
\begin{gather*}
T\phi_k^{(n+1)}=\lambda_k\psi_k^{(n+1)}, \qquad T\psi_k^{(n+1)}=\lambda_k\phi_k^{(n+1)}.
\end{gather*}
\end{Lemma}
\begin{proof}
We prove by induction. Obviously, $\phi_k^{(n+1)}$, $\psi_k^{(n+1)}$ are rational functions of $\phi_j$, $\psi_j$. Assume $T\phi_k^{(n)}=\lambda_k\psi_k^{(n)}$, $T\psi_k^{(n)}=\lambda_k\phi_k^{(n)}$ for certain $n$. Then,
\[
T\phi_k^{(n+1)}=\lambda_k\big(\lambda_k\phi_k^{(n)}\big)-\lambda_n\big(\lambda_n\phi_n^{(n)}\big)\big(\lambda_n\psi_n^{(n)}\big)^{-1}
\lambda_k\psi_k^{(n)}=\lambda_k\psi_k^{(n+1)}.
\]
Similarly, $T\psi_k^{(n+1)}=\lambda_k\phi_k^{(n+1)}$.
\end{proof}

\begin{Theorem}\label{ncdsG:rDt}
\begin{subequations}
\begin{gather}
w^{(n)} =\prod_{j=1}^n\big({-}\lambda_j^{-1}\big)\cdot\begin{vmat} \phi_1 & \phi_2 & \cdots & \phi_n & 1 \\
T\phi_1 & T\phi_2 & \cdots & T\phi_n & 0 \\
\vdots & \vdots & \ddots & \vdots & \vdots \\
T^{n-1}\phi_1 & T^{n-1}\phi_2 & \cdots & T^{n-1}\phi_n & 0 \\
T^n\phi_1 & T^n\phi_2 & \cdots & T^n\phi_n & \fbox{$0$}
\end{vmat}w, \label{ncdsG:Casoratian_w} \\
v^{(n)} =\prod_{j=1}^n\big({-}\lambda_j^{-1}\big)\cdot\begin{vmat} \psi_1 & \psi_2 & \cdots & \psi_n & 1 \\
T\psi_1 & T\psi_2 & \cdots & T\psi_n & 0 \\
\vdots & \vdots & \ddots & \vdots & \vdots \\
T^{n-1}\psi_1 & T^{n-1}\psi_2 & \cdots & T^{n-1}\psi_n & 0 \\
T^n\psi_1 & T^n\psi_2 & \cdots & T^n\psi_n & \fbox{$0$}
\end{vmat}v, \label{ncdsG:Casoratian_v}\\
\phi_k^{(n+1)} =\begin{vmat} \phi_1 & \phi_2 & \cdots & \phi_n & \phi_k \\
T\phi_1 & T\phi_2 & \cdots & T\phi_n & T\phi_k \\
\vdots & \vdots & \ddots & \vdots & \vdots \\
T^{n-1}\phi_1 & T^{n-1}\phi_2 & \cdots & T^{n-1}\phi_n & T^{n-1}\phi_k \\
T^n\phi_1 & T^n\phi_2 & \cdots & T^n\phi_n & \fbox{$T^n\phi_k$}
\end{vmat}, \\
\psi_k^{(n+1)} =\begin{vmat} \psi_1 & \psi_2 & \cdots & \psi_n & \psi_k \\
T\psi_1 & T\psi_2 & \cdots & T\psi_n & T\psi_k \\
\vdots & \vdots & \ddots & \vdots & \vdots \\
T^{n-1}\psi_1 & T^{n-1}\psi_2 & \cdots & T^{n-1}\psi_n & T^{n-1}\psi_k \\
T^n\psi_1 & T^n\psi_2 & \cdots & T^n\psi_n & \fbox{$T^n\psi_k$}
\end{vmat}.
\end{gather}
\end{subequations}
Here, quasideterminants~{\rm \cite{GGRW}} are used $($see Appendix~{\rm \ref{apd:qdet}}$)$. When $n=0$, \eqref{ncdsG:Casoratian_w} and \eqref{ncdsG:Casoratian_v} read
\[
w^{(0)}=1\cdot\big|\fbox{$1$} \big| w, \qquad v^{(0)}=1\cdot\big|\fbox{$1$} \big| v,
\]
respectively.
\end{Theorem}

Before proceeding to the proof, we prepare the following lemma.

\begin{Lemma}\label{qdet_lem}
Let $C=(c_{ij})$, $C'=(c'_{ij})$ be $n\times n$ matrices where $c_{ij},c'_{ij}\in\Mat(N,\C)$. Assume $c_{ij}=c'_{ij}$ for $1\le i\le n$, $1\le j\le n-1$. Then,
\begin{gather}
|C|_{p_1n}^{-1}|C'|_{p_1n}=|C|_{p_2n}^{-1}|C'|_{p_2n}. \label{qdet_lem:eq1}
\end{gather}
\end{Lemma}
\begin{proof}
By the column homological relation (\propref{qdet:homrel}), we have
\begin{subequations}
\begin{gather}
|C^{p_2n}|_{p_1j}^{-1}|C|_{p_1n} =-|C^{p_1n}|_{p_2j}|C|_{p_2n}, \label{qdet_lem:eq2} \\
|C'^{p_2n}|_{p_1j}^{-1}|C'|_{p_1n} =-|C'^{p_1n}|_{p_2j}|C'|_{p_2n}. \label{qdet_lem:eq3}
\end{gather}
\end{subequations}
By the assumption, we have $C^{p_2n}=C'^{p_2n}$, $C^{p_1n}=C'^{p_1n}$. Therefore, we obtain \eqref{qdet_lem:eq1} by multiplying the inverse of \eqref{qdet_lem:eq2} to \eqref{qdet_lem:eq3} from the left.
\end{proof}

\begin{proof}[Proof of \thmref{ncdsG:rDt}]
We prove by induction. The case $n=0$ is trivial.

Assume $w^{(n-1)}$, $v^{(n-1)}$, $\phi_k^{(n)}$, $\psi_k^{(n)}$ have the above expressions for certain $n>0$. Then,
\begin{gather*}
w^{(n)} =\lambda_n^{-1}\big(T\phi_n^{(n)}\big)\big(\phi_n^{(n)}\big)^{-1}w^{(n-1)} \\
\hphantom{w^{(n)}}{} =-\prod_{j=1}^n\lt(-\lambda_j^{-1}\rt)\cdot\begin{vmat}   T\phi_1 & \cdots & T\phi_n \\
\vdots & \ddots & \vdots \\
T^n\phi_1 & \cdots & \fbox{$T^n\phi_n$}
\end{vmat}\begin{vmat}   \phi_1 & \cdots & \phi_n \\
\vdots & \ddots & \vdots \\
T^{n-1}\phi_1 & \cdots & \fbox{$T^{n-1}\phi_n$}
\end{vmat}^{-1} \\
\hphantom{w^{(n)}=}{} \times\begin{vmat}   \phi_1 & \cdots & \phi_{n-1} & 1 \\
T\phi_1 & \cdots & T\phi_{n-1} & 0 \\
\vdots & \ddots & \vdots & \vdots \\
T^{n-1}\phi_1 & \cdots & T^{n-1}\phi_{n-1} & \fbox{$0$}
\end{vmat}w.
\end{gather*}
By \lemref{qdet_lem}, we obtain
\begin{gather*}
\begin{vmat} \phi_1 & \cdots & \phi_n \\
\vdots & \ddots & \vdots \\
T^{n-1}\phi_1 & \cdots & \fbox{$T^{n-1}\phi_n$}
\end{vmat}^{-1}\begin{vmat} \phi_1 & \cdots & \phi_{n-1} & 1 \\
T\phi_1 & \cdots & T\phi_{n-1} & 0 \\
\vdots & \ddots & \vdots & \vdots \\
T^{n-1}\phi_1 & \cdots & T^{n-1}\phi_{n-1} & \fbox{$0$}
\end{vmat} \\
\qquad {} =\begin{vmat} \phi_1 & \cdots & \fbox{$\phi_n$} \\
\vdots & \ddots & \vdots \\
T^{n-1}\phi_1 & \cdots & T^{n-1}\phi_n
\end{vmat}^{-1}\begin{vmat} \phi_1 & \cdots & \phi_{n-1} & \fbox{$1$} \\
T\phi_1 & \cdots & T\phi_{n-1} & 0 \\
\vdots & \ddots & \vdots & \vdots \\
T^{n-1}\phi_1 & \cdots & T^{n-1}\phi_{n-1} & 0
\end{vmat}.
\end{gather*}
We def\/ine an $(n-1)\times(n-1)$ matrix $A_0$ by
\[
A_0=\begin{pmat} T\phi_1 & \cdots & T\phi_{n-1} \\
\vdots & \ddots & \vdots \\
T^{n-1}\phi_1 & \cdots & T^{n-1}\phi_{n-1}
\end{pmat}.
\]
With $A_0$, the above quasideterminants are rewritten as
\begin{gather*}
\begin{vmat} T\phi_1 & \cdots & T\phi_n \\
\vdots & \ddots & \vdots \\
T^n\phi_1 & \cdots & \fbox{$T^n\phi_n$}
\end{vmat} =\begin{vmat} {} & & & T\phi_n \\
& A_0 & & \vdots \\
& & & T^{n-1}\phi_n \\
T^n\phi_1 & \cdots & T^n\phi_{n-1} & \fbox{$T^n\phi_n$}
\end{vmat}, \\
\begin{vmat} \phi_1 & \cdots & \fbox{$\phi_n$} \\
\vdots & \ddots & \vdots \\
T^{n-1}\phi_1 & \cdots & T^{n-1}\phi_n
\end{vmat} =\begin{vmat} \phi_1 & \cdots & \phi_{n-1} & \fbox{$\phi_n$} \\
& & & T\phi_n \\
& A_0 & & \vdots \\
& & & T^{n-1}\phi_n
\end{vmat}, \\
\begin{vmat} \phi_1 & \cdots & \phi_{n-1} & \fbox{$1$} \\
T\phi_1 & \cdots & T\phi_{n-1} & 0 \\
\vdots & \ddots & \vdots & \vdots \\
T^{n-1}\phi_1 & \cdots & T^{n-1}\phi_{n-1} & 0
\end{vmat} =\begin{vmat} \phi_1 & \cdots & \phi_{n-1} & \fbox{$1$} \\
& & & 0 \\
& A_0 & & \vdots \\
& & & 0
\end{vmat},
\end{gather*}
and also, we have a trivial identity
\[
0=\begin{vmat} {} & & & 0 \\
& A_0 & & \vdots \\
& & & 0 \\
T^n\phi_1 & \cdots & T^n\phi_{n-1} & \fbox{$0$}
\end{vmat}.
\]
By the invariance under row and column permutations (\propref{qdet:perm}) and Sylvester's identity (\propref{qdet:Sylvester}), we can combine these four quasideterminants into one to obtain
\begin{gather*}
w^{(n)} =\prod_{j=1}^n\big({-}\lambda_j^{-1}\big)\cdot\begin{vmat} \phi_1 & \cdots & \phi_{n-1} & \phi_n & 1 \\
& & & T\phi_n & 0 \\
& A_0 & & \vdots & \vdots \\
& & & T^{n-1}\phi_n & 0 \\
T^n\phi_1 & \cdots & T^n\phi_{n-1} & T^n\phi_n & \fbox{$0$}
\end{vmat}w.
\end{gather*}
Similarly for $v^{(n)}$.

For $\phi_k^{(n+1)}$, we have
\begin{gather*}
\phi_k^{(n+1)} =T\phi_k^{(n)}-\big(T\phi_n^{(n)}\big)\big(\phi_n^{(n)}\big)^{-1}\phi_k^{(n)} \\
\hphantom{\phi_k^{(n+1)}}{}
 =\begin{vmat} T\phi_1 & \cdots & T\phi_{n-1} & T\phi_k \\
\vdots & \ddots & \vdots & \vdots \\
T^n\phi_1 & \cdots & T^n\phi_{n-1} & \fbox{$T^n\phi_k$}
\end{vmat}-\begin{vmat} T\phi_1 & \cdots & T\phi_{n-1} & T\phi_n \\
\vdots & \ddots & \vdots & \vdots \\
T^n\phi_1 & \cdots & T^n\phi_{n-1} & \fbox{$T^n\phi_n$}
\end{vmat} \\
\hphantom{\phi_k^{(n+1)} =}{}
\times\begin{vmat} \phi_1 & \cdots & \phi_{n-1}& \phi_n \\
T\phi_1 & \cdots & T\phi_{n-1} & T\phi_n \\
\vdots & \ddots & \vdots & \vdots \\
T^{n-1}\phi_1 & \cdots & T^{n-1}\phi_{n-1} & \fbox{$T^{n-1}\phi_n$}
\end{vmat}^{-1} \\
\hphantom{\phi_k^{(n+1)} =}{}
\times\begin{vmat} \phi_1 & \cdots & \phi_{n-1} & \phi_k \\
T\phi_1 & \cdots & T\phi_{n-1} & T\phi_k \\
\vdots & \ddots & \vdots & \vdots \\
T^{n-1}\phi_1 & \cdots & T^{n-1}\phi_{n-1} & \fbox{$T^{n-1}\phi_k$}
\end{vmat}.
\end{gather*}
With the same technique for $w^{(n)}$, we obtain
\[
\phi_k^{(n+1)}=\begin{vmat} \phi_1 & \cdots & \phi_n & \phi_k \\
\vdots & \ddots & \vdots & \vdots \\
T^n\phi_1 & \cdots & T^n\phi_n & \fbox{$T^n\phi_k$}
\end{vmat}.
\]
Similarly for $\psi_k^{(n+1)}$.
\end{proof}

\subsection{Noncommutative ultradiscrete sine-Gordon equation}\label{sec:ncudsG}

\subsubsection{Ultradiscretization}

We perform ultradiscretization of ncdsG by the parametrization
\begin{gather*}
a=\mu_Ae^{\wt As}, \qquad b=\mu_Be^{\wt As}, \qquad \wt A,\wt B<0.
\end{gather*}
Assuming
\begin{gather*}
a\ud A, \qquad b\ud B, \qquad w\ud W, \qquad v\ud V,
\end{gather*}
we obtain
\begin{subequations}
\begin{gather}
W_{,lm}W_{,m}^{-1}\ominus W_{,l}W^{-1}\oplus AB\big(V_{,m}W^{-1}\ominus W_{,lm}V_{,l}^{-1}\big) \bals\minf, \label{ncudsG:ncudsG1} \\
V_{,lm}V_{,m}^{-1}\ominus V_{,l}V^{-1}\oplus AB\big(W_{,m}V^{-1}\ominus V_{,lm}W_{,l}^{-1}\big) \bals\minf. \label{ncudsG:ncudsG2}
\end{gather}
\end{subequations}
We call the pair \eqref{ncudsG:ncudsG1} and \eqref{ncudsG:ncudsG2} the noncommutative ultradiscrete sine-Gordon equation (ncudsG). Because $\uMat(N,\uC)$ can be realized by $\uMat(2N,\uR)$, we use $\uMat(N,\uR)$ as the underlying algebra for simplicity.

\subsubsection{1-soliton solution}

In order to ultradiscretize solutions for ncdsG, we introduce
\begin{gather}
p_j=\frac{1+a\lambda_j}{1-a\lambda_j}, \qquad q_j=\frac{1+b\lambda_j^{-1}}{1-b\lambda_j^{-1}}. \label{ncudsG:pqlambda}
\end{gather}
These solve the dispersion relation
\begin{gather}
(1-ab)(1+p_jq_j)=(1+ab)(p_j+q_j), \label{ncdsG:disprel}
\end{gather}
and any solution of \eqref{ncdsG:disprel} is parametrized by $\lambda_j$ through \eqref{ncudsG:pqlambda} unless $ab=1$. As in the commutative case, \eqref{ncdsG:disprel} is ultradiscretized to
\begin{gather}
0\oplus P_jQ_j\bals P_j\oplus Q_j, \label{ncudsG:disprel}
\end{gather}
where $p_j\ud P_j$, $q_j\ud Q_j$.

We can directly discretize the 1-soliton solution \eqref{ncdsG:1sol1}, \eqref{ncdsG:1sol2} to obtain
%\begin{subequations}
\begin{gather*}
w \ud W\bals (0\ominus F_1)(0\oplus F_1)^{-1}, \qquad
v \ud V\bals (0\oplus F_1)(0\ominus F_1)^{-1},
\end{gather*}
%\end{subequations}
where
\begin{gather*}
F_j=P_j^lQ_j^mC_j, \qquad c_j\ud C_j\in\uMat(N,\uR).
\end{gather*}
This relation is valid, but inadequate to determine $W$, $V$ in many cases. For simplicity, we assume $N=2$ hereafter. If we write $W=\big(W^{\iota\kappa}\big)$, $F_j=\big(F_j^{\iota\kappa}\big)$, the $(1,2)$-th element of $(0\ominus F_1)(0\oplus F_1)^{-1}$ is given by
\[
\frac{\ominus F_1^{12}\lt(\lt(0\ominus F_1^{11}\rt)\oplus\lt(0\oplus F_1^{11}\rt)\rt)}{\det(0\oplus F_1)}=\frac{\ominus F_1^{12}\lt(0\oplus\bal{\lt(F_1^{11}\rt)}\rt)}{\det(0\oplus F_1)},
\]
and $\uabs{F_1^{11}}$ exceeds $0$ for large $\pm l$ or $\pm m$. Then this element is balanced and $W^{12}$ cannot be determined. Therefore, we need more precise expressions to ultradiscretize.

\begin{figure}[t]
\centering
\includegraphics[width=6.5cm]{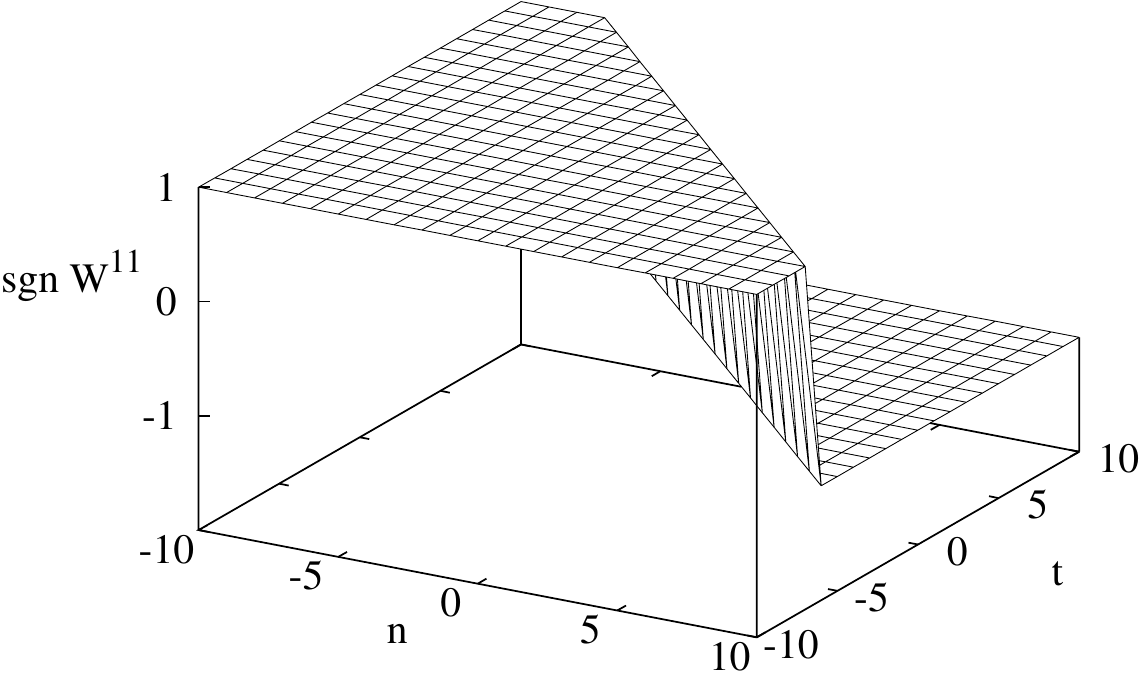} \qquad
\includegraphics[width=6.5cm]{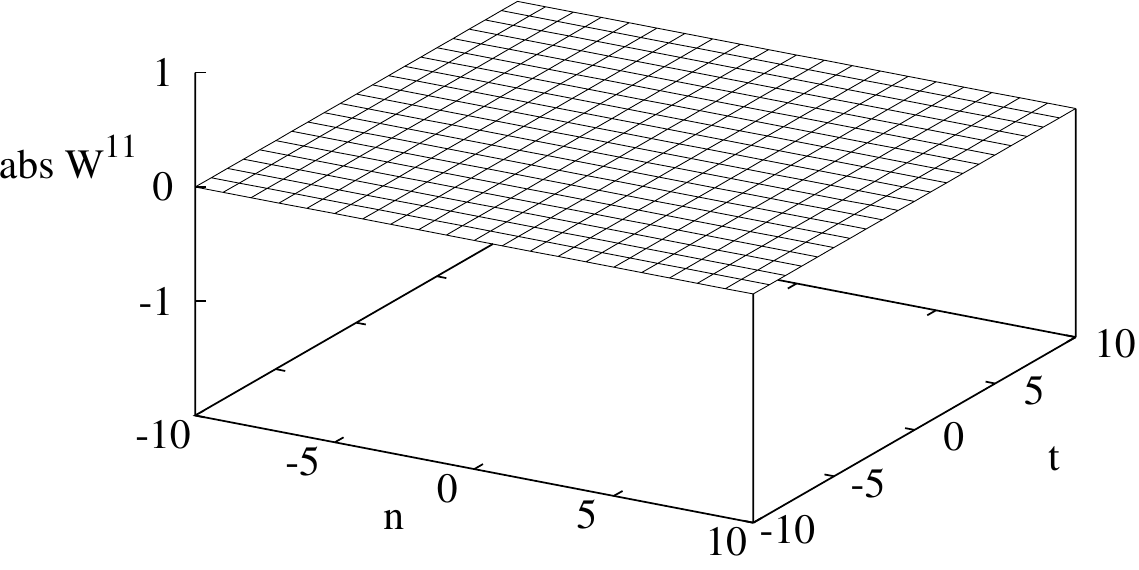}

\includegraphics[width=6.5cm]{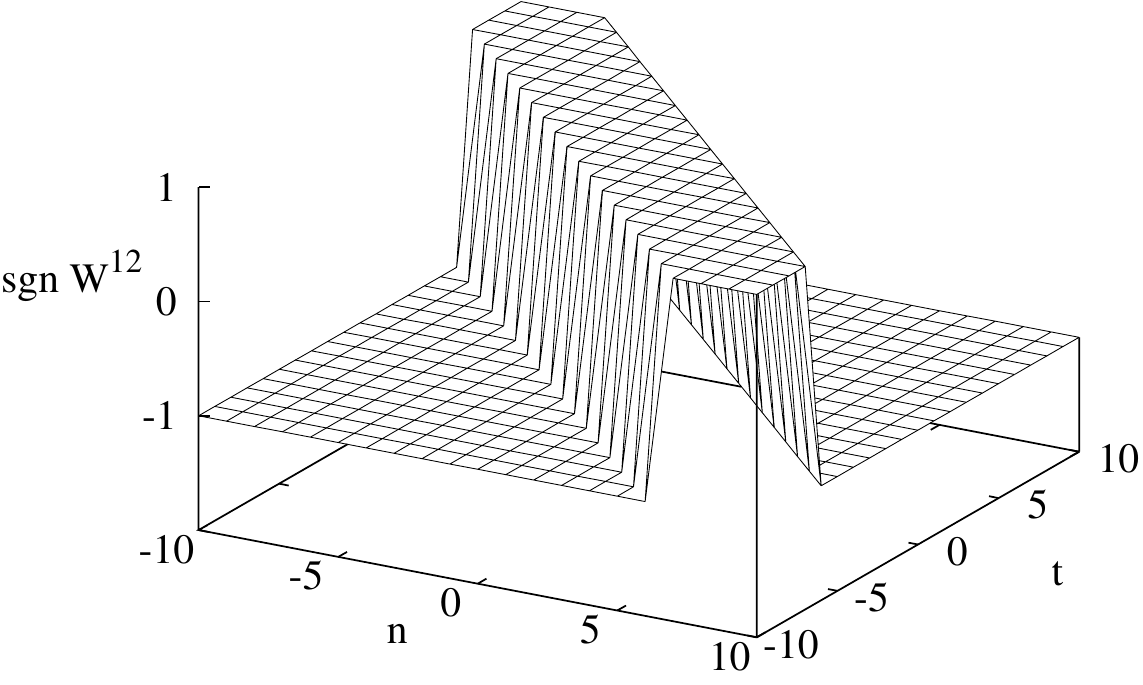} \qquad
\includegraphics[width=6.5cm]{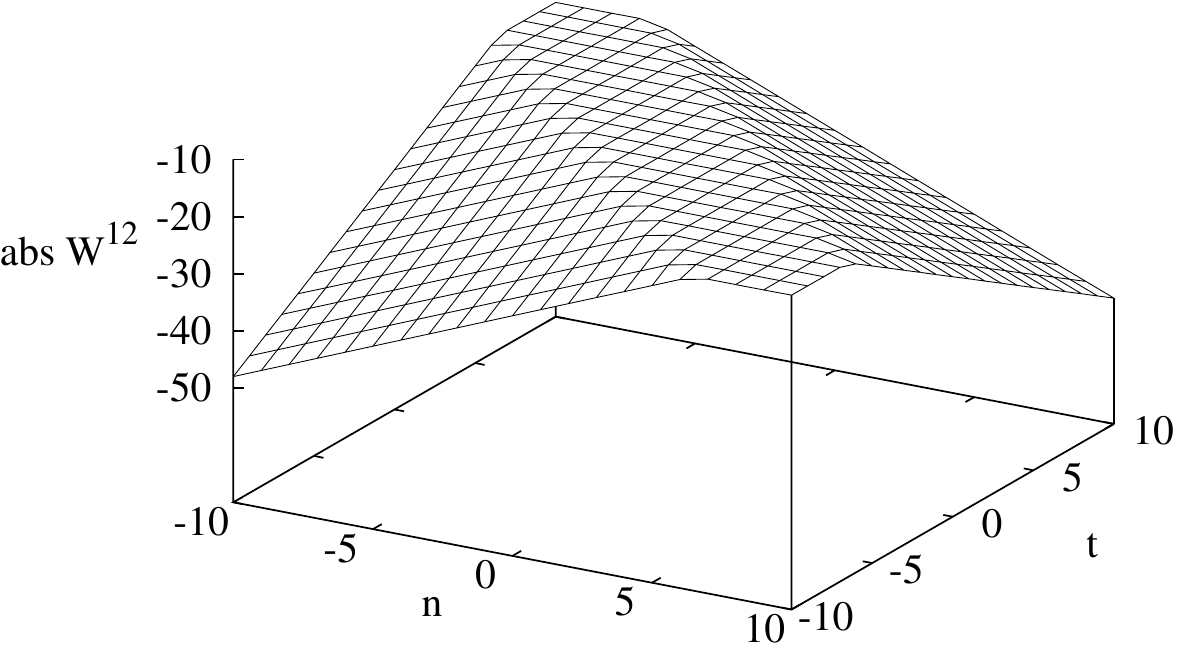}

\includegraphics[width=6.5cm]{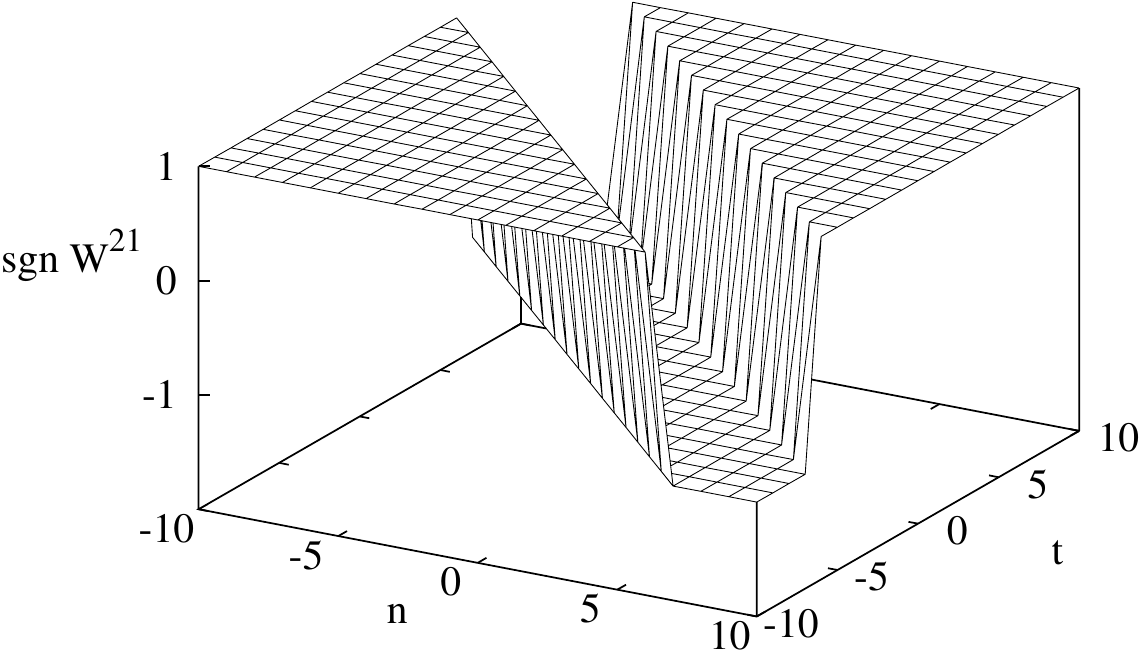} \qquad
\includegraphics[width=6.5cm]{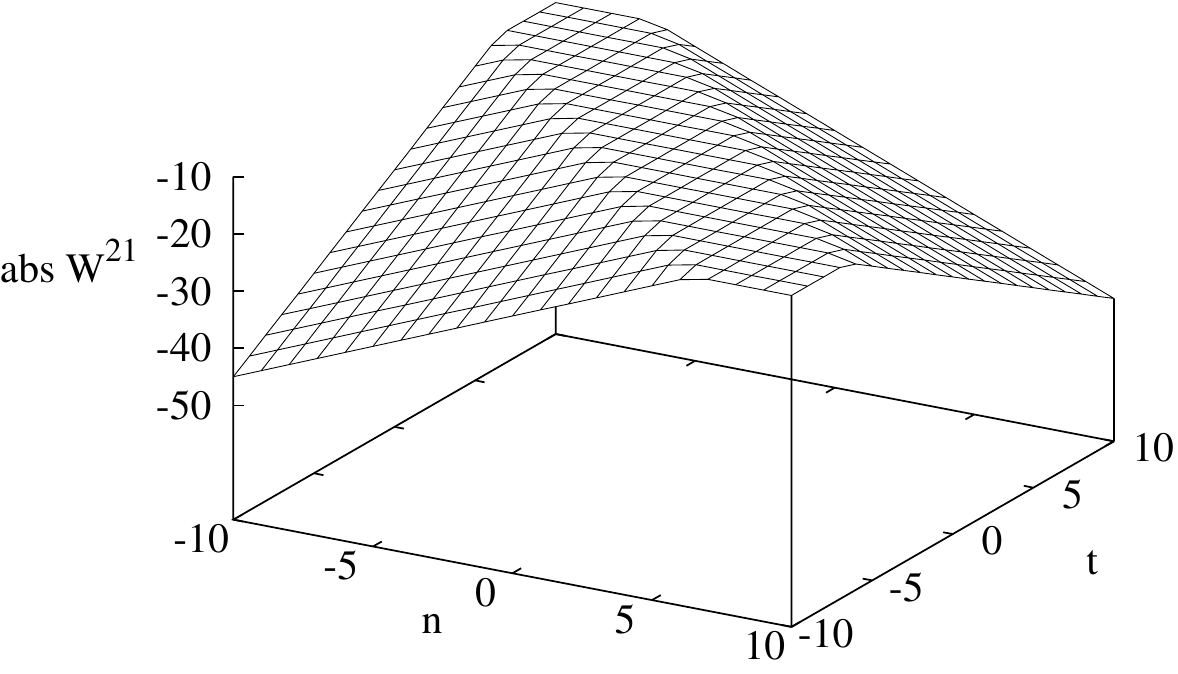}

\includegraphics[width=6.5cm]{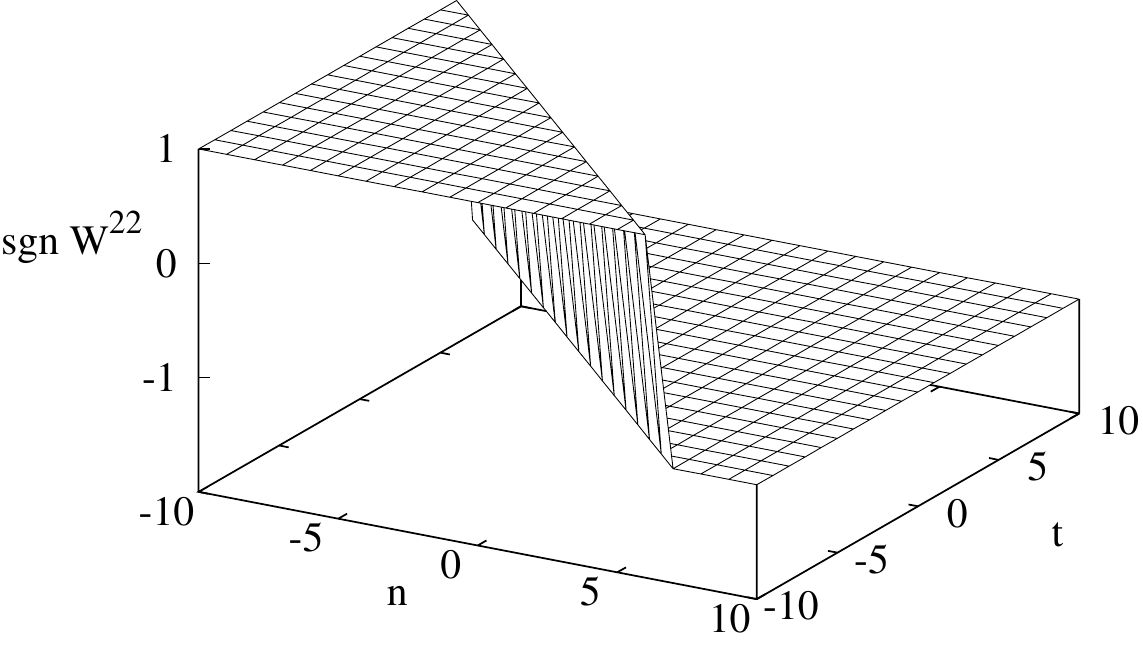} \qquad
\includegraphics[width=6.5cm]{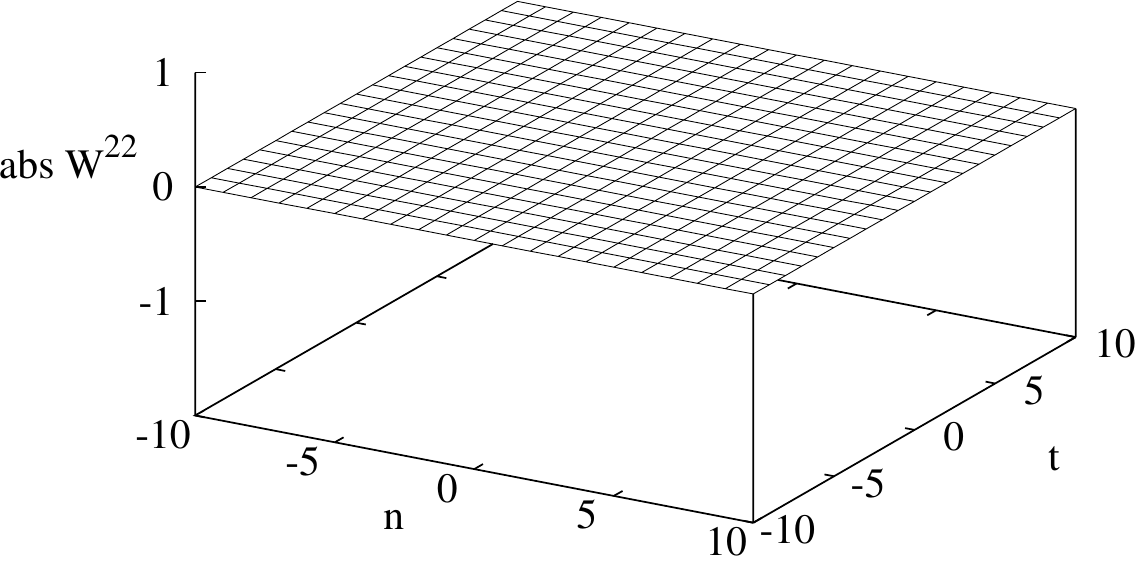}
\caption{1-soliton solution for ncudsG.}
\label{ncudsG:fig:1sol}
\end{figure}

Def\/ine
\begin{gather*}
g_j=(1+f_j)(1-f_j)^{-1}, \qquad h_j=(1-f_j)(1+f_j)^{-1}.
\end{gather*}
Of course, $w=h_1$, $v=g_1$ is a 1-soliton solution for ncdsG. Writing $f_j=\big(f_j^{\iota\kappa}\big)$, we have
%\begin{subequations}
\begin{gather*}
g_j =\bpm
\dfrac{\lt(1+f_j^{11}\rt)\lt(1-f_j^{22}\rt)+f_j^{12}f_j^{21}}{\det(1-f_j)} &
\dfrac{2f_j^{12}}{\det(1-f_j)} \vspace{2mm}\\
\dfrac{2f_j^{21}}{\det(1-f_j)} &
\dfrac{\lt(1+f_j^{22}\rt)\lt(1-f_j^{11}\rt)+f_j^{21}f_j^{12}}{\det(1-f_j)}
\epm, \\
h_j =\bpm
\dfrac{\lt(1-f_j^{11}\rt)\lt(1+f_j^{22}\rt)+f_j^{12}f_j^{21}}{\det(1+f_j)} &
\dfrac{-2f_j^{12}}{\det(1+f_j)} \vspace{2mm}\\
\dfrac{-2f_j^{21}}{\det(1+f_j)} &
\dfrac{\lt(1-f_j^{22}\rt)\lt(1+f_j^{11}\rt)+f_j^{21}f_j^{12}}{\det(1+f_j)}
\epm.
\end{gather*}
%\end{subequations}
By ultradiscretization, we obtain
\begin{subequations}
\begin{gather}
g_j\ud G_j \bals\bpm \dfrac{\lt(0\oplus F_j^{11}\rt)\lt(0\ominus F_j^{22}\rt)\oplus F_j^{12}F_j^{21}}{\det(0\ominus F_j)} & \dfrac{F_j^{12}}{\det(0\ominus F_j)} \vspace{2mm}\\
\dfrac{F_j^{21}}{\det(0\ominus F_j)} & \dfrac{(0\oplus F_j^{22})(0\ominus F_j^{11})\oplus F_j^{21}F_j^{12}}{\det(0\ominus F_j)}
\epm, \label{ncudsG:G} \\
h_j\ud H_j \bals\bpm \dfrac{\lt(0\ominus F_j^{11}\rt)\lt(0\oplus F_j^{22}\rt)\oplus F_j^{12}F_j^{21}}{\det(0\oplus F_j)} & \dfrac{\ominus F_j^{12}}{\det(0\oplus F_j)} \vspace{2mm}\\
\dfrac{\ominus F_j^{21}}{\det(0\oplus F_j)} & \dfrac{(0\ominus F_j^{22})(0\oplus F_j^{11})\oplus F_j^{21}F_j^{12}}{\det(0\oplus F_j)}
\epm. \!\!\!\label{ncudsG:H}
\end{gather}
\end{subequations}
We can choose $C_j, P_j, Q_j\in\uZ$ such that $\big(0\oplus F_j^{11}\big)\big(0\ominus F_j^{22}\big)$ is always even and $F_j^{12}F_j^{21}$ odd. Then all the elements on the r.h.s.\ of~\eqref{ncudsG:G}, \eqref{ncudsG:H} are signed and $G_j$, $H_j$ are completely determined. Fig.~\ref{ncudsG:fig:1sol} shows $W=H_1$ with
\[
A=B=-1, \qquad C_1=\bpm \ominus(-7) & -8 \\ \ominus(-5) & \ominus7 \epm, \qquad P_1=2, \qquad Q_1=0.
\]

\subsubsection{2-soliton solution}

\begin{figure}[t]
\centering

\includegraphics[width=6.5cm]{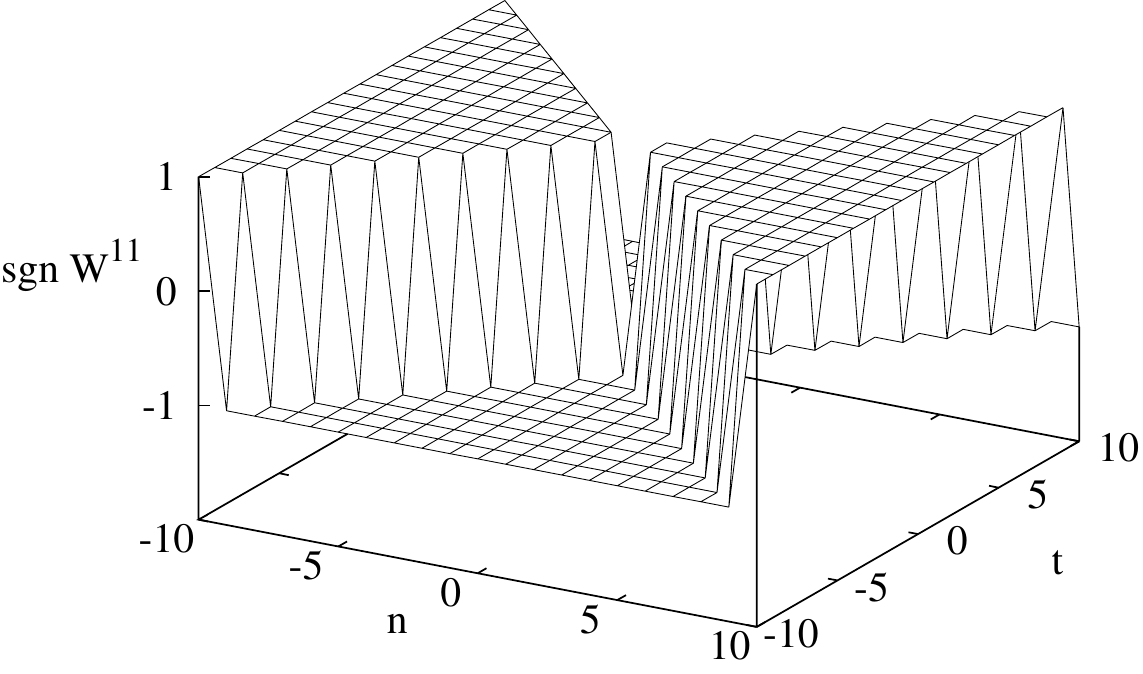} \qquad
\includegraphics[width=6.5cm]{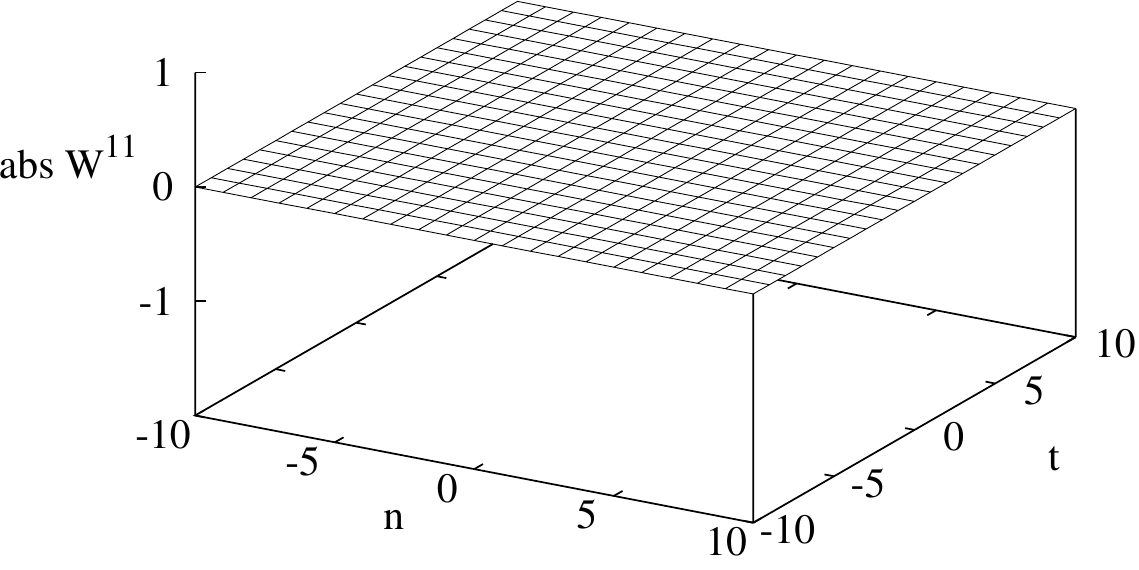}

\includegraphics[width=6.5cm]{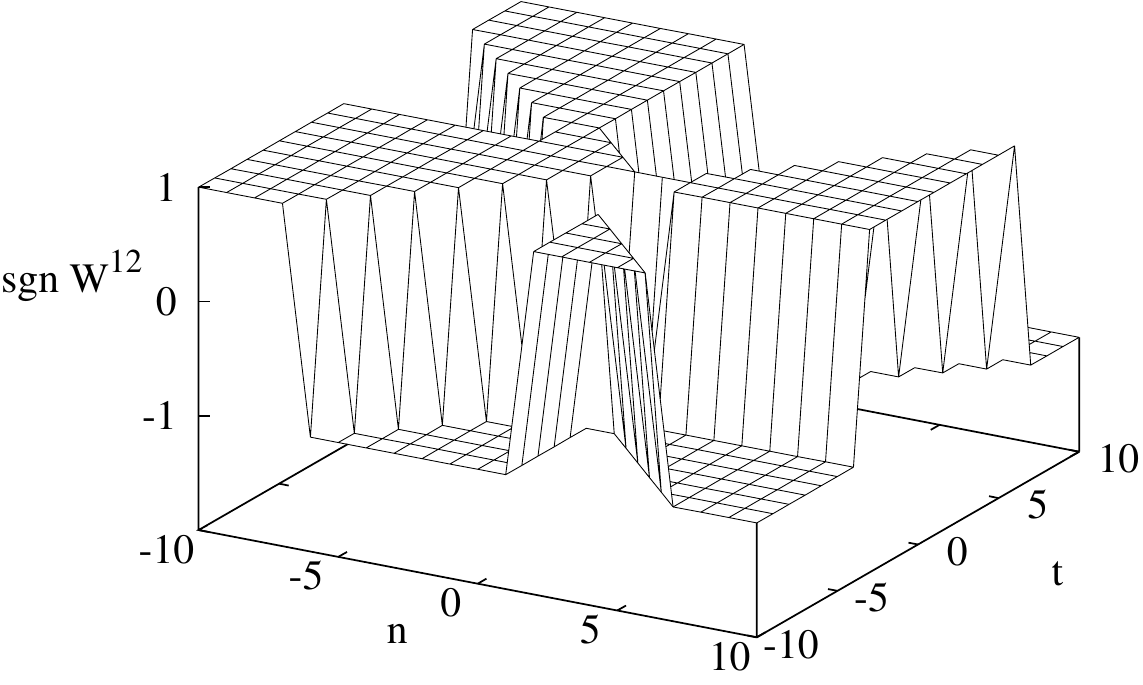} \qquad
\includegraphics[width=6.5cm]{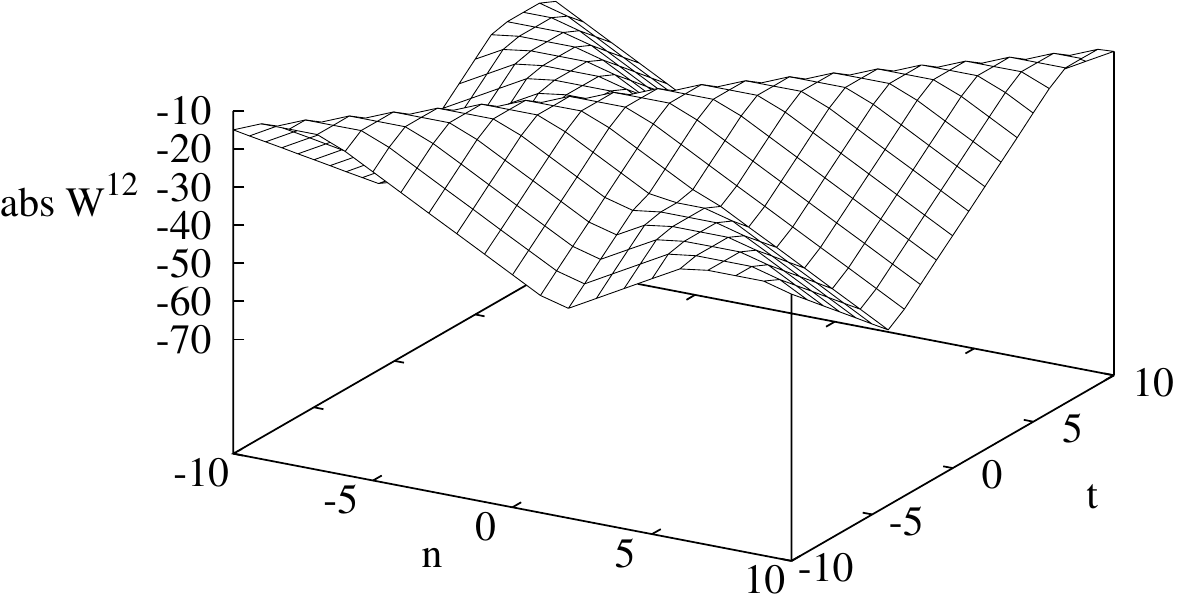}

\includegraphics[width=6.5cm]{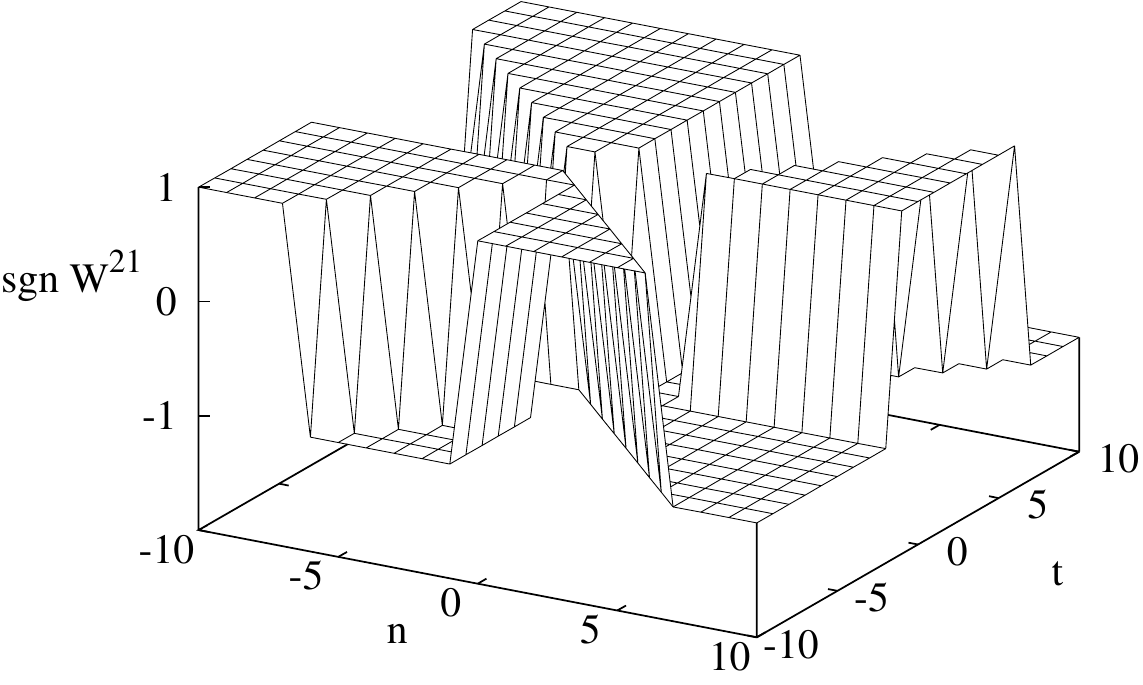} \qquad
\includegraphics[width=6.5cm]{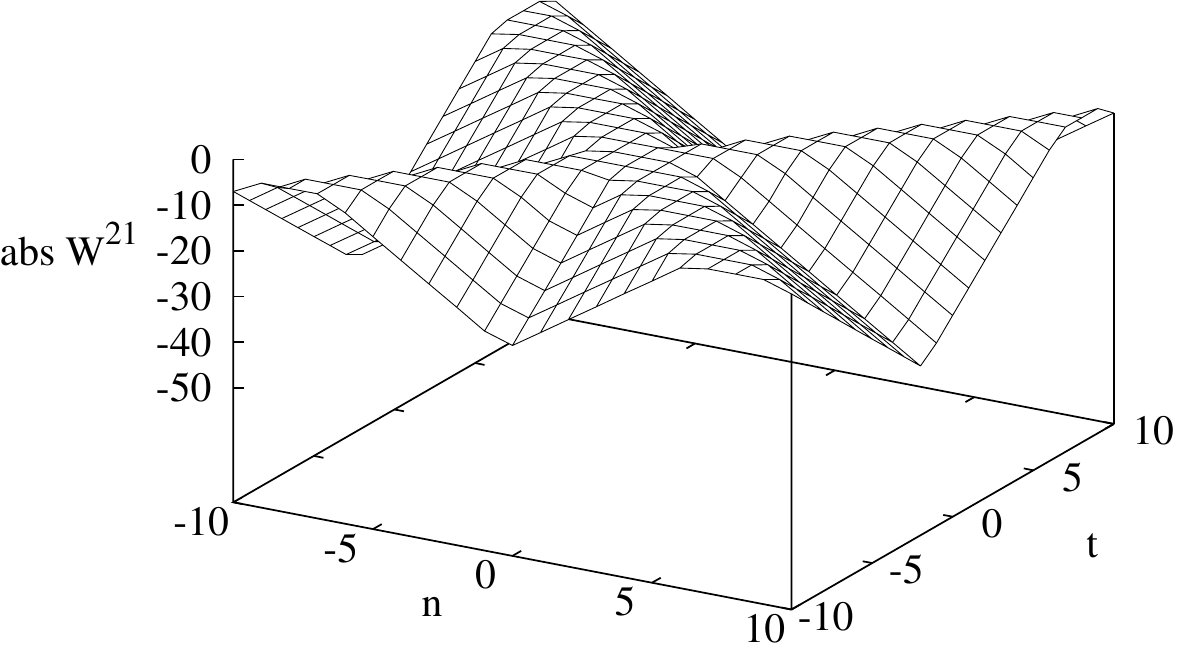}

\includegraphics[width=6.5cm]{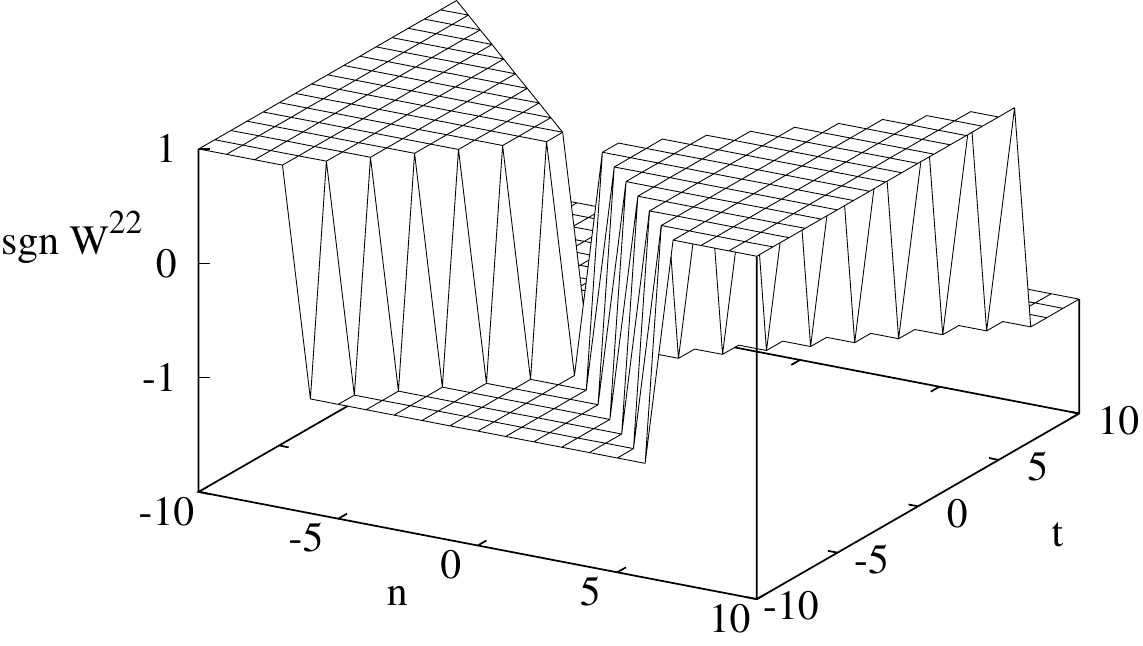}\qquad
\includegraphics[width=6.5cm]{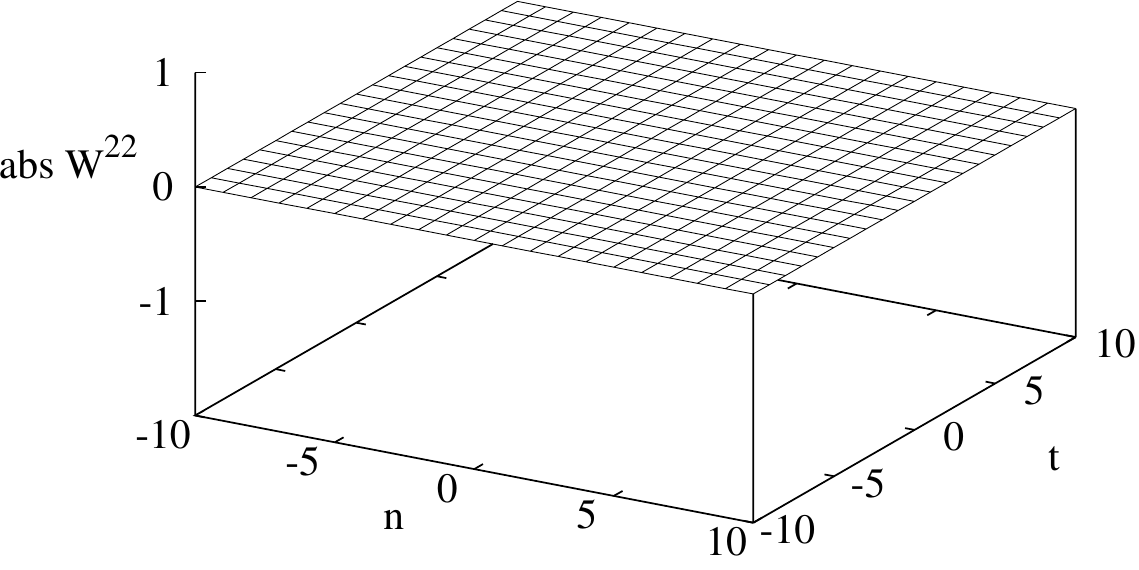}

\caption{2-soliton solution for ncudsG.}
\label{ncudsG:fig:2sol}
\end{figure}

Ultradiscretization of \eqref{ncdsG:2sol1}, \eqref{ncdsG:2sol2} gives
%\begin{subequations}
\begin{gather*}
W \bals\lt(L_2G_2\ominus L_1G_1\rt)\lt(L_2G_1\ominus L_1G_2\rt)^{-1}, \\
V \bals\lt(L_2H_2\ominus L_1H_1\rt)\lt(L_2H_1\ominus L_1H_2\rt)^{-1},
\end{gather*}
%\end{subequations}
where
\begin{gather*}
\lambda_j\ud L_j.
\end{gather*}
In order to determine the value of $L_j$, we examine the relation
\begin{gather*}
\lambda_j=\frac{p_j-1}{a(p_j+1)}=\frac{b(q_j+1)}{q_j-1}.
\end{gather*}
By the dispersion relation~\eqref{ncudsG:disprel}, we have $P_j=0$ or $Q_j=0$. When $Q_j=0$, $q_j$ behaves like a constant with regard to the ultradiscretization parameter $s$ and $p_j$ cannot behave like one. Therefore, we have
%\begin{subequations}
\begin{gather*}
L_j=\frac{P_j\ominus 0}{A(P_j\oplus 0)}=\begin{cases}A^{-1}, &   \uabs{P_j}>0 , \\ \ominus A^{-1}, & \uabs{P_j}<0 .
\end{cases}
\end{gather*}
Similarly, when $P_j=0$, we have
\begin{gather*}
L_j=\frac{B(Q_j\oplus 0)}{Q_j\ominus 0}=\begin{cases}B, & \uabs{Q_j}>0 , \\ \ominus B, &  \uabs{Q_j}<0 .\end{cases}
\end{gather*}
%\end{subequations}
If we choose $P_2=Q_1=0$, we have $\uabs{L_1}>\uabs{L_2}$ and thus
%\begin{subequations}
\begin{gather*}
W\bals G_1G_2^{-1}, \qquad V\bals H_1H_2^{-1}.
\end{gather*}
Similarly, if $P_1=Q_2=0$,
\begin{gather*}
W\bals G_2G_1^{-1}, \qquad V\bals H_2H_1^{-1}.
\end{gather*}
%\end{subequations}
Fig.~\ref{ncudsG:fig:2sol} shows behavior of $W=G_1G_2^{-1}$ with parameters $A=B=-1,$
\[
C_1=\bpm \ominus2 & \ominus(-13) \\ \ominus11 & 15 \epm, \qquad C_2=\bpm -3 & \ominus(-15) \\ \ominus(-7) & \ominus(-13) \epm, \qquad P_1=Q_2=4.
\]
These are chosen so that every elements involved are signed.

\section{Conclusion and discussion}\label{sec:conclusion}

We have proposed an ultradiscrete analogue of the sine-Gordon equation and constructed signed 1-soliton and 2-soliton solutions utilizing~$\uR$. The traveling-wave, kink-antikink, and kink-kink solutions, which contain ultradiscrete complex numbers, do exist and their correspondence to those for the discrete sine-Gordon equation is quite clear. When the range of solutions are restricted to $\uR$, even deterministic time evolution is possible.

As stated in \secref{sec:intro}, another ultradiscretization of the sine-Gordon equation has been given by Isojima et al.~\cite{Isojima2004,IS2009}. There, only $\tau$ is ultradiscretized, no complex solutions are dealt with, and time evolution is not possible. Our formulation looks better in these respects. Also, ultradiscretization via the parametrization \eqref{udsG:ud_param} can be considered as another aspect of continuum limit, but ultradiscretization in~\cite{Isojima2004} cannot since they choose a parametrization such that $\delta\to\pm1$.

We have also proposed a noncommutative discrete analogue of the sine-Gordon equation and revealed its relation to other integrable systems including the noncommutative discrete KP equation. Also, multisoliton solutions are constructed by a repeated application of Darboux transformations. And f\/inally, a noncommutative ultradiscrete analogue of the sine-Gordon equation and its signed 1-soliton and 2-soliton solutions are derived by ultradiscretization with~$\uR$.

\appendix

\section{Symmetrized max-plus algebra and ultradiscretization}\label{apd:ud}

We make extensive use of the symmetrized max-plus algebra $\uR$ in the main part of the paper. Therefore we describe basic def\/initions and properties of $\uR$ here. For details, see Baccelli et al.~\cite{BCOQ}.

\subsection{Symmetrized max-plus algebra}\label{sec:smp}

\subsubsection{Pair of the max-plus algebra}

Let $\Rmax=\R\cup\{\minf\}$. $\Rmax$ has the obvious total order. Def\/ine $\oplus$ and $\otimes$ by
\begin{gather*}
x\oplus y=\max(x, y), \qquad x\otimes y=x+y
\end{gather*}
for $x,y\in\Rmax$. With these operations, $\Rmax$ becomes a commutative dioid called the max-plus algebra. The null element is $\minf$ and the unit element is $0$. We extend $\oplus$ and $\otimes$ over~$\Rmax^2$ by
\begin{gather*}
(x_1,x_2)\oplus(y_1,y_2) =(x_1\oplus y_1, x_2\oplus y_2), \\
(x_1,x_2)\otimes(y_1,y_2) =(x_1y_1\oplus x_2y_2, x_1y_2\oplus x_2y_1).
\end{gather*}
Then $\Rmax^2$ is a commutative dioid with null element $(\minf, \minf)$ and unit element $(0,\minf)$. $\Rmax$ is embedded into $\Rmax^2$ by $x\MapsTo (x,\minf)$.

Def\/ine minus sign $\ominus$ by
\begin{gather*}
\ominus(x_1,x_2)=(x_2,x_1)
\end{gather*}
for $x=(x_1,x_2)\in\Rmax^2$. We write $x\ominus y$ for $x\oplus (\ominus y)$, which is regarded as subtraction. Def\/ine absolute value $\uabs{\,}:\Rmax^2\to\Rmax$ by
\begin{gather*}
\uabs{(x_1,x_2)}=x_1\oplus x_2.
\end{gather*}
Def\/ine balance operator $\bal{}$ by
\begin{gather*}
\bal{(x_1,x_2)}=(x_1,x_2)\ominus(x_1,x_2)=(x_1\oplus x_2,x_1\oplus x_2).
\end{gather*}

\subsubsection{Symmetrized max-plus algebra}

It is natural to consider the balance relation $\bals$ def\/ined by
\begin{gather*}
(x_1,x_2)\bals(y_1,y_2) \ \EquivTo \ x_1\oplus y_2=x_2\oplus y_1.
\end{gather*}
$\bals$ is ref\/lexive and symmetric, but \textit{not} transitive. Therefore, we introduce another relation $\Rel$ def\/ined by
\begin{gather*}
(x_1,x_2) \Rel (y_1,y_2) \ \EquivTo \ \begin{cases}
(x_1,x_2)\bals (y_1,y_2), &  \mbox{when $x_1\ne x_2$ and $y_1\ne y_2$}, \\
(x_1,x_2)=(y_1,y_2), &  \mbox{otherwise}.
\end{cases}
\end{gather*}
$\Rel$ is an equivalence relation compatible with the operations $\oplus$, $\otimes$, $\ominus$, $\uabs{\,}$, $\bal{}$, and the relation $\bals$. Thus, we can def\/ine the quotient structure
\begin{gather*}
\uR=\Rmax^2/\!\Rel.
\end{gather*}
This is called the symmetrized max-plus algebra~\cite{MaxPlus, BCOQ}. Usually this is denoted by $\mathbb{S}$, but we use $\uR$ to imply it is somehow a whole set of \textit{ultradiscrete real numbers}. We will also introduce $\uZ, \uC$ later.

\begin{Proposition}
We have three kinds of equivalence classes:
\begin{gather*}
\overline{(x,\minf)} =\{(x,t): \mbox{$t\in\Rmax$ and $x>t$}\}, \\
\overline{(\minf,x)} =\{(t,x): \mbox{$t\in\Rmax$ and $t < x$}\}, \\
\overline{(x,x)} =\{(x,x)\}.
\end{gather*}
\end{Proposition}

$\Rmax$ is embedded into $\uR$ by $x\MapsTo\overline{(x,\minf)}$. Def\/ine
\begin{gather*}
\ominus\Rmax=\big\{\overline{(\minf,x)}:x\in\Rmax\big\}, \qquad \bal{\Rmax}=\big\{\overline{(x,x)}:x\in\Rmax\big\}.
\end{gather*}
Then $\uR$ has a decomposition
\begin{gather*}
\uR=\Rmax\cup\ominus\Rmax\cup\bal{\Rmax},
\end{gather*}
and $\overline{(\minf,\minf)}$ is the only element which belongs to any two of the three sets. Thus, we simply write $x$ for $\overline{(x,\minf)}$, $\ominus x$ for $\overline{(\minf,x)}$, and $\bal{x}$ for $\overline{(x,x)}$.

Def\/ine sign function $\sgn x$ by
\begin{gather*}
\sgn x=\begin{cases}0, & x\in\R , \\ \bal0, & x\in\bal{\Rmax} , \\ \ominus 0, & x\in\ominus\R .\end{cases}
\end{gather*}
$x\in\uR$ is said to be positive if $\sgn x=0$, negative if $\sgn x=\ominus0$, and balanced if $\sgn x=\bal0$.

Def\/ine $\uR^\vee=\Rmax\cup\ominus\Rmax$. $x\in\uR$ is said to be signed if $x\in\uR^\vee$.

\begin{Proposition}
Let $\uR^\otimes$ denote the whole set of invertible elements in $\uR$. Then,
\begin{gather*}
\uR^\otimes=\uR^\vee\setminus\{\minf\}=\uR\setminus\bal{\Rmax}.
\end{gather*}
\end{Proposition}

Def\/ine $\uZ,\uZ^\vee\subset\uR$ by
\begin{gather*}
\uZ=\{\minf\}\cup\Z\cup\ominus\Z\cup\bal{\Z}, \qquad \uZ^\vee=\uZ\cap\uR^\vee
\end{gather*}
with obvious notations. $\uZ$ is a subdioid of $\uR$ and can be regarded as a whole set of \textit{ultradiscrete integers}. $x\in\uZ$ is said to be even if $\uabs{x}$ is even, odd if $\uabs{x}$ is odd. We do not def\/ine whether $\minf$ is even or odd. We have of course
\begin{gather*}
\uZ^\otimes=\uZ^\vee\setminus\{\minf\}.
\end{gather*}

\subsubsection{Properties of balance relation}

We make much use of $\bals$, rather than $\Rel$, since members of $\bal{\Rmax}$ can be regarded as a kind of null elements by virtue of the following proposition.

\begin{Proposition}
For any $x\in\uR$,
\begin{gather*}
x\bals\minf \ \EquivTo \ x\in\bal{\Rmax}.
\end{gather*}
\end{Proposition}

\begin{Proposition}\label{smp:bal:prop2}
For any $x\in\uR$ and $t\in\Rmax$,
\begin{gather*}
\mbox{$x\bals\bal{t}$ and $x\not\in\bal{\Rmax}$} \ \EquivTo \ \uabs{x}\le t.
\end{gather*}
\end{Proposition}

\begin{Proposition}
For any $x,y\in\uR$, we have
\begin{gather*}
x\bals y \ \EquivTo \ x\ominus y\bals\minf.
\end{gather*}
\end{Proposition}

\begin{Proposition}
For any $x,y,z,w\in\uR$, we have
\begin{gather*}
\mbox{$x\bals y$ and $z\bals w$} \ \LeadsTo \ x\oplus z\bals y\oplus w, \\ %\label{bals:prop2} \\
x\bals y \ \LeadsTo \ xz\bals yz. %\label{bals:prop3}
\end{gather*}
\end{Proposition}

\begin{Proposition}[weak substitution]
\begin{gather*}
\mbox{$x\bals y$, $cy\bals z$, and $y\in\uR^\vee$} \ \LeadsTo \ cx\bals z.
\end{gather*}
\end{Proposition}

\begin{Corollary}[weak transitivity]
\begin{gather*}
\mbox{$x\bals y$, $y\bals z$, and $y\in\uR^\vee$}\ \LeadsTo \ x\bals z.
\end{gather*}
\end{Corollary}

\begin{Proposition}[reduction of balances]\label{smp:bal:red}
\begin{gather*}
\mbox{$x\bals y$ and $x,y\in\uR^\vee$} \ \LeadsTo \ x=y.
\end{gather*}
\end{Proposition}

\subsubsection{Matrices and determinants}

Let $\uMat(N,\uR)$ denote the whole set of $N\times N$ matrices over $\uR$. Def\/ine addition $\oplus$ by
\begin{gather*}
(a_{ij})\oplus(b_{ij})=(a_{ij}\oplus b_{ij})
\end{gather*}
and multiplication $\otimes$ by
\begin{gather*}
(a_{ij})\otimes(b_{ij})=(c_{ij}), \qquad c_{ij}=\bigoplus_k a_{ik}\otimes b_{kj}
\end{gather*}
for any $(a_{ij}), (b_{ij})\in\uMat(N,\uR)$. Then $\uMat(N,\uR)$ becomes a dioid, noncommutative when $N>1$. $\ominus$, $\bal{}$, and $\bals$ are of course def\/ined by
\begin{gather*}
\ominus(a_{ij})=(\ominus a_{ij}), \qquad \bal{(a_{ij})}=(\bal{a_{ij}}), \\
(a_{ij})\bals(b_{ij}) \ \EquivTo \ \text{$a_{ij}\bals b_{ij}$ for any $i$, $j$}
\end{gather*}
respectively. $(a_{ij})\in\uMat(N,\uR)$ is said to be signed if all the elements are signed. The whole set of signed elements in $\uMat(N,\uR)$ is denoted by $\uMat(N,\uR)^\vee$. $\uR$ is embedded into $\uMat(N,\uR)$ by
\begin{gather*}
x\MapsTo\begin{pmat}
x      & \minf  & \cdots & \minf  \\
\minf  & x      & \ddots & \vdots \\
\vdots & \ddots & \ddots & \minf  \\
\minf  & \cdots & \minf  & x
\end{pmat}.
\end{gather*}

For any permutation $\sigma\in S_N$, def\/ine $\sgn(\sigma)$ by
\begin{gather*}
\sgn(\sigma)=\begin{cases}0, & \text{when $\sigma$ is even,} \\ \ominus 0, &  \text{when $\sigma$ is odd.}\end{cases}
\end{gather*}
And def\/ine the determinant of a matrix $A=(a_{ij})\in\uMat(N,\uR)$ by
\begin{gather*}
\det A=\bigoplus_\sigma\sgn(\sigma)\bigotimes_i a_{i\sigma(i)}.
\end{gather*}
$\det A$ is also denoted by $|A|$ or $|a_{ij}|$.

\begin{Proposition}
\begin{gather*}
|\tr A|=|A|
\end{gather*}
where $\tr A$ denotes transposition of $A$.
\end{Proposition}

\begin{Proposition}
\begin{gather*}
 \begin{vmat} v_1 & \cdots & \lambda v_j\oplus u & \cdots & v_N\end{vmat}
  =\lambda\begin{vmat} v_1 & \cdots & v_j & \cdots & v_N\end{vmat}\oplus\begin{vmat} v_1 & \cdots & u & \cdots & v_N\end{vmat}
\end{gather*}
where $v_j=\tr(a_{1j},\ldots,a_{Nj})$ and $u=\tr(u_1,\ldots,u_N)$.
\end{Proposition}

\begin{Proposition}
For any permutation $\sigma\in S_N$,
\begin{gather*}
|a_{i\sigma(j)}|=\sgn(\sigma)|a_{ij}|.
\end{gather*}
\end{Proposition}

\begin{Corollary} If $v_j=v_k$ for some $j\ne k$, then
\begin{gather*}
\begin{vmat} v_1 & \cdots & v_N\end{vmat}\bals\minf.
\end{gather*}
\end{Corollary}

Let $\cof_{ij}(A)$ denote the cofactor of $a_{ij}$ in $|A|$, which by def\/inition satisf\/ies
\begin{gather*}
|A|=\bigoplus_i a_{ij}\otimes\cof_{ij}(A)
\end{gather*}
for any $j$. Def\/ine the adjacent matrix of $A$ by
\begin{gather*}
\adj A=(b_{ij}), \qquad b_{ij}=\cof_{ji}(A).
\end{gather*}

\begin{Theorem}
\begin{gather*}
A\otimes\adj A\bals|A|, \qquad \adj A\otimes A\bals|A|.
\end{gather*}
\end{Theorem}

If $|A|\in\uR^\otimes$, def\/ine $A^{-1}$ by
\begin{gather*}
A^{-1}=|A|^{-1}\adj A.
\end{gather*}
This is not a multiplicative inverse in general, but plays a similar role with regard to $\bals$. Therefore we use the notation $A^{-1}$.

\subsubsection{Ultradiscrete complex numbers}

It is well known that we can construct complex numbers by $2\times 2$ real matrices, using
\[
i=\bpm 0 & -1 \\ 1 & 0 \epm
\]
as the imaginary unit. Here we try to construct \textit{ultradiscrete complex numbers} in a similar way.

Let $\uMat(N,\uR)$ denote the algebra of $N\times N$ matrices whose elements are in $\uR$. Def\/ine $I\in\uMat(2,\uR)$ by
\begin{gather*}
I=\pM{\minf}{\ominus 0}{0}{\minf}.
\end{gather*}
We have
\begin{gather*}
I^2=\pM{\minf}{\ominus 0}{0}{\minf}\pM{\minf}{\ominus 0}{0}{\minf}=\pM{\ominus 0}{\minf}{\minf}{\ominus 0}=\ominus 0.
\end{gather*}
Def\/ine $\uC\subset\uMat(2,\uR)$ by
\begin{gather*}
\uC=\{x\oplus yI\;|\;x,y\in\uR\}.
\end{gather*}

\begin{Proposition}
$\uC$ is a commutative subdioid of $\uMat(2,\uR)$.
\end{Proposition}
\begin{proof}
Obviously $\uC$ includes $\minf$ and $0$. For any $a\oplus bI, c\oplus dI\in\uC$, we have
\begin{gather*}
(a\oplus bI)\oplus(c\oplus dI) =(a\oplus c)\oplus(b\oplus d)I\in\uC, \\
(a\oplus bI)\otimes(c\oplus dI) =(ac\ominus bd)\oplus(ad\oplus bc)I\in\uC.
\end{gather*}
And $\uC$ is commutative because $I^0$ and $I^1$ are commutative.
\end{proof}

When $z\in\uC$ is expressed as $z=x+yI$ where $x,y\in\uR$, we write
\begin{gather*}
\uRe z=x, \qquad \uIm=y.
\end{gather*}
The whole set of signed elements of $\uC$ is denoted by $\uC^\vee$.

If $\det(x\oplus yI)=x^2\oplus y^2\in\uR^\otimes$, we have
\[
(x\oplus yI)^{-1}=\frac{x\ominus yI}{x^2\oplus y^2}
\]
and
\[
(x\oplus yI)(x\oplus yI)^{-1}=0\oplus\frac{\bal{(xy)}}{x^2\oplus y^2}I\bals 0.
\]

\subsection{Ultradiscretization with negative numbers}

Ultradiscretization with negative numbers is presented in De Schutter et al.~\cite{DeSchutter}. Here we reformulate it in a similar, but more convenient form for our purpose.

Let $f(s)$ and $g(s)$ be real functions. We say $f(s)$ is asymptotically equivalent to $g(s)$ if there exists a real number $s_0$ such that $g(s)\ne0$ for any $s>s_0$ and
\begin{gather*}
\lim_{s\to\infty}\frac{f(s)}{g(s)}=1.
\end{gather*}
We also say $f(s)$ is asymptotically equivalent to $0$ if there exists a real number $s_1$ such that $f(s)=0$ for any $s>s_1$. Asymptotic equivalence is an equivalence relation and denoted by $f(s)\sim g(s)$.

We are interested in asymptotic equivalence to exponential functions. If
\[
f(s)\sim\mu_F e^{\wt{F}s}, \qquad  \mu_F\in\R^\times, \qquad \wt F\in\R ,
\]
we write
\begin{gather*}
f(s)\ud F, \qquad F=S(\mu_F)\otimes\wt{F}\in\uR^\vee,
\end{gather*}
where
\begin{gather*}
S(\mu)=\begin{cases}0, & \mu>0 , \\ \ominus0, & \mu<0 .\end{cases}
\end{gather*}
We regard $0\sim\mu e^{(\minf)s}$ for some $\mu\in\R^\times$ and $0\ud\minf$ as a convention.

It is very important here to notice that $\mu_F$ is \textit{not} restricted to positive numbers, unlike the usual ultradiscretization procedure.

\begin{Proposition}[ultradiscretization of addition]\label{ud:add}
Let $f(s),g_1(s),\ldots,g_n(s)$ be real functions satisfying
\[
f(s)=\sum_{k=1}^n g_k(s)
\]
and
\[
f(s)\ud F, \qquad g_k(s)\ud G_k.
\]
Then,
\[
F\bals\bigoplus_{k=1}^n G_k.
\]
\end{Proposition}

\begin{Remark}
$f(s)\ud F$ and $g(s)\ud G$ do not imply $f(s)+g(s)\ud F\oplus G$ because $f(s)+g(s)$ might be no longer asymptotically equivalent to exponential functions. But if $f(s)$, $g(s)$ can be expressed by power series in $\delta=\mu_De^{\wt Ds}$ where $\wt D<0$, this is not a problem because $f(s)+g(s)$ can also be expressed by a power series in $\delta$.
\end{Remark}

\begin{Proposition}[ultradiscretization of multiplication]\label{ud:mul}
Let $f(s)$, $g(s)$, $h(s)$ be real functions satisfying
\[
f(s)=g(s)h(s)
\]
and
\[
g(s)\ud G, \qquad h(s)\ud H.
\]
Then,
\[
f(s)\ud F=G\otimes H.
\]
\end{Proposition}

\begin{Corollary}[ultradiscretization of polynomials]\label{ud:polynomial}
Let real functions $f(s)$, $g_{kl}(s)$ satisfy
\[
f(s)=\sum_{k=1}^n\prod_{l=1}^{m_k}g_{kl}(s)
\]
and
\[
f(s)\ud F, \qquad g_{kl}(s)\ud G_{kl}.
\]
Then,
\[
F\bals\bigoplus_{k=1}^n\bigotimes_{l=1}^{m_k}G_{kl}.
\]
\end{Corollary}

\subsection{Ultradiscretization of matrices and complex numbers}

We also reformulate ultradiscretization of matrices in~\cite{DeSchutter}. Extension to complex numbers is straightforward.

Consider a matrix-valued function $f(s)=(f_{ij}(s)):\R\to\Mat(N,\R)$. If
\[
f_{ij}(s)\ud F_{ij},
\]
we write
\begin{gather*}
f(s)\ud F=(F_{ij})\in\uMat(N,\uR).
\end{gather*}
This is a componentwise property; there is no exponential functions of matrices.

\begin{Proposition}\label{ud:matrix}
Let matrix-valued functions $f(s)$, $g_{kl}(s)$ satisfy
\[
f(s)=\sum_{k=1}^n\prod_{l=1}^{m_k}g_{kl}(s)
\]
and
\[
f(s)\ud F, \qquad g_{kl}(s)\ud G_{kl}.
\]
Then,
\begin{gather*}
F\bals\bigoplus_{k=1}^n\bigotimes_{l=1}^{m_k}G_{kl}.
\end{gather*}
\end{Proposition}

Considering $2\times 2$-matrix construction of complex numbers, we have
\[
i=\pM{0}{-1}{1}{0} \ud I=\pM{\minf}{\ominus 0}{0}{\minf}.
\]
Let $f(s)=u(s)+v(s)i$ where $u(s)$, $v(s)$ are real functions. If
\[
u(s)\ud U, \qquad v(s)\ud V,
\]
we have of course
\begin{gather*}
f(s)\ud F=U\oplus VI\in\uC.
\end{gather*}

\begin{Proposition}
Let complex-valued functions $f(s)$, $g_{kl}(s)$ satisfy
\[
f(s)=\sum_{k=1}^n\prod_{l=1}^{m_k}g_{kl}(s)
\]
and
\[
f(s)\ud F, \qquad g_{kl}(s)\ud G_{kl}.
\]
Then,
\[
F\bals\bigoplus_{k=1}^n\bigotimes_{l=1}^{m_k}G_{kl}.
\]
\end{Proposition}

\section{Quasideterminants}\label{apd:qdet}

Quasideterminants~\cite{GGRW} are noncommutative extension of determinants, or, more precisely, determinants divided by cofactors. Here we describe the def\/inition and some properties required for \thmref{ncdsG:rDt}. See~\cite{GGRW} for more detail.

Let $R$ be a ring and $\Mat(N,R)$ be the whole set of $N\times N$ matrices over $R$. $R$ is not commutative in general. For any $(a_{ij}),(b_{ij})\in\Mat(N,R)$, def\/ine addition by
\begin{gather*}
(a_{ij})+(b_{ij})=(a_{ij}+b_{ij})
\end{gather*}
and multiplication by
\begin{gather*}
(a_{ij})(b_{ij})=(c_{ij}), \qquad c_{ij}=\sum_{k=1}^N a_{ik}b_{kj}.
\end{gather*}
Ordering of multiplication is important here.

For any $A=(a_{ij})\in\Mat(N, R)$, def\/ine the $(p,q)$-th quasideterminant $|A|_{pq}$ by
\begin{gather*}
|A|_{pq}=a_{pq}-r_p^q\lt(A^{pq}\rt)^{-1}c_q^p,
\end{gather*}
where $r_p^q$ is the $p$-th row of $A$ without the $q$-th element, $c_q^p$ is the $q$-th column of $A$ without the $p$-th element, and $A^{pq}$ is $A$ without the $p$-th row and the $q$-th column. $|A|_{pq}$ is also written as
\begin{gather*}
|A|_{pq}=\begin{vmatrix}
a_{11} & \cdots & a_{1N} \\
\vdots & \fbox{$a_{pq}$} & \vdots \\
a_{N1} & \cdots & a_{NN}
\end{vmatrix}.
\end{gather*}
For example, we have
\[
\vM{{}\fbox{$a_{11}$}}{a_{12}}{a_{21}}{a_{22}}=a_{11}-a_{12}a_{22}^{-1}a_{21}, \qquad \vM{a_{11}}{{}\fbox{$a_{12}$}}{a_{21}}{a_{22}}=a_{12}-a_{11}a_{21}^{-1}a_{22}
\]
and
\begin{gather*}
\begin{vmatrix}
\fbox{$a_{11}$} & a_{12} & a_{13} \\
a_{21} & a_{22} & a_{23} \\
a_{31} & a_{32} & a_{33}
\end{vmatrix} =a_{11}-\bpm a_{12} & a_{13} \epm \bpm a_{22} & a_{23} \\ a_{32} & a_{33} \epm^{-1} \bpm a_{21} \\ a_{31} \epm \\
\hphantom{\begin{vmatrix}
\fbox{$a_{11}$} & a_{12} & a_{13} \\
a_{21} & a_{22} & a_{23} \\
a_{31} & a_{32} & a_{33}
\end{vmatrix}}{}
=a_{11}-a_{12}\left(a_{22}-a_{23}a_{33}^{-1}a_{32}\right)^{-1}a_{21}  -a_{13}\left(a_{23}-a_{22}a_{32}^{-1}a_{33}\right)^{-1}a_{21} \\
\hphantom{\begin{vmatrix}
\fbox{$a_{11}$} & a_{12} & a_{13} \\
a_{21} & a_{22} & a_{23} \\
a_{31} & a_{32} & a_{33}
\end{vmatrix}=}{}
 -a_{12}\left(a_{32}-a_{33}a_{23}^{-1}a_{22}\right)^{-1}a_{31}
 -a_{13}\left(a_{33}-a_{32}a_{22}^{-1}a_{23}\right)^{-1}a_{31}.
\end{gather*}

\begin{Proposition}
If we write $A^{-1}=(b_{ij})$, we have
\begin{gather*}
b_{ij}=|A|_{ji}^{-1}.
\end{gather*}
\end{Proposition}

\begin{Proposition}\label{qdet:perm}
Quasideterminants are invariant under row and column permutations. $($If the row or column contains the  {box}, it is moved together.$)$
\end{Proposition}

\begin{Proposition}[homological relations]\label{qdet:homrel}
For $p_1\ne p_2$, $q_1\ne q_2$, $i\ne p$, $j\ne q$, we have the row homological relation
\begin{gather*}
|A|_{pq_1}\lt|A^{pq_2}\rt|_{iq_1}^{-1}+|A|_{pq_2}\lt|A^{pq_1}\rt|_{iq_2}^{-1}=0
\end{gather*}
and the column homological relation
\begin{gather*}
\lt|A^{p_2q}\rt|_{p_1j}^{-1}|A|_{p_1q}+\lt|A^{p_1q}\rt|_{p_2j}^{-1}|A|_{p_2q}=0.
\end{gather*}
\end{Proposition}

\begin{Proposition}[Sylvester's identity]\label{qdet:Sylvester}
For any $A=(a_{ij})\in\Mat(N,R)$, define $(N-k)\times(N-k)$ matrix $A_0$ by
\[
A_0=(a_{ij}), \qquad  k+1\le i,j\le N
\]
and $k\times k$ matrix $C$ by
\[
C=(c_{ij}), \qquad c_{ij}=\bvm \fbox{$a_{ij}$} & a_{i(k+1)} & \cdots & a_{iN} \\ a_{(k+1)j} & & & \\ \vdots & & A_0 & \\ a_{Nj} & & & \evm.
\]
Then
\begin{gather*}
|A|_{pq}=|C|_{pq}.
\end{gather*}
\end{Proposition}

\subsection*{Acknowledgements}

The author would like to express his gratitude to Professor Tetsuji Tokihiro, who provided precise advices with a f\/ine prospect. The author is also grateful to Professor Ralph Willox, who provided helpful comments for ref\/ining the results. In addition, the author thanks the anonymous referees for carefully reading the paper and giving many suggestions.

\pdfbookmark[1]{References}{ref}
\LastPageEnding

\end{document}